\documentclass[preprint]{aastex}
\usepackage{graphicx, natbib, txfonts}
\bibpunct{(}{)}{;}{a}{}{,}

\newif\ifgras

\grasfalse

\shorttitle{\it Scanamorphos}
\shortauthors{H. Roussel}

\begin{document}

\title{Scanamorphos\,\thanks{This name is a portmanteau word, formed from ``scan''
and ``anamorphosis'' (reversible transformation of an image by a mathematical or optical
operator).} ~: a map-making software for Herschel\,\thanks{Herschel is an ESA space
observatory with science instruments provided by European-led Principal Investigator
consortia and with important participation from NASA.} ~ and similar scanning bolometer arrays}

\author{H. Roussel}

\affil{Institut d'Astrophysique de Paris, Universit\'e Pierre et Marie Curie (UPMC), CNRS (UMR 7095),
75014 Paris, France\label{inst1}}

\email{roussel@iap.fr}

\begin{abstract}
{\it Scanamorphos} is one of the public softwares available to post-process scan observations
performed with the Herschel photometer arrays. This post-processing mainly consists in
subtracting the total low-frequency noise (both its thermal and non-thermal components),
masking
high-frequency artefacts such as
cosmic ray hits, and projecting the data onto a map. Although it was developed
for Herschel, it is also applicable with minimal adjustment to scan observations made
with some other imaging arrays subjected to low-frequency noise,
provided they entail sufficient redundancy; it was successfully
applied to P-Art\'emis, an instrument operating on the APEX telescope. Contrary to
matrix-inversion softwares and high-pass filters,
{\it Scanamorphos} does not assume any particular noise model, and does not apply any Fourier-space
filtering to the data, but is an empirical tool using purely the redundancy built in the observations --
taking advantage of the fact that each portion of the sky is sampled at multiple times
by multiple bolometers. It is an interactive software in the sense that the user is allowed
to optionally visualize and control results at each intermediate step, but the processing
is fully automated. This paper describes the principles and algorithm of {\it Scanamorphos}
and presents several examples of application.
\end{abstract}
 
\keywords{Astronomical instrumentation; Data analysis and techniques}

\noindent
{\footnotesize withdrawn from A\&A June 29, 2012 / submitted to PASP July 9 /
report 1 received August 17 / re-submitted August 31 /} \\
\hspace*{0.7cm} {\footnotesize referee 1 abandons / report 2 received November 19 / re-submitted November 30 / report 3 received February 19, 2013 /} \\
\hspace*{0.7cm} {\footnotesize re-submitted April 3 / report 4 received July 11 / re-submitted July 26 / accepted July 27} \\
\hspace*{0.7cm} {\footnotesize I have addressed 7 referee reports in total
(including those from A\&A) !} \\

\section{Introduction}

The software described here was developed with the aim to process scan observations
made with the photometers of the Herschel space telescope \citep{Pilbratt10},
subunits of the PACS \citep{Poglitsch10} and SPIRE instruments \citep{Griffin10, Swinyard10},
but has broader applicability. Due to the mapping efficiency of the scan mode
(also called on-the-fly mapping in ground-based astronomy)
for fields larger than a few arcminutes,
as opposed to the raster and jiggle modes, this is the preferred acquisition mode for
extended Galactic regions, nearby galaxies and high-redshift surveys, in particular.
Hence, scans are used extensively and represent a large fraction of the total observing
time of Herschel, and the corresponding observing templates have been commissioned
first.\footnote{
They are described in the observers' manuals at this URL:
\url{http://herschel.esac.esa.int/Documentation.shtml}\,.
}
New observing templates based on very small scans have then been introduced
after the Science Demonstration Phase to observe compact sources, in place of those based
on chop-nod observations, that were found less sensitive.

During the course of a scan observation, the
{\ifgras \bf \fi
astronomical signal received by
}
each bolometer is
superimposed on a drift, occurring on timescales much longer than the sampling time
interval. For small fields of view, the decontamination
can be accomplished by chopping and nodding at frequencies
higher than those characteristic of the drifts. This technique is not applicable to
scan observations, and is appropriate only when reference positions devoid of
emission from the target or background sources can be defined.
The dual-beam scanning technique \citep{Emerson79}, with a processing
improved by \citet{Motte07} to better remove atmospheric fluctuations, can be
exploited to subtract only the thermal part of the drift, leaving the data contaminated
by the non-thermal instrumental drifts. Simple scans with multiple detectors
have the advantage
of automatically providing the redundancy necessary to extract all the drift components,
while spending all the time on source. A single scan already provides some level of redundancy,
but most frequently at least two non-parallel scans are combined, ensuring that the
drifts are well characterized \citep{Waskett07, Kovacs08a},
that transient signals can be suppressed, and that the
desired sensitivity is achieved.

Low-frequency noise manifests itself most clearly by the presence of stripes in maps,
oriented parallel to the scanning direction. But its removal is far from being only
a cosmetic improvement, since this noise also decreases the sensitivity to low-brightness
sources and alters the photometric and morphological source properties. It is natural
to assign the task of removing low-frequency noise to the map-making module, because
temporal and spatial information are interdependent and both necessary to characterize
brightness drifts.

Several processing tools are available, within or outside the Herschel pipeline
\citep[HIPE;][]{Wieprecht09, Dowell10, Ott10}.
{\it Scanamorphos} is a tool offered to the general user, coded in the 
Interactive Data Language (IDL),
mainly for ease of programming and for the efficiency of multi-dimensional array
processing that it affords. It is intended
as an interactive tool, which means that users have the latitude to choose some
map parameters and to visualize the results of intermediate processing steps.
The algorithm principles are relatively intuitive and thus easy to understand.
Its capabilities include the removal
of additive brightness drifts caused by low-frequency thermal noise and flicker noise,
the masking of cosmic ray hits left by the pipeline deglitching and of brightness
discontinuities caused by glitches or electronic instabilities in PACS data, the
projection of the data onto a spatial grid, and the production of the associated error
and weight maps. The correction for all other instrumental effects is made beforehand
by successive pipeline modules.

Bolometric imaging has been used extensively in the (sub)millimeter wavelength
range, both on the ground or in balloons with medium and then large arrays and
in space with single feedhorns, but
Herschel and Planck are the first space missions to host bolometer arrays.
Two different designs
have been adopted for the photometer arrays of PACS (equiped with filters centered
at 70, 100 and 160\,$\mu$m) and for the photometer arrays of SPIRE (operating at
central wavelengths of 250, 350 and 500\,$\mu$m). The PACS arrays are composed of
filled bolometer matrices, while the SPIRE arrays are made of pairs of bolometers
and conical feedhorns, providing a sparse spatial coverage in staring mode.
The instantaneous field of view is of the order of $3.5\arcmin \times 1.75\arcmin$
for PACS and $4\arcmin \times 8\arcmin$ for SPIRE.
Bolometer arrays operating in the far-infrared and submillimeter domain contain
increasing numbers of pixels and thus generate ever larger volumes of data. For
Herschel, the number of bolometers per array varies between 43 (longest-wavelength
SPIRE array) and 2048 (shortest-wavelength PACS array).
As the algorithm described here makes use of the spatial redundancy built in
the scan observations, the knowledge of the full history of all bolometers is
needed at the same time, making the code a heavy consumer of computer memory
for long observations.
Extensive tests have been performed to determine memory specifications and
execution times for given observation areas and depths (Sect.\,\ref{tests}).
For deep observations
of very large fields, such as cosmology surveys, the software has the ability
to automatically slice the field into several overlapping areas, that are
processed successively and then mosaicked.

{\it Scanamorphos} requires only minimal knowledge of the instrumental
characteristics for most of the processing, which should make it easily adaptable
to other bolometer arrays.
This was demonstrated by the results obtained with the ground-based P-Art\'emis
instrument \citep{Andre08, Minier09}, the prototype of Art\'emis \citep{Talvard06},
that will be fitted with filters centered at 200, 350 and 450\,$\mu$m. Since its
architecture is directly inspired from that of PACS, data acquired with P-Art\'emis
were instrumental in testing and refining the code before the launch of Herschel.
In all generality, however, the subtraction of the thermal low-frequency noise (atmospheric
and instrumental) may require introducing additional noise terms that were not
necessary for Herschel \citep[see][]{Kovacs08b}, and the observing strategy (scan pattern
and sampling rate) has to be adequate for the chosen algorithm.

Sections \ref{main} and \ref{algo} describe the principles and algorithm of the code.
Section \ref{simul} presents the analysis of two PACS simulations to demonstrate
the properties of {\it Scanamorphos}. Section \ref{tests} presents the results of
tests performed on flight Herschel data, providing visual examples of operation.
Appendix \ref{spire_gains} gives an overview of the calibration of gain corrections
for SPIRE, and Appendix \ref{options} of the inputs and options of {\it Scanamorphos}.
We conclude and give some practical details on the distribution in Sections\,\ref{conclu}
and \ref{distrib}.

\section{Principles and prerequisites}
\label{main}

Spatial redundancy is a general feature of scan observations. Within the portion
of the sky observed with full coverage, each position is sampled by several
bolometers, at many different epochs that are distributed over a large time interval.
This provides an efficient means to remove brightness drifts without the
need for chopped observations. Even though the redundancy built in standard
scan observations with Herschel is minimal (a tiny fraction of the field
is covered by any given bolometer), this is still sufficient to reconstruct
brightness drifts accurately, as we shall demonstrate in this paper.

Since the power spectral density of the low-frequency noise has a broad overlap
with that of real extended emission, and is therefore very difficult to determine
empirically, {\it Scanamorphos} does not use any assumption in this respect. This
is at odds with maximum-likelihood softwares based on matrix inversion,
and more specifically MADmap \citep{Ashdown07, Cantalupo10},
initially developed for cosmic microwave background experiments and implemented
in the PACS and SPIRE pipelines. MADmap deals only with noise that is uncorrelated
between detectors, assumes that the noise is Gaussian, piecewise stationary and circulant,
and needs as input the noise spectral density, considered a stable calibration product
(one per array), that has to be derived from deep observations of blank fields.
In the event that the noise spectral properties are not accurately calibrated,
vary with time or are altered by the pre-processing, then matrix-inversion
algorithms will not produce optimal results, even though they update the noise
properties by using an iterative approach.
{\ifgras \bf \fi
The effectiveness of a given class of algorithms will largely depend on the nature
of the data and the acquisition mode. For pointed observations (as opposed to
an all-sky survey),
}
a modular and empirical approach,
such as described by \cite{Kovacs08b} for the CRUSH software (see also references therein),
allows more flexibility: the signal can be weighted and masked as appropriate,
without having to interpolate missing data, and other instrumental effects can be dealt
with at optimal points along the processing chain, if necessary interleaved with the
low-frequency noise subtraction. Another benefit of a software exploiting the redundancy
is that it can be used to calibrate gains or flatfields from science observations,
as shown in Appendix\,\ref{spire_gains}.

In addition, the software ought to be able to process any data set containing
both compact sources and extended structures on arbitrary spatial scales, while
preserving their brightness distribution. Observations of nearby galaxies, or
Galactic star formation regions, generate such data sets with a wide range of
source characteristic lengths. With ground-based submillimeter instruments,
extended emission is in practice extremely difficult to restore and to separate
from the constantly varying atmospheric emission, unless more sophisticated
observing strategies are designed. Simple scan observations performed on the
ground are thus tailored to restore mostly compact or little-extended sources
(i.e. smaller than the instantaneous field of view of the detector array).
With space observatories, however, there is no such essential difference between
compact and extended emission; the only contamination of the astronomical signal
comes from the telescope and the detectors themselves, and since this noise varies
on much longer timescales than atmospheric emission, it can be separated from the
sky signal.

Herschel is not an absolute photometer, and all maps are therefore, at best, the
superposition of an accurate representation of the sky and a global offset.
The way of estimating the background level will in general depend on the exact
astronomical application envisioned, especially in complex fields such as Galactic
star formation regions, and should therefore not be determined by the map-making
software, but by the map user.

Some other algorithms implemented in the Herschel pipelines are used extensively.
PhotProject (for PACS) employs a high-pass filter. It is
therefore useable only for compact sources, or else limited to subtract only the
lowest-frequency component of the noise, so that extended emission will be filtered
as little as possible. The Destriper (for SPIRE) performs the drift subtraction
exclusively on the longest timescales, i.e. the same task as the subunit
of {\it Scanamorphos} described in Section\,\ref{baselines}.

\subsection{Important scan characteristics}

During a scan observation, the telescope slews across the sky at approximately
constant speed, while the detectors are read out at fixed time intervals,
producing, after appropriate processing and calibration, samples of brightness
as a function of time, with attached sky coordinates. We will hereafter refer
to such brightness time series as simply ``series''.

The scan pattern, i.e. the shape of the trajectory followed by the telescope,
is indifferent
for the subtraction of the drifts, except on the longest timescales of each scan
(see Sect.\,\ref{baselines}),
as long as it provides sufficient coverage and redundancy.
A rectangular snake pattern is used for Herschel and P-Art\'emis, while a spiral
pattern will be implemented for Art\'emis. Since there is often confusion
between a scan and a scan leg (or subscan), that has to be lifted for the rest
of this paper to be clear, we wish to emphasize the difference between them.
A scan leg is a continuous segment
of data acquired at constant scanning angle with respect to a fixed direction
on the sky, and at constant speed. A scan is also a continuous segment of data,
but comprises several parallel legs as well as the data acquired at non-zero
acceleration while slewing from one leg to the next (turnaround data).
It is best to have access to turnaround data whenever they exist in order to
avoid edge effects. These data are indeed available for Herschel, but by default
are not included in level-1 data for SPIRE
(their inclusion requires the use of a dedicated option).

For the SPIRE instrument, the optimum scanning strategy, that is the basis of the
default scan mode proposed to observers, is described by \citet{Waskett07}.
\footnote{
Its main features were already in place in the early years of the
mission, as shown by ESA document SCI-PT-RS-07725 ``FIRST pointing modes'' (April 2000).
}
For PACS, the choice of observing parameters such as the position angle of the
arrays with respect to the scan direction is not as critical as for SPIRE, due
to the fact that the arrays are filled. Constraints stemming from the array
structure nevertheless exist (some position angles have to be avoided, because
of the gaps between the subarrays).

Except for non-standard observations or preliminary reductions of incomplete datasets,
at least two scans are combined for each field. The scanning directions should be
separated by an angle of at least 20 degrees to ensure that the low-frequency noise
is most efficiently separated from the astronomical signal \citep{Waskett07}.
The low-frequency noise series are in this case superposed to signal series that
are intrinsically different in successive scans, making the two easier to disentangle.

The critical parameters for our purpose are the scan speed, the sampling rate
and the onboard compression rate, that directly affect the available redundancy
and the number of samples per beam crossing. The nominal scan speed is
$30\arcsec / {\rm s}$ for SPIRE. For PACS, it is $20\arcsec / {\rm s}$.
The fast scan speed of $60\arcsec / {\rm s}$, available for both instruments,
should only be used for very large areas, as it does not ensure optimum redundancy
and results in appreciably distorted point spread functions due to insufficient
sampling (for PACS). The sampling rate of SPIRE is 18.6\,Hz. Onboard compression
is necessary for PACS, so as to be able to store and downlink the immense data
volume generated by this instrument; from an initial sampling rate of 40\,Hz,
data frames are averaged to a final rate of 10\,Hz. In parallel mode, in which
both instruments acquire data simultaneously, the sampling rate is also 10\,Hz
for SPIRE and the red band of PACS (at 160\,$\mu$m), but 5\,Hz for the blue and
green bands of PACS (at 70 and 100\,$\mu$m).

For P-Art\'emis, the bolometer array has the same architecture as one PACS matrix
(a filled square array with 16 pixels on a side), but the scanning strategy differs,
especially in that it was chosen to offset the telescope
by only a few arcseconds between two scan
legs (about one pixel), making the available redundancy much higher, to increase
the sensitivity. Simulations and flight acquisition of Herschel scans were therefore
essential to test the robustness of the algorithm with respect to limited spatial
redundancy.

The redundancy in each observation can be conveniently quantified by the coverage map
or the weight map (see Sect.\,\ref{weight_error_maps}), excluding the edges. We shall
take as unit of redundancy the number of samples per square FWHM per scan pair (the
two scans having distinct orientations, most often nearly orthogonal). The redundancy
depends on the structure and position angle of the array with respect to the scan
direction, on the separation between adjacent scan legs, on the scan speed and sampling
rate. For the Herschel observations presented as illustrations in Section\,\ref{tests},
it varies between 75 and 530. Nominal SPIRE observations at $30\arcsec / {\rm s}$ have
redundancies of 150 to 180, with $3 \sigma$ dispersions of 30 to 40\%\,. Nominal PACS
observations at $20\arcsec / {\rm s}$ have redundancies of 350 (at 70\,$\mu$m) to 530
(at 100\,$\mu$m), with $3 \sigma$ dispersions of 35 to 55\%\,. For the PACS blue and
green bands in parallel mode, it is divided by two, thus on the same order as for
nominal SPIRE observations. With a pixel size equal to the default for the final projection,
i.e. FWHM/4, this fiducial value translates to about 10 samples per pixel per scan pair.

\subsection{Low-frequency noise}

The low-frequency noise has two main components. A first part comes from small
temperature fluctuations of the cryogenic bath. As the bath temperature varies
coherently across the array, these fluctuations produce for each bolometer
a noise that is strongly correlated with the noise of the other bolometers.
In principle, this noise is calibratable for SPIRE, because the arrays contain
some thermistors and blind bolometers that can be used as temperature probes,
unaffected by temperature variations due to the sky signal. In practice, however,
the thermal drifts have other dependencies that are currently not accounted for
in the parameterization that is used to correct for them in the pipeline.
The option to choose one or the other method does not exist for PACS, that lacks
blind detectors, so the removal of correlated noise by the map-making software
is a genuine necessity for this instrument.

The second component, that can be assimilated to flicker noise, is non-thermal
and uncorrelated from bolometer to bolometer. It is determined by detector
physics and readout electronics and cannot be calibrated. The knee frequency is
defined as the frequency where the power spectral density of the flicker noise
is equal to that of the white noise. In laboratory tests and flight operations,
it has been measured to be less than 0.1 Hz for SPIRE \citep{Griffin10}, but
more than 1 Hz for PACS \citep{Poglitsch10}.

The brightness drifts are expected to come only from these two sources of noise
and therefore to be purely additive.
It is assumed that multiplicative instrumental effects such as gains and flatfields
are perfectly stable, which was so far verified.

\subsection{Pre-processing and interface with the Herschel pipeline}
\label{preproc}

In the Herschel data processing chain, map-making utilities take as input level-1 data
(flux-calibrated time series with the associated pointing information) and produce
level-2 data, i.e. maps with all instrumental effects removed as well as possible.
The pre-processing by the pipeline includes the following steps (not mentioned here
in their actual order): 1) from level 0 to level 0.5, bad channels are masked and
samples with underflow/overflow are flagged, ADUs are converted to voltages and times
are attached to each sample; from level 0.5 to level 1, the pointing of each bolometer
is computed, the data are deglitched and converted to brightnesses (in Jy per beam for SPIRE
or Jy per array pixel for PACS), corrections are made for electrical and optical
crosstalk, for the electrical filter response and the bolometer time response,
and, for SPIRE only, the thermal drifts are subtracted by using
the smoothed series of thermistors located on the detector array as inputs to a drift model.

When post-processing SPIRE data with {\it Scanamorphos}, it is possible to bypass
the thermal drift correction within the pipeline, since this correction is also
dealt with by {\it Scanamorphos}. Below, a comparison is made between the results from
the two algorithms, to demonstrate the usefulness of {\it Scanamorphos} in removing
thermal drifts in both instruments.
Users are encouraged to compare both approaches and to
select the most effective for their data, or possibly to use the {\it Scanamorphos}
module coping with thermal drifts as a second-order correction, if the redundancy
is very limited (i.e. if the observation consists of only one scan or two scans).

Before correcting for the low-frequency noise, another pre-processing step
is necessary for Herschel data. The conversion of the voltage of each bolometer
to an in-beam flux density is made with reference to a fixed voltage, representing
what would be measured on a perfectly dark sky in the same conditions as those
of the observation. In practice, this fixed voltage cannot be accurately calibrated,
which means that the signal of each bolometer is the superposition of the true
sky signal, the noise, and a large offset (varying strongly from bolometer to
bolometer). These offsets are removed within {\it Scanamorphos} as part of zero-order
baselines (one constant per bolometer per scan leg is derived). The constants
are computed so as to be minimally contaminated by both extended and compact sources;
but any small error in their determination can be corrected afterwards by a
linear baseline removal module making use of the redundancy instead of simple fits
(i.e. a destriper; see below). The pipeline also includes a module dealing with these
offsets and performing either baseline subtraction or high-pass filtering (between levels 1 and 2),
and it is strongly recommended to bypass it in view of a subsequent {\it Scanamorphos} processing.

Level-1 data can be accessed outside the pipeline by first saving them on disk in
Flexible Image Transport System (FITS) files \citep{Wells81, Hanisch01},
with a format specific to Herschel, and then converting these files
with an IDL utility to a format readable by the map-making software. The conversion
utility is included in the {\it Scanamorphos} tree.

Invalid bolometers are rejected by this data formatting task.
The SPIRE 250\,$\mu$m array contains a total of 139 detectors
(7 always de-activated), the 350\,$\mu$m array 88 detectors (1 always de-activated),
and the 500\,$\mu$m array 43 detectors (1 always de-activated).

The PACS blue array (operating at 70 and 100\,$\mu$m) contains 2048 pixels, and the
red array (operating at 160\,$\mu$m) 512 pixels. They also contain a very small fraction
of dead pixels. In addition, for a few rows of the blue array, the signal is seen to
oscillate between an upper state and a lower state. This is caused by electronic
instabilities in the multiplexing circuit. The few 16-pixel blocks that are the most
frequently affected by such artefacts have been identified from a database of nearby
galaxy observations, and are entirely masked before the processing begins.

\section{Algorithm}
\label{algo}

\subsection{Overview}
\label{algooverview}

Our algorithm does not strictly speaking separate the low-frequency noise
into thermal and non-thermal components,
{\ifgras \bf \fi
since it cannot discriminate between different physical origins.
}
It uses a different but mathematically
equivalent decomposition: the thermal noise is replaced with the
average noise (the same time function for all bolometers), and the flicker
noise by the complement (an independent time function for each bolometer).
This alternative decomposition is easier to handle algorithmically.
We also make a practical distinction between long-timescale and short-timescale
drifts, but only because they are subtracted by different means.

Here are the main computations and corrections performed in sequential order: \\
1) computation of map coordinates from sky coordinates; optionally slicing of the
   dataset into several spatial blocks (for very large data volumes; see Sect.\,\ref{slicing}) \\
2) computation of scan speed as a function of time and tagging of nominal data and
   non-zero acceleration data (on the edges of the map);
masking of scan speed anomalies due to position errors \\
3) computation of space and time grids adapted to the drift subtraction problem
   (Sect.\,\ref{space}) \\
4) division of the signal by the relative gains (SPIRE only; Appendix\,\ref{spire_gains}) \\
5) computation of high-frequency noise (Sect.\,\ref{highf}) \\
6) subtraction of linear baselines, effectively removing brightness drifts with
   timescales comparable to or larger than the scan leg duration (Sect.\,\ref{baselines});
   and, for PACS only, masking of brightness discontinuities (see Sect.\ref{jumps}) \\
7) update of high-frequency noise, first-pass glitch masking (Sect.\,\ref{glitches}),
   and detection of asteroids (Sect.\,\ref{asteroids}) \\
8) subtraction of the average drift (on timescales smaller than the scan leg duration)
   and residual glitch masking (Sect.\,\ref{averdrift}, \ref{avermatrix} and \ref{glitches}) \\
9) subtraction of individual drifts (on timescales smaller than the scan leg duration)
   and residual glitch masking (Sect.\,\ref{indivdrifts} and \ref{glitches}) \\
10) projection of signal, error, total drifts and weight maps on a fine pixel grid (Sect.\,\ref{proj}) \\
11) optionally mosaicking of the spatial blocks processed separately (Sect.\,\ref{slicing}).

In what follows, it is unavoidable to mention some of the software options in order to
highlight some branch points of the algorithm. They are presented in Appendix\,\ref{options}.

\subsection{Spatial and temporal grids}
\label{space}

To determine drifts from redundant data, the underlying principle is the following:
at a given position on the sky, the astronomical signal received by the detectors
is the same at any time, and the signal read out from various bolometers at
various times differs only by the low-frequency noise affecting the detectors
at these particular times plus the high-frequency noise
(that includes glitches, physical white noise and quantization noise, also white):
\begin{equation}
R(t, b_i) = G(b_i) \times S(p) + LF(t, b_i) + HF_0(t, b_i)
\end{equation}
where $R(t, b_i)$ is the signal recorded at time $t$ by the bolometer $b_i$\,,
$S(p)$ is the sky brightness in the pixel $p$ sampled by $b_i$ at time $t$,
the gain $G(b_i)$ is the relative beam area of bolometer $b_i$ (for SPIRE only;
see Appendix \ref{spire_gains}), $LF(t, b_i)$ the total low-frequency instrumental
noise affecting bolometer $b_i$, and $HF_0$ the high-frequency noise.
Defining the average drift as~
$\bar{D}(t) = \frac{1}{N_b} \sum_{i=1}^{N_b} \frac{LF(t, b_i)}{G(b_i)}$~
(where $N_b$ is the number of valid bolometers)
and the individual drift of bolometer $b_i$ as~
$D_{b_i}(t) = \frac{LF(t, b_i)}{G(b_i)} - \bar{D}(t)$~ we obtain
\begin{equation}
R(t, b_i) / G(b_i) = S(p) + \bar{D}(t) + D_{b_i}(t) + HF(t, b_i)
\label{eq:sample}
\end{equation}
where the new $HF$ term is simply the original high-frequency
noise divided by the gain. Our decomposition of the low-frequency noise allows us
to ignore any differential gains between signal and noise, as long as they are constant.

It is then necessary to define
precisely the finite region of the sky within which the intrinsic signal will
be considered uniform, to compare all the samples taken within this region
and derive the drifts from them. Ideally, this region should be much smaller
than the point spread function to avoid distorting compact sources. In practice
however, the sampling rate is too limited to provide sufficient statistics
within regions smaller than the beam FWHM; another consideration to take into
account is the efficiency of the code in terms of execution time. We thus adopt
the beam FWHM as the spatial scale of signal invariance in our drift removal
procedure. In this case, a way to detect and protect compact sources has to be
implemented. In addition, we must ensure that the low-frequency noise is stable
during the time needed to cross this region.

Since the flicker noise becomes smaller than the white noise above a knee
frequency, a limiting timescale can be defined below which its effects
will be negligible and impossible to correct.
This timescale should be longer than the beam FWHM crossing time to obtain
an optimal correction.
For the SPIRE bands at 250, 350 and 500\,$\mu$m respectively, at the nominal
scan speed of 30~arcsec~s$^{-1}$, the FWHM corresponds to minimum timescales of
0.6, 0.8 and 1.2\,s, or maximum frequencies of 1.7, 1.2 and 0.8\,Hz.
Laboratory measurements of the SPIRE noise knee frequencies are all well below
these limits, of the order of 0.1\,Hz.
For the PACS bands at 70, 100 and 160\,$\mu$m, at the nominal scan speed
of 20~arcsec~s$^{-1}$, the minimum timescales are about half those of SPIRE,
0.28, 0.34 and 0.57\,s, corresponding to maximum frequencies of 3.6, 2.9 and 1.8\,Hz.
The PACS noise knee frequencies are of the order of 1\,Hz or higher,
for all applied polarization voltages tested in the laboratory \citep{Billot08}.
Then, in practice, the knee
timescale may be shorter than the beam FWHM crossing time in some configurations
for PACS (see below).

We thus define two spatial grids during the processing: a grid for mapping,
with a pixel size equal to half the beam FWHM by default, and a coarse grid
for drifts removal (for steps 8 and 9 in Sect.\,\ref{algooverview}),
with a pixel sized by the beam FWHM by default, i.e. our
chosen stability length. The algorithm uses an iterative process, and a map using
the latest drift correction is projected onto the mapping grid before each
iteration. Then, at the end of the processing, the data are projected onto
a final grid, with a default pixel size set to a fourth of the beam FWHM
(changeable by the user).

We require at least 6 samples per stability length $l_s$ to enable the computation
of simple statistics. The sampling rate is too low in some cases to meet this
requirement with a stability length of FWHM (for PACS, or for SPIRE in parallel
mode). The pixel size of the coarse grid is automatically increased if needed,
in increments of FWHM/2. The coarse spatial grid is in fact duplicated after
offsetting it by half a pixel in each coordinate, to provide better sampling
of the drifts.

Similarly, two temporal grids are used during the processing: the native grid
corresponding to the sampling times, and a coarse grid adapted to the drifts
computation. The time step of the coarse grid is the crossing time of the
stability length defined above:
$T_c~=~l_s~/~v_{\rm scan}$,
where $v_{\rm scan}$ designates the scan speed,
which defines the coarse time grid:
$t_c~=~k~T_c$, $k = 0, 1, ... , n_c$.

In unfrequent cases, and when the spatial grid makes an angle close to $45^{\circ}$
with a scan direction, it was noticed that the subtraction of the low-frequency noise
did not perform optimally, leaving some residual short-timescale striping.
This may be due to the fact that in this configuration, a larger fraction
of the data is unusable for the drifts computation, since bolometers scanning
a pixel corner, instead of crossing a pixel along a diagonal or a side, are
rejected for lack of sufficient statistics.
Because it is not possible to predict if a given orientation will be unfavorable
for the drifts subtraction or not, all the processing is systematically done
in the optimal orientation determined by the code (e.g., for the simplest case of
observations made of orthogonal scans,
with the x axis parallel to the direction of the first scan).
The spatial grid chosen by the user is enforced only for the final projection.

\subsection{High-frequency noise}
\label{highf}

The high-frequency noise standard deviation serves as a benchmark to quantify
the low-frequency noise amplitude, and to weight the data by its inverse square.
It is measured, for each bolometer and each scan, from the binned average of the
spectral density between 2.5 and 5\,Hz for SPIRE (where spectral densities reach
a plateau), and between 3 and 5\,Hz for PACS (for which the true white noise
cannot be estimated, since spectral densities do not reach their minimum at 5\,Hz).
For convenience, we call these estimates ``white noise'', although this is an
inaccurate designation in the general case.

\begin{figure*}[!ht]
%
%
\hspace*{-1.5cm} \includegraphics[width=9.5cm]{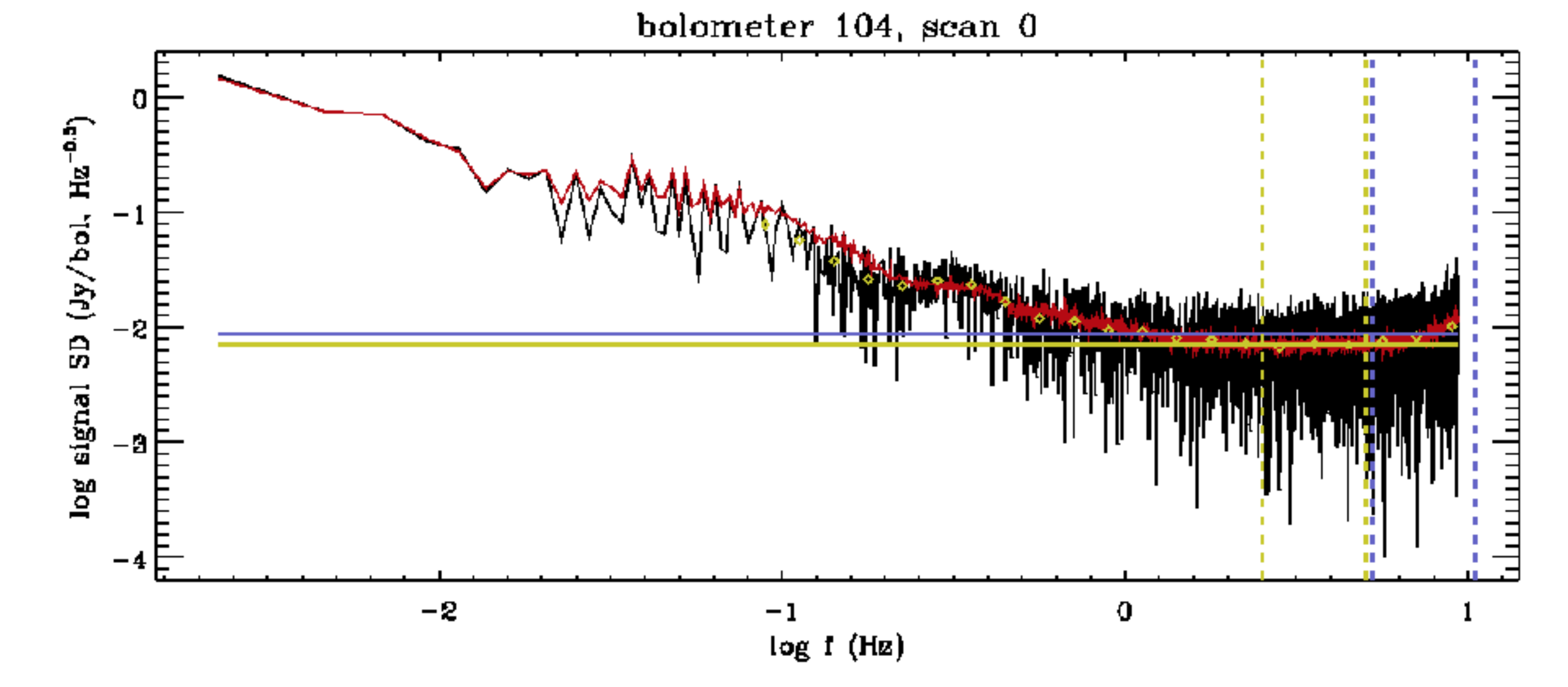}
\hspace*{-0.5cm} \includegraphics[width=9.5cm]{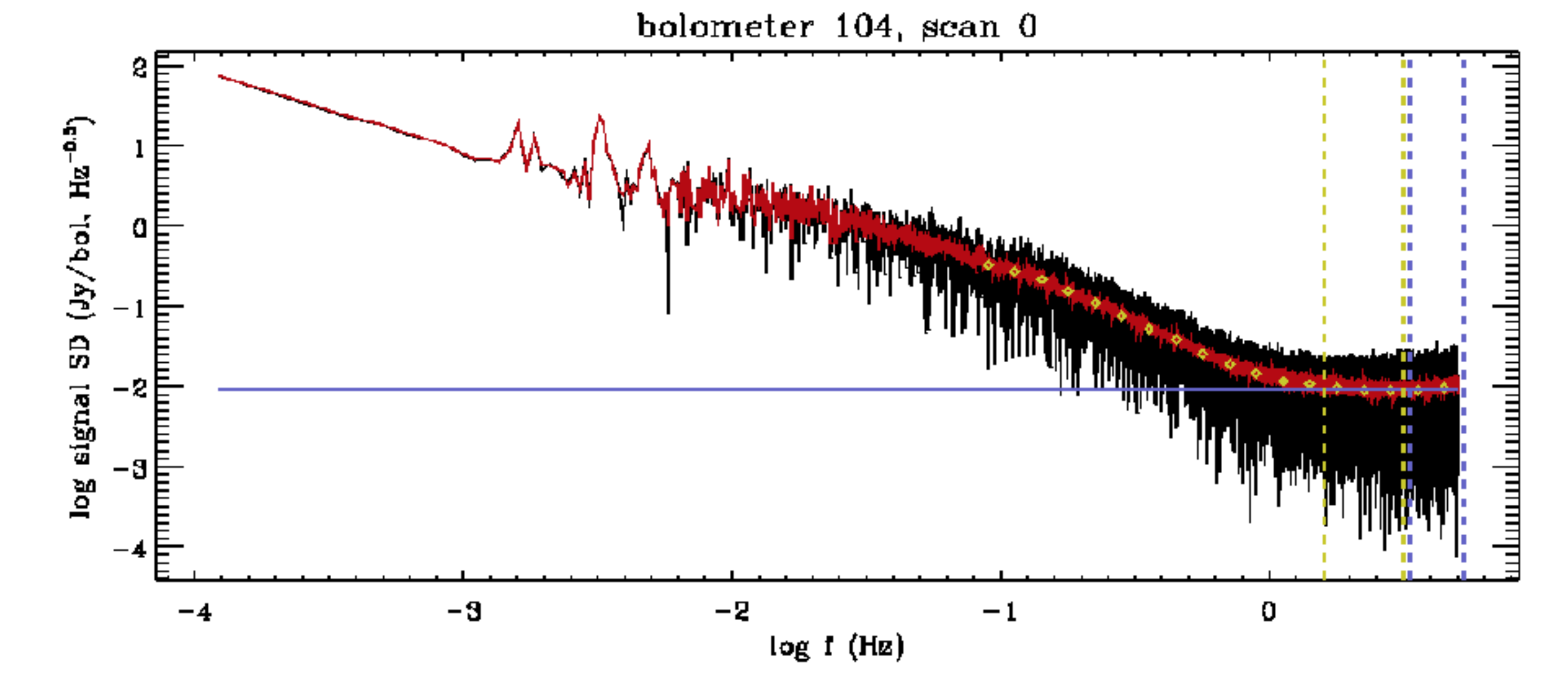} \\
\hspace*{-1.5cm} \includegraphics[width=9.5cm]{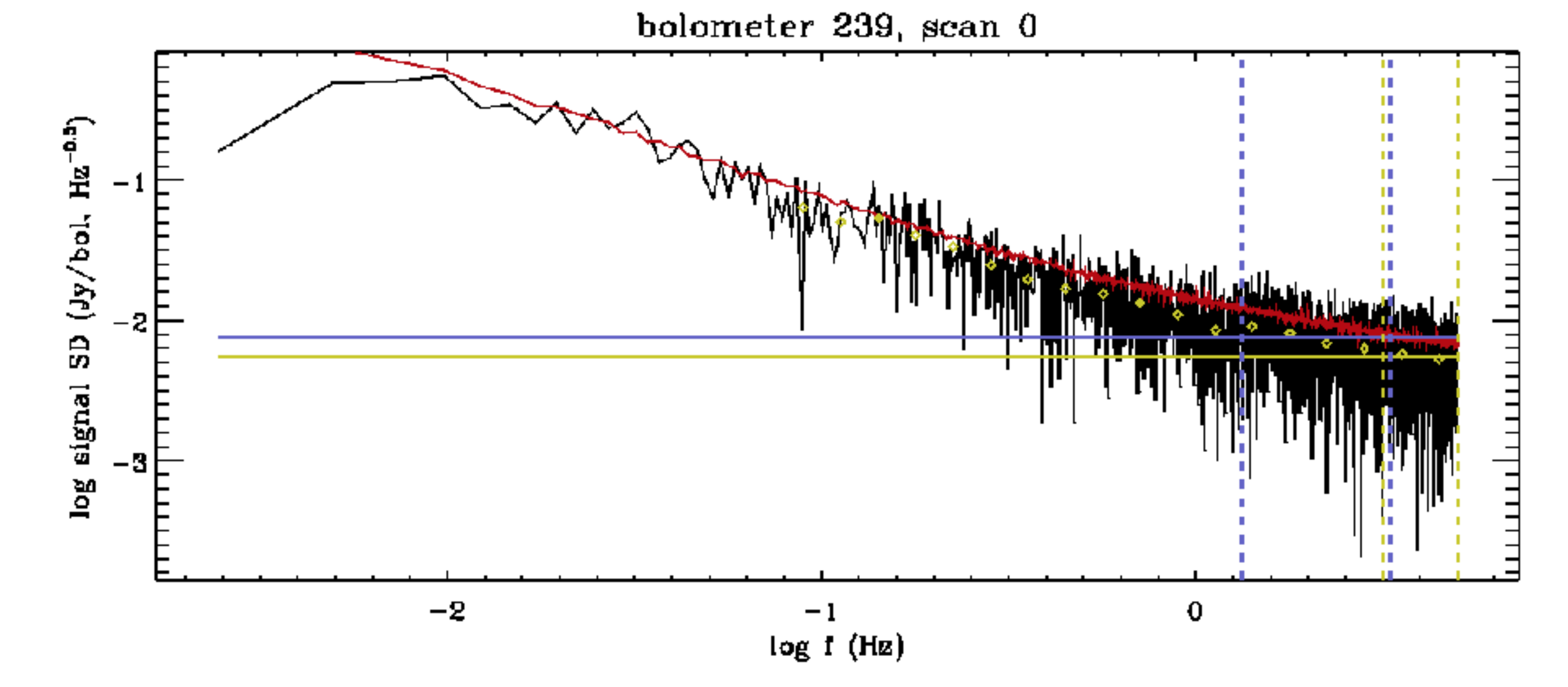}
\hspace*{-0.5cm} \includegraphics[width=9.5cm]{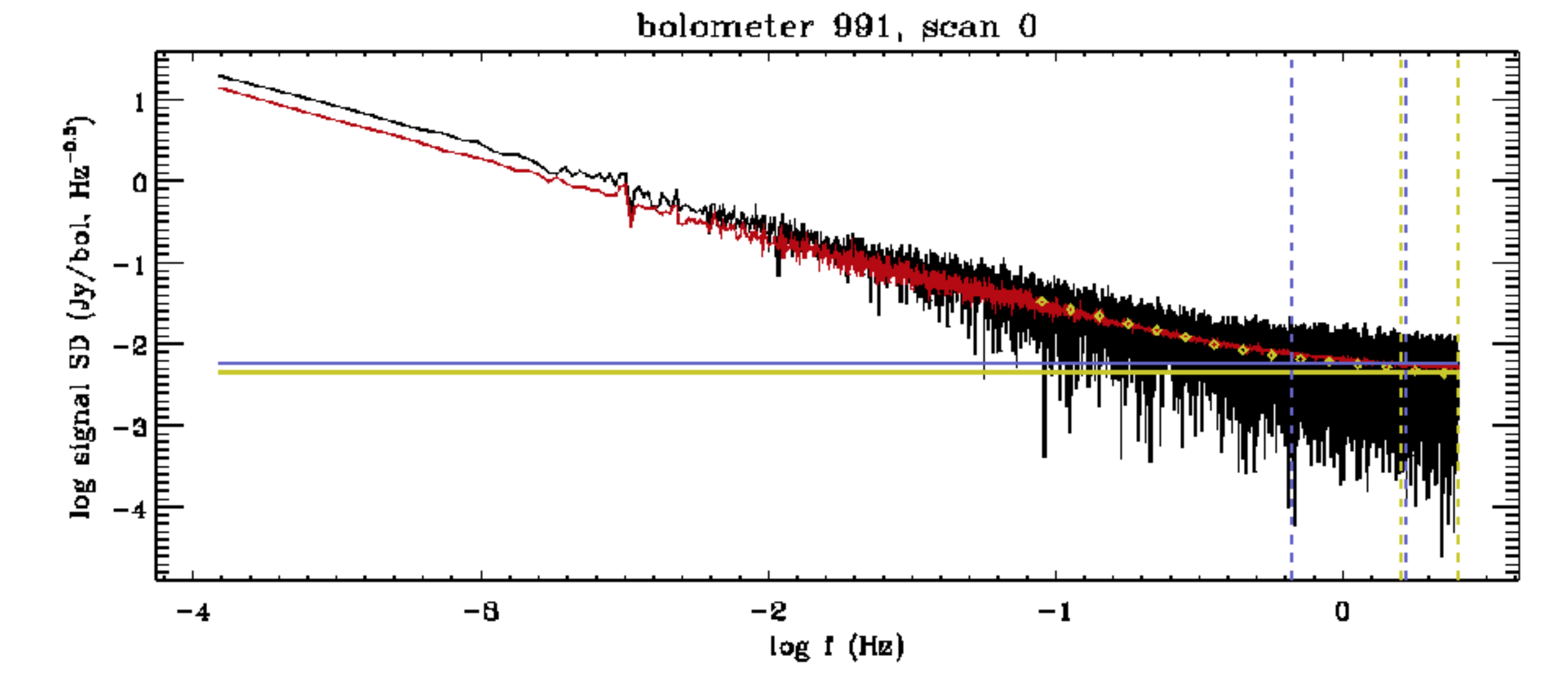}
\caption{Example of determination of the high-frequency noise from spectral densities computed
after interpolating compact sources and glitches, at the nominal SPIRE sampling rate
(top left, from the NGC\,6946 250\,$\mu$m observation), at the parallel-mode sampling rate
for SPIRE (top right, from the Rosette 250\,$\mu$m observation); and for PACS in nominal
mode (bottom left, from the NGC\,4559 160\,$\mu$m observation), and in parallel mode
for the blue array (bottom right, from the Rosette 70\,$\mu$m observation).
{\it SD} stands for spectral density, and {\it f} for frequency.
The vertical lines show the frequency ranges within which the average of the spectral
density is computed, and the horizontal lines the derived noise spectral density.
The white noise is coded in yellow, and the threshold noise in blue (see Sect.\,\ref{highf}).
The yellow diamonds represent the binned spectral density. The high-frequency noise is computed
individually for each bolometer. The red curve shows the average spectral density of all bolometers,
for comparison.
}
\label{fig:wnoise}
\end{figure*}

Prior to the computation of spectral densities, real high-frequency signal - that
may contaminate our noise estimates - is masked in the following way: whenever
the absolute difference between the signal at a given time and the signal between
one and three time steps later exceeds a given threshold (computed from the dispersion
of the absolute differences in the observation), a total interval of six samples around
that time are temporarily masked out. Because the Fourier transform needs continuous
time series, the masked samples are then interpolated, and an artificial Gaussian noise is
added to the interpolates. The mapping of the signal that has been removed in this way
shows that the procedure is effective at removing both bright compact sources and glitches
(an example is shown in Fig.\,\ref{fig:rosette_maps}).

For the removal of drifts, it is necessary to use a second noise quantification
that serves to define a threshold for the detection of compact sources, steep brightness
gradients and glitches (see Section \ref{averdrift}). For SPIRE, since some pipeline modules
using Fourier transforms amplify the high-frequency noise, the spectral densities
rise again beyond 5\,Hz. To take this into account in our noise threshold, we measure
the average noise standard deviation between 5 and 10\,Hz. We call this estimate
(or the ``white noise'' estimate if it is larger) the ``threshold noise''. Similarly,
we measure such a quantity for PACS, between 1.3 and 3\,Hz. All the frequency windows
are modified in parallel mode, when the sampling rate is lower. Illustrations of the
high-frequency noise computation are shown in Fig.\,\ref{fig:wnoise}.
After computing the high-frequency noise of all bolometers, in each scan, those with
very deviant noise values are masked and not further used.

For PACS, the brightness quantization noise is in some cases a significant fraction
of the high-frequency noise. We estimate it in the following way: for each bolometer
and each scan, the quantization step $q$ is defined as the minimum brightness where
the histogram of $| B(t_{i+1}) - B(t_i) |$, computed on a very fine brightness grid
(excluding zero), is populated, $B(t_i)$ being the brightness recorded at the $i^{\rm th}$
time sample. Assuming that the quantization error is uniformly distributed between
0 and $q$, the quantization noise variance is then $q^2 / 3$.
When this term dominates the high-frequency noise (which happened in the early stages
of the mission when using a very low gain), it may limit our ability to subtract
the low-frequency noise down to acceptable levels.

\subsection{Long-timescale drift subtraction}
\label{baselines}

Scan observations do not allow to
separate the brightness drifts from the signal on spatial scales similar to or larger
than the map size, i.e. on timescales of the order of the scan leg duration of larger.
Such drifts, if left uncorrected, may introduce artificial brightness gradients.
For thermal drifts, that are usually coherent over the duration of several scan legs
or whole scans, these gradients will be perpendicular to the scan direction
(see an example in Sect.\,\ref{pacs_rosette}).
The simplest way to remedy this problem is to subtract linear baselines, computed
either on a scan basis, or on a scan leg basis, but this requires particular care,
in order to keep the baselines immune to real structures that are smaller than the
field of view, and to preserve as much as possible brightness gradients that could
be real. These baselines in fact cover several distinct instrumental artefacts:
the primary component consists of flux calibration offsets, as mentioned above
(Sect.\,\ref{preproc}), and the rest comprises the long-timescale component of 
both the thermal drift and the flicker noise.

We argue that baselines should not assume the form of polynomials of any order higher
than 1 (this is an option offered in the pipeline), whenever possible. With an order
of 2 or higher, the shape of structures smaller than the field of view is altered,
in an uncontrolled way. In {\it Scanamorphos}, the subtraction of the drifts on timescales
smaller than the field of view crossing time are left to specific modules using the redundancy,
and thus operating in a controlled way, such that flux and morphology are conserved.

The baseline-subtraction algorithm described here was implemented specifically for Herschel,
and needs to be adapted if scans cannot be sliced into distinct scan legs.
We need to distinguish three categories of fields, for which the algorithm will
be slightly different. Extragalactic fields are the easier case, because the sky
has at most very mild brightness gradients, that are insignificant with respect to
the effects of a thermal drift (any field where a flat sky can be isolated from
sources belongs to the same category, even if located within the Galaxy).
The second case is that of bright Galactic fields (e.g. star formation regions),
where sky gradients can become comparable to the artificial gradients induced by
the drifts; their separation requires special steps. Finally, another modification
to the general algorithm is needed for very small fields, as explained at the end
of this section.

\paragraph{Extragalactic or faint fields:}
~
First, simple baselines are computed by the process outlined below in points 1 to 3,
iterated three times: \\
1) On a scan basis, we fit a linear function to the signal averaged over all valid
bolometers, and subtract it from the signal of each bolometer. In practice, this
removes a major part of the average low-frequency noise. This step can modify the
general brightness gradient in the scan-perpendicular direction, but not in the
scan-parallel direction. \\
2) On a scan leg basis, one offset (or zero-order baseline) is subtracted for each
valid bolometer. This step takes care of flux-calibration offsets, and can also include
a very small contribution from the low-frequency noise, since the offsets are allowed
to vary from one scan leg to the next. Each offset is determined as the median of the series. \\
3) At the third iteration, zero-order baselines in step 2 are replaced with linear
baselines. A distinction is made between fields with structure extending over scales
similar to the map size, and fields where the extended emission is confined to the
central part of the map: if the former case applies,
then the baselines are constrained such that their average over all
valid bolometers reduces to the zeroth order, so that in practice the general brightness
gradient is not altered. An example of such a case is a field with strong cirrus emission
on one side of the map (see Sect.\,\ref{spire_n6822}).

The fits in steps 1 and 3 and the offsets in step 2 are robust, in the sense that a
mechanism is in place to iteratively exclude sources from the brightness series,
both in the time domain and in the spatial domain. For the latter, a new map is
created after each step, and sources above a given threshold are masked; the mask
is then transferred to the time domain. The applied brightness threshold is determined
automatically, and is chosen so as to avoid masking predominantly the edges, in case
significant structure exists outside the central part of the map
(see an example in Sect.\,\ref{spire_rosette}).

The discrimination between the two types of structure in step 3 is made automatically,
and requires no user input.
It is based on the same source mask as above, except that the brightness threshold is
fixed. The map is divided into a central part (one ninth of the surface area) and the
periphery. The fractional area of the source mask located within the central part decides
of which case applies: if it is smaller than a given threshold, and if the total area of
the mask is significant, then we consider that there is extended emission throughout the map.

\noindent
4) Once these simple baselines are subtracted, they are refined by using the redundancy
in the observation instead of fits (which provide a useful pre-correction but are
unphysical). This complementary destriping ensures that the most accurate solution
is found, by minimizing the deviations between different bolometers and between
successive scans. This is an iterative process using the same source mask as
above. At each iteration, and for each scan leg, the difference between the series
of each valid bolometer and the series simulated from
a reference map (both associated with the same sky coordinates)
is computed and fitted
by a linear function, and this linear baseline is subtracted from the data; then
the map is updated and the process repeated until convergence.
The reference map is first built from scans taken in the direction orthogonal to
the scan leg under consideration, and then in later iterations from all scans
(see below for a detailed explanation).

The destriping algorithm assumes that each scan leg is separated from the next by a
spatial offset larger than the FWHM or a few times the FWHM. While this is always
true for nominal observations, this condition is not verified for fine scans used
for calibration purposes. When processing fine scans, the destriping module should
therefore be deactivated.

\paragraph{Galactic or complex bright fields:}
For observations of such fields, if the relevant option is used ({\it /galactic}),
the baseline subtraction is modified with the aim
to protect as well as possible complex structures of0.3ten extending over much larger
scales than the field of view. The modified algorithm is applicable only when
sources are bright enough to be visible in the map built from raw data
(or rather when only the flux calibration offsets have been subtracted). \\
1) On a scan basis, linear baselines are replaced with simple offsets, so that the natural
brightness gradients of molecular clouds, usually dominant over gradients induced
by low-frequency noise, are left intact. The offsets are computed for each valid bolometer
as the median of the series, and in practice eliminate the flux calibration offsets. \\
2) Still on a scan basis, linear baselines are computed by using the redundancy: the
difference between the average series of the valid bolometers and the average series
simulated from a reference map is computed and fitted by a linear function, and this
linear baseline is subtracted from the data; then the map is updated and the process
repeated for a total of four times. For each scan, the reference map is built from
the scans acquired in the (near-)orthogonal direction only, in the first two iterations.
In the last two, once the scans have been rectified, the reference map includes all the scans.
Natural gradients are preserved for this reason: when subtracting a linear baseline
from a whole scan, only the gradient perpendicular to the scan direction can be modified;
any gradient parallel to the scan direction would create a periodic signal (with a
period equal to the duration of two scan legs), and would not create any linear
component.
Any real gradient will be present in both the fitted scan and
the reference map, and thus cancelled when subtracting the series simulated from
the reference map, before doing the fit.
{\ifgras \bf \fi
As to the artificial gradients caused by the thermal drift, they are to first order
perpendicular to the scan direction (i.e. parallel to the displacement from one scan leg
to the next), since low frequencies are dominant. Thus they have a distinct orientation
in the fitted scan and in the reference map, when the latter is built from scans
acquired at a sufficiently distinct angle: the drift in a given scan will be left
intact by the subtraction of the reference series and can be fitted by a linear function
and removed. \\
}
3) The destriping is repeated, this time for each bolometer independently,
on smaller segments of four scan legs, i.e. at a better time resolution but still
ensuring immunity to scan-parallel gradients. The scheme to build the reference map
is the same as in step 2. \\
4) Then, the destriping on a scan leg basis is made in the same way as for extragalactic
fields. \\
A source mask is also defined and used in the same way as above in all steps.

\paragraph{Very small fields:}
For such observations, the destriping and average drift subtraction on small scales
become impossible, because there are not enough resolution elements across the map.
This can lead to an ill-determined thermal drift, in case it is significantly non-linear,
which occasionally happens. To solve this problem,
for small fields whose source mask covers a small fraction of the total area,
we allow third-order baselines in place of linear baselines
(on a scan basis).
The order change
is however effected only when the new fit is of much better quality than the linear fit,
as measured by the $\chi^2$ value.

\subsection{Short-timescale drift subtraction}
\label{lowfnoise}

We now discuss the removal of drifts on timescales shorter than the scan leg
duration, which can be derived without any assumptions on the sky structure,
contrary to those discussed in the previous section.

\subsubsection{Average drift}
\label{averdrift}

In a first step,
{\ifgras \bf \fi
as mentioned in Section\,\ref{algooverview},
}
thermal drifts are considered to be uniform over the whole
array; any component that differs from the average is treated in the second
step, as part of the uncorrelated drifts. We thus have to determine a single
series for the thermal noise.

For each pixel of the coarse grid
(Sect.\,\ref{space}),
all the samples recorded while the beam center
lies within this pixel are searched; an efficient IDL function is used to perform
this task quickly and store the sample indices beforehand.
Then, the samples are divided into separate bolometer crossings, for each of
which the brightness mean and the brightness mean absolute deviation are computed,
and to each of which an average time is attached.
Bolometers scanning only pixel corners, with insufficient statistics, are discarded.

At this point, the presence of a compact source, a steep brightness gradient, or
an abnormal level of high-frequency noise within the pixel, is tested in a
straightforward way: if the mean absolute deviation of a bolometer crossing is
below the average threshold noise of this bolometer (defined in Section\,\ref{highf}),
then it is deemed suitable for drift determination; otherwise, the assumption
of a uniform astronomical signal is not valid.
We then distinguish two cases: 1) If more than 70 to 80\% of the bolometer crossings
are suitable for drift computation, then they are effectively used, and only those
bolometer crossings with an elevated mean absolute deviation are discarded.
2) In the opposite case, then a significant portion of the coarse pixel area is assumed
to contain non-uniform signal, and all the bolometer crossings are discarded.
If the sampling rate allows the statistics to remain sufficient at smaller scales,
then the algorithm switches to a finer spatial grid, to which the same test is applied;
if the test is again negative, all the bolometer crossings are again rejected.

The measured brightness difference between each pair of suitable bolometer crossings
is in principle equal to the difference of the drifts between the two sampled
times.
Calling the first bolometer $b_i$, crossing the pixel $p$ at the average time $t_1$,
and the second bolometer $b_j$, crossing $p$ at the average time $t_2$,
\begin{eqnarray}
\label{eq:def_delta}
& & \hspace*{-7.1ex} \delta(t_1, t_2) = R(t_1, b_i) - R(t_2, b_j) \\
& &   = \bar{D}(t_1) - \bar{D}(t_2) + D_{b_i}(t_1) - D_{b_j}(t_2) + HF(t_1, b_i) - HF(t_2, b_j) \nonumber
\end{eqnarray}
The $S(p)$ term in Equ.\,\ref{eq:sample} cancels because the astronomical signal is assumed
both invariant in time and uniform within $p$, making the different trajectories of $b_i$
and $b_j$ within $p$ irrelevant. The software is in fact able to cope with mild non-uniformity
within $p$, since
the current map is used to cancel, whenever possible, brightness differences purely caused by
small pointing differences, as explained in Section\,\ref{point}.
Then, by linear interpolation, $\delta(t_{c 1}, t_{c 2})$
(on the coarse time grid defined in Section\,\ref{space})
is obtained from $\delta(t_1, t_2)$\,.

Each term $\delta(t_{c 1}, t_{c 2})$ is stored in a matrix where the first dimension represents
the smaller of the two times, and the second dimension the larger. Drift differences are
coadded in this matrix as they are computed, regardless of the position on the
sky or the bolometers involved. Their weights, inversely proportional
to the square white noise of each bolometer, are stored symmetrically across the
matrix diagonal. This ensures, together with the use of a coarse time grid adapted
to the minimum drifts timescale, that memory usage is minimized.

One obtains by coadditions
\begin{equation}
\delta(t_{c 1}, t_{c 2}) = \bar{D}(t_{c 1}) - \bar{D}(t_{c 2}) + {\rm WM}_{p, i, j}~ [D_{b_i}(t_{c 1}) - D_{b_j}(t_{c 2})]
\label{eq:aver}
\end{equation}
where WM designates the weighted mean and the sum runs on all pixels $p$ and all
pairs of bolometer indices ($i$, $j$) such that $b_i$ is crossing $p$ at time $t_{c 1}$
and $b_j$ is crossing the same pixel $p$ at time $t_{c 2}$.
The high-frequency noise term rapidly vanishes and, given that the individual drifts
are all independent of each other, the WM term above also
{\ifgras \bf \fi
becomes negligible
}
if the redundancy
is sufficient.
{\ifgras \bf \fi
Taking as an example the 70\,$\mu$m PACS observations of the Rosette nebula discussed
in Section\,\ref{pacs_rosette}, made of 2 scans in parallel mode at the scan speed of
$20\arcsec / {\rm s}$ (thus with relatively low redundancy), the median number of contributing
bolometer crossings for a given ($t_{c 1}$, $t_{c 2}$) time pair is $N_{\rm coadd} \simeq 380$\,,
with a very large spread. The residuals of the WM term in $\delta(t_{c 1}, t_{c 2})$ are thus
of the order of $\frac{1}{\sqrt N_{\rm coadd}} \simeq 5$\% of the average amplitude of the
individual drifts (which is larger than the amplitude of the average drift at any given
frequency for PACS). However, the average drift series is obtained from the
$\delta(t_{c 1}, t_{c 2})$ terms in a complicated fashion (see next section), involving
many more coadditions. Although a rigorous analytic estimate cannot be made, the final
residuals due to the WM terms are very small compared with the amplitude of the average
drift, as proven by the simulations in Section\,\ref{simul}.
}
Let us remark that the probability law of the individual drifts
{\ifgras \bf \fi
does not have to be
}
assumed Gaussian;
it only has to be symmetric about zero.
Finally, $\delta(t_{c 1}, t_{c 2}) \simeq \bar{D}(t_{c 1}) - \bar{D}(t_{c 2})$.

Once all pixels have been covered, the matrix is read so as to restore from
the drift differences the average drift series (see next section), that is
subtracted from the signal series of each bolometer. Since it is impossible to
define the absolute zero of the drift, it is previously offset so that its time
average is null. Then a new map is projected.

This process is iterated until convergence, which is reached when the drift amplitude
has become smaller than the white noise level for most bolometers. Here, the amplitude
of the drift is defined as three times its standard deviation.

\subsubsection{From drift differences to the absolute drift series}
\label{avermatrix}

The conversion of the drift differences matrix to the absolute drift series
is not trivial. First, the matrix is scanned several times in each direction,
progressively populating the drift series. One entry time is chosen and arbitrarily
assigned a null drift. Then, each drift difference between two times of which one
has already an entry in the drift series is used to assign an absolute drift to the
second time index. At the same time as the drift series is built, a corresponding
weight series is computed, and used to weight appropriately the drifts as they are
coadded. The matrix is scanned until the drift series converges.

This process does not provide a unique solution.
Let us assume that the average drift $\bar{D}(t)$ is replaced with $\bar{D}(t) + E(t)$,
where $E(t)$ is an excess drift, such that the projection of this series produces the same
map for each successive scan (i.e. $E(t)$ is an oscillatory function, with a period
equal to that of the spatial coordinates, just like the average true sky series). Then
the $\delta(t_1, t_2)$ term in Equation\,\ref{eq:def_delta} is unchanged, because it is
computed within a single pixel $p$, where $E(t_1)$ and $E(t_2)$ take the same value.
The drift series obtained by scanning the drift differences matrix is thus
the superposition of the true drift series and an oscillatory component.
The
{\ifgras \bf \fi
average
}
drift has no reason to repeat itself exactly in successive scans.
We thus extract this periodic component from the average drift and remove it,
before subtracting the average drift from the data. This is achieved by first
mapping the average drift for each scan separately, and extracting the component
common to all these maps, that we call the excess drift. Then, the series of the
excess drift is simulated from its map, and finally subtracted from the original
drift.

We now briefly explain the method used to extract the excess drift from
the drift maps. This method has to provide reliable results even when the observation
is composed of only two scans, i.e. only one period of the excess drift. We exploit
the fact that, by definition, the excess drift varies on much longer timescales
than the true drift (that varies on timescales shorter than the duration of a scan leg).
This implies that the spatial variations of the excess drift in the drift maps are
much smoother than those of the true drift. For each drift map (one per scan),
a map of the local variance is computed (using as zero moment a boxcar-averaged
version of the drift map). Then the excess drift map is defined as the weighted average
of all the drift maps, the weight for each one being the inverse variance map.
Obviously, the error on the excess average drift will go down as the number of scans
increases. When
a large number of scans are combined,
the excess drift map is simply
defined as the median of all the drift maps.

\subsubsection{Individual drifts}
\label{indivdrifts}

In a second step, the
{\ifgras \bf \fi
individual
}
drifts are removed. At this stage, by definition,
the drift affecting the average signal of all valid bolometers should be close to zero
at all times. Using the same scheme as previously and protecting non-uniform signal
in the same way, this allows us to equate the drift
of an individual bolometer to the difference between its signal and the average
bolometer signal, at any given position on the sky. This average signal is weighted
by the inverse square white noise of each bolometer, and considered to represent the
intrinsic sky brightness. Calling the bolometer of interest $b_i$, crossing a pixel
$p$ at time $t$, and $b_j$ all the valid bolometers crossing $p$ at times $t_k$:
\begin{eqnarray}
& & \hspace*{-6.7ex} \Delta(t, b_i) = R(t, b_i) - {\rm WM}_j [R(t_k, b_j)] \nonumber \\
& &   = \bar{D}(t) + D_{b_i}(t) + HF(t, b_i) - {\rm WM}_j [\bar{D}(t_k) + D_{b_j}(t_k)]
\end{eqnarray}
By the same argument as before, the $S(p)$ term in Equ.\,\ref{eq:sample} cancels and
the high-frequency noise term of the weighted mean vanishes. Since the average
drift has been subtracted beforehand, and interpolating linearly on the coarse time grid,
we have
\begin{equation}
\Delta(t_c, b_i) = D_{b_i}(t_c) + HF(t_c, b_i) - WM_j [D_{b_j}(t_{c k})]\,.
\label{eq:indiv}
\end{equation}
Given sufficient redundancy, the rightmost term once again
{\ifgras \bf \fi
becomes negligible
}
because the
individual drifts are uncorrelated.
{\ifgras \bf \fi
Taking the same example of the Rosette nebula as above (Sect.\,\ref{averdrift}),
the average redundancy is $N_{\rm coadd} \simeq 650$ bolometer crossings per coarse pixel
(with size $l_s = 4.5$ FWHM). The residuals of the WM term are thus of the order of
$\frac{1}{\sqrt N_{\rm coadd}} \simeq 4$\% of the average amplitude of the individual drifts
for this particular example. The other examples in Table\,\ref{tab:runs} have much higher
redundancies, thus lower residuals. Finally, one obtains
}
$\Delta(t_c, b_i) \simeq D_{b_i}(t_c) + HF(t_c, b_i)$.
We can neglect the high-frequency noise term (binned on the coarse time grid),
as long as the drift dominates.

A value of zero is assigned to the drifts over the time steps when they are undefined
(corresponding to compact sources, steep gradients and glitches).
At each iteration, a residual map -- equal to the difference between the map produced
at the previous step and the current map -- allows to immediately assess whether
compact sources have been altered or not, and by which percentage.
The interpretation of the residual map is
however not straightforward, since artefacts at the location of bright sources can
result from small pointing and gain errors. Intermediate results can also be visualized
in the form of selected signal and drift series.
The convergence criterion is the same as for the average drift.

Although combining two scans or more is desirable, this algorithm is able
to remove some of the low-frequency noise even when only one scan is available.
This allows data quality and depth to be assessed on a single-scan basis before
multi-scan observations are fully completed.

The noise spectral density of each bolometer, computed from the noise series
derived by the software, is an optional output, and can be compared with the
noise spectral density that is part of the pipeline calibration products,
allowing an independent test of the assumptions on the noise used by the
pipeline. An example obtained from the processing of flight data is shown
below (Sect.\,\ref{spire_atlas}).

\subsubsection{Correction for pointing differences within the stability length}
\label{point}

As mentioned in Section\,\ref{averdrift}, for the computation of both the average
drift and the individual drifts, we have to cope with mild gradients within the
stability length, i.e. non-uniformities of the signal that are too shallow to produce
a significant increase of the mean absolute deviations of the bolometer crossings.
Given that each bolometer follows a different trajectory within a coarse pixel,
and since the stability length is large compared with the angular resolution, especially
for PACS data, these mild gradients may contaminate the drift estimate if ignored.
The strategy to follow depends on the prominence of the low-frequency noise:
in SPIRE data, where the signal is never dominated by the short-timescale drifts,
the correction for pointing differences is always applied; but in PACS data,
where the short-timescale drifts are of much higher amplitude, the correction
should not be applied in all circumstances, because it might be dominated by noise
rather than real spatial variations.

In each coarse pixel where no compact
source was detected, we have to decide whether the correction for pointing differences
will be applied to all the bolometer crossings, or not at all. We consider two sets
of series: the uncorrected series of recorded samples $R(t, b_i)$, and the series corrected for
pointing differences $\tilde{R}(t, b_i) = R(t, b_i) - M(p(t, b_i))$, where $M(p(t, b_i))$ is
the signal of bolometer $b_i$ simulated from the current map, built on a fine grid
of pixels $p$ much smaller than the stability length. Brightness gradients, which
affect all bolometers in the same way, are present in both $R(t, b_i)$ and $M(p(t, b_i))$,
and can thus be cancelled by computing the difference. If the signal within the coarse
pixel is not dominated by the drifts but by sources, then $M(p(t, b_i))$ is a fair
representation of the sky emission; otherwise, it cannot be used to subtract gradients.

For SPIRE, $R(t, b_i)$ is always replaced with $\tilde{R}(t, b_i)$ before computing
the drifts. For PACS, to choose between the two alternatives, we try to quantify the non-uniformity
of the fine map at the location of the coarse pixel, and compare it with the expected
white noise $\Sigma$ at the same location (estimated from the square root of the inverse
weight map). If the local brightness is more than a few times $\Sigma$ above a
local background or above the global background,
then the pointing correction is applied; in the opposite case, the signal is deemed
perfectly uniform and the pointing correction is not applied.

\subsubsection{Time resolution for the individual drifts}
\label{timeres}

For PACS data, we have serendipitously found that the accuracy of the individual drifts
could be improved by first doing the correction on much longer timescales than allowed
by the stability length (Sect.\,\ref{space}), and then gradually increasing the time
resolution until reaching the minimum drift timescale. At the first iteration,
the size of the coarse pixel is left unchanged, but the drift correction is binned into
time intervals that are 27 times larger than the step $T_c$ of the coarse time grid.
The time resolution is increased by a factor of 3 at each of the following three iterations.

\begin{figure*}[!ht]
\hspace*{-1.3cm}
\hspace*{0.5cm} \includegraphics[width=18cm]{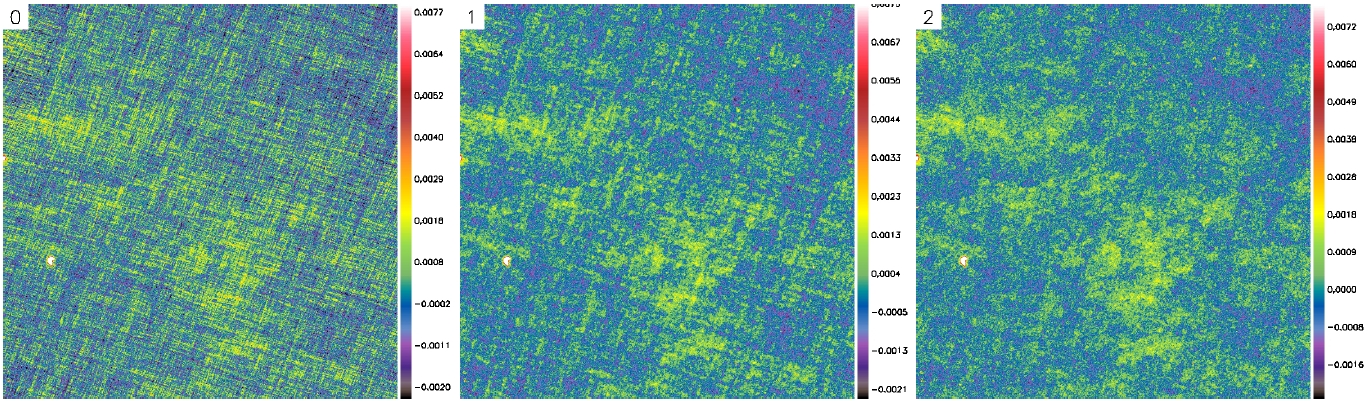}
\caption{A subfield of the HEVICS survey \citep{Davies10}, about 2.2 degrees on a side,
from 160-micron data processed in three different ways: 0) No drifts (on timescales shorter
than the scan leg duration) have been subtracted. 1) The iterations to subtract the
individual drifts all have the minimum time step $T_c$\,. 2) They have successive time steps
of $27 \times T_c$, $9 \times T_c$, $3 \times T_c$ and $T_c$. The display emphasizes
very diffuse structures.
}
\label{fig:timescales}
\end{figure*}

This modification greatly reduces the residual noise in diffuse areas of large PACS maps,
and the scan pattern becomes indiscernible, as shown in Figure\,\ref{fig:timescales}.
We conjecture the following reason. The algorithm relies on the
assumption that the coaddition of the individual drifts affecting all the bolometer
crossings (the rightmost term in Equ.\,\ref{eq:indiv}) is close to zero, because
the drifts are uncorrelated. This assumption may not be perfectly verified, causing
a residual noise. Averaging a great number of drift values within a large time
interval will average out this noise. Since the drifts have more power at lower
frequencies, the first iteration with a time step of $27 \times T_c$ removes a major
part of them, and it becomes easier to verify our initial assumption on smaller and
smaller timescales.

\subsection{Detection of other artefacts}

\subsubsection{Masking of brightness discontinuities}
\label{jumps}

\begin{figure}[!ht]
%
%
\hspace*{-1.5cm} \includegraphics[width=4.2cm, angle=90]{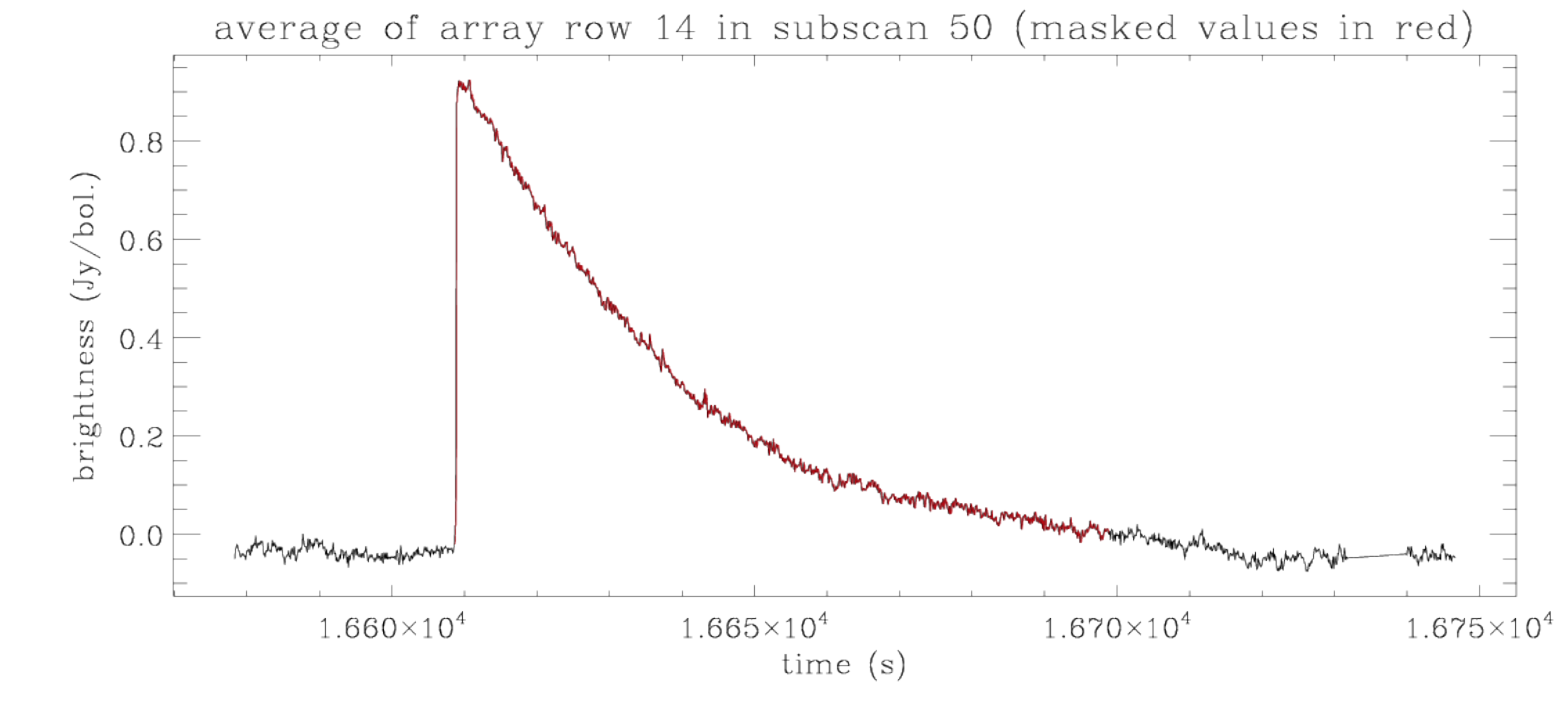}
\hspace*{-0.2cm} \includegraphics[width=4.2cm, angle=90]{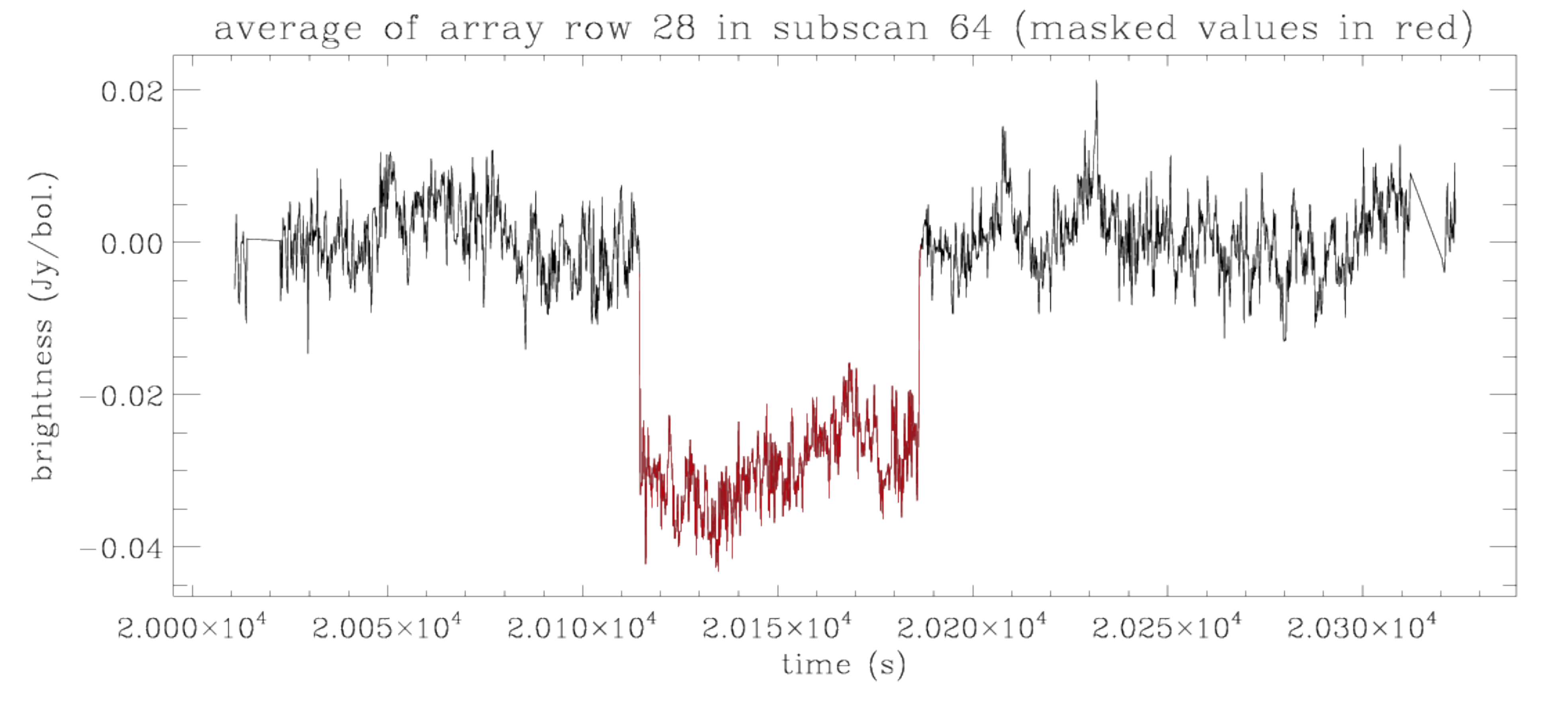}
\caption{Example of brightness discontinuities in PACS data
(left: caused by a glitch; right: caused by an electronic instability).
The average series of an
array row (16 contiguous pixels) is plotted over a particular scan leg. All the bolometers
in the row are affected by the artefact. Overplotted in red are the data that have been
masked by the dicontinuity detection module.
}
\label{fig:pacs_jump}
\end{figure}

During tests conducted on various PACS datasets, the presence of brightness discontinuities
affecting whole array rows (blocks of 16 pixels) or, less frequently, individual bolometers,
was noticed. These artefacts are caused by glitches, and also often by electronic instabilities
in the multiplexing circuit, that may come and go during the course of an observation.
They are currently not corrected for in the pipeline, but they can impact the quality of the
maps, leaving bright or dark bands. Since they are high-frequency features followed by a slow
decay or a plateau, they can be corrected neither by the deglitching modules nor by the drift
subtraction modules. Examples of such artefacts are shown in Fig.\,\ref{fig:pacs_jump}.

We have therefore developed a module to detect these
discontinuities and mask the subsequent frames. This module handles the average series
of each 16-pixel block, and then the series of individual bolometers of each block.
Spatial information is used jointly with temporal information,
making the detections more robust and more sensitive: the series simulated from the
current map are subtracted from the observed series. Candidate discontinuities are
found whenever the absolute difference between the signal at a given time and the signal
three time steps later exceeds a predefined threshold. A candidate is confirmed if it can
be unambiguously distinguished from a real brightness variation caused by a compact source
or a steep gradient of emission.
To this effect, the median brightness and the standard deviation of the series are computed
within four time windows, on each side of the potential discontinuity. A diagnostic is made
based on these values
to rule out compact sources. In addition, to avoid false detections caused by steep gradients
(such as present in molecular clouds or edge-on disks), this diagnostic is combined with a
thresholding based on the dispersion of the signal in the neighbourhood of each detection.
If the candidate is confirmed, subsequent frames of the array row
are masked, until the end of the scan leg or until their average brightness gets back
to the pre-discontinuity brightness to within the white noise level.


\subsubsection{Masking of glitches}
\label{glitches}

The software includes the capability to filter glitches remaining after the
pipeline processing,
whether a deglitching module was run in HIPE or not. They are not interpolated
but all affected samples are permanently masked.
The detection of glitches shares a portion of its algorithm
with modules in charge of computing both the high-frequency noise, and the average
and individual brightness drifts. For this reason, the deglitching task is embedded
within these modules, and thus split into several steps.

The computation of the high-frequency noise from the spectral densities is made
after replacing high-frequency contaminants (both compact sources and glitches) with
interpolates (see Sect.\,\ref{highf}). The samples that are filtered in this process
are further examined to separate glitches from real signal. Glitches must simultaneously
show a significant deviation from the median signal at their location, and exceed
three or four times this reference signal in absolute value (to avoid flagging the sidelobes
of a bright compact source). An example of the effectiveness of this approach is
shown in Sect.\,\ref{spire_atlas}. This is especially useful for SPIRE, because
the pipeline includes Fourier-filtering tasks, that can strongly amplify undetected
glitches in some instances, as shown in Fig.\,\ref{fig:ampl_glitch}. This problem
may be alleviated by using second-level deglitching, or can
be dealt with by {\it Scanamorphos}.

The computation of the brightness drifts makes use of some simple statistics on
bolometer crossings, in order to filter out those with large standard deviations,
to protect compact sources (see Sect.\,\ref{averdrift}). These statistics can also
be used to detect glitches. The separation between glitches and real signal with
a strong brightness gradient is made by estimating gradients and standard deviations
just before and just after the suspect samples.
This second deglitching is bypassed when the stability length, determining the pixel
size of the grid used for drifts computation, is large compared with the beam FWHM.

\subsubsection{Detection of transient sources}
\label{asteroids}

Asteroids or other transient sources, that are present at a given location in one scan
but have moved or vanished in the other scans, are also detected.
The method used to detect them and distinguish them from glitches is based on the fact that
they produce a coherent signal for all bolometers as the latter cross the source, while glitches
do not. They are dealt with at the same time as glitches in step 7 of Section\,\ref{algooverview}.
They will thus be ignored if the deglitching is deactivated.

If asteroids are detected, a transient-signal map is produced at the end of the processing,
using the same spatial grid as the other final maps. In this map, only the affected portions
of the bolometer series are projected, and not those when the bolometers were not illuminated
by the asteroids. In this way, it is possible to locate them and to obtain an estimate of their
true brightness, whereas they appear dimmer in the nominal signal map (if they are present
in one of two scans at a given position, their brightness will be roughly divided by two).

\subsection{Projection}
\label{proj}

At the end of the processing, a final projection produces a sky map, a weight map,
an error map and a low-frequency noise map, stored together in a FITS data cube.
The low-frequency noise map includes the sum of the drifts subtracted at each
step. The signal is weighted by the inverse square white noise of each bolometer.
The gnomonic projection is used.

The brightness unit is different for both instruments. The flux calibration of SPIRE
is valid for point sources, and voltages are converted to Jy per beam by the
pipeline. To obtain extended-source flux calibration, the maps have to be divided
by the average
beam area. Since the associated uncertainties are larger than those of the point-source
flux calibration, and since beam areas may continue to be refined in the future,
it is best to apply this correction outside the map-making software. SPIRE maps
produced by {\it Scanamorphos} are thus in Jy per beam. For PACS, the data are initially
in Jy per array pixel, and are converted to Jy per map pixel by {\it Scanamorphos}.
The default pixel size, approximately one fourth of the FWHM, better samples the
point response function (PRF) than that
recommended for the projection within the SPIRE pipeline
(between 0.33 and 0.4 times the FWHM, depending on the wavelength).

In the SPIRE pipeline, each sample is projected entirely to the closest pixel.
Since pixel sizes are a large fraction of the FWHM, this method introduces
a bias between the beam center and the projection center. This bias varies as a function
of the source position relative to the pixel grid, and is up to $a / \sqrt{2}$, where
$a$ is the pixel size. We use instead a projection method akin to the drizzling technique
\citep{Fruchter02}, where the imprint of each sample on the map is a uniform disk
of the same area as the default pixel for the final projection (i.e. ${\rm FWHM}^2 / 16$),
and of the same center as the beam. This method does not introduce any bias in the
projection center, and reduces the standard deviation of the sky with respect to
HIPE maps, by a significant amount (in some configurations,
by more than 30\% if the low-frequency noise
has been perfectly removed). This comes at the expense of a slightly larger PRF:
the FWHM is increased by 1.5\%
(and the area by 3\%). Note that the projection affects only the PRF profile,
and has no incidence on the beam area to be used to convert units of Jy per beam
to units of Jy per pixel.

For PACS, our projection also uses the simplifying assumption of axisymmetric
bolometers, whereas the bolometers in reality consist of square pixels
(with small distortions). The bolometer
imprint is chosen to have a slightly smaller area than the true array pixel, just like
in the drizzle projection. It is in principle possible to save the corrected signal
series and to perform the final projection within the pipeline for optimum accuracy,
but the PRF has not been found to be appreciably modified by our projection.

\subsubsection{Weight and error maps}
\label{weight_error_maps}

The weight map is defined differently from the coverage map produced by HIPE.
It is built by coadding the weights, using the same projection as the sky map,
and is normalized by the average of the weights (the average is computed over
all the valid bolometers and all the scans). For the nominal data acquired at
constant scan speed, the weights are defined as the inverse square white noise
values. For turnaround data, acquired at lower or higher scan speed, these weights
are decreased so that edges of the map, where both nominal and turnaround data
can contribute, are not dominated by low-velocity data.

The error map produced by {\it Scanamorphos} is defined as the error on the mean brightness
in each pixel. It is built using the (unbiased) weighted variance instead of the
simple unweighted variance, since weights are used for the projection of the
sky map. It does not include any errors associated with the different processing
steps in the pipeline, because these errors are currently not propagated.
It should however be useful to estimate the random photometric errors, and
to filter glitches that may have escaped detection. Elevated errors will be found
both at the location of bright compact sources and in pixels affected by unmasked
glitches.

Artefacts that may remain when the redundancy is low can also be spotted by comparing
the signal map with the ``clean'' map in the fifth plane of the cube, that is weighted
to exclude noisy scans in each pixel. The latter map is produced in general
only when there are two or three scans, and is intended as a diagnostic for the
presence of artefacts in the signal map (features that are absent in the "clean" map).
It should not be used for scientific purposes, only as an aid to mask bad data.

\subsection{Spatial slicing}
\label{slicing}

For deep observations of wide fields, the software has the ability to slice the
data into several partially overlapping fields, in order to minimize memory
and computing time usage. This is suitable for observations without very extended
bright emission, since it cannot be ensured that extended sources are covered by a single
sub-field. For each sub-field, all the available redundancy is used as if it were
an entire observation. In the end, the different maps are stitched by matching
the brightness levels in the parts of the overlap areas where coverage is nominal.

In the various tests that we conducted, we have not found it necessary to reduce the
data volume when the processing was done on a machine with a total memory of 48 Gb,
but this functionality was used for some deep observations processed on a machine
with a total memory of 8 Gb.

It may sometimes be desirable to do the processing in several spatial blocks, even though
the available memory is sufficient to process the entire observation at once. If the observation
is long and the subtraction of the
short-timescale
average drift is required, limitations on the size of
the drift differences matrix can make the minimum timescale for the average drift computation
longer than that for the individual drifts computation. It is then possible to increase the
time resolution of the average drift by decreasing the data volume, which is achieved by
slicing the field of view into several blocks.

\section{Simulations of PACS observations}
\label{simul}

We now quantify the effect of the processing with {\it Scanamorphos} on the data,
using simulations.
The instrumental artefacts of PACS are more difficult to remove
than those of SPIRE, chiefly because the low-frequency noise has a much greater
amplitude with respect to both the signal and the high-frequency noise, and
because the stability lengths are greater with respect to the FWHM (Sect.\,\ref{space}).
Additional complexities arise from the presence of frequent brightness discontinuities
(see Sect.\,\ref{jumps}),
a sampling rate divided by two in parallel mode for the blue array, and a quantization noise
that is in general a large fraction of the high-frequency noise (Sect.\,\ref{highf}).
Consequently, our simulations are restricted to the PACS instrument, for which
it is more meaningful to demonstrate the response of {\it Scanamorphos}.
\footnote{
{\ifgras \bf \fi
A supplementary document, demonstrating the conservation of very extended diffuse
emission (which cannot be realistically simulated because of the high quantization noise
in staring observations), can be found at this location:
}
\url{http://www2.iap.fr/users/roussel/herschel/report\_cirrus\_pacs\_mips.pdf}\,.
}

\subsection{Star formation region}
\label{simu_n6334}

Since no artificially-made input will ever capture the true complexity of the sky,
we simulated an input sky using a map produced from real observations of a Galactic
star formation region, NGC\,6334, studied by \cite{Russeil12}.
These observations were made in parallel mode for a
duration of 7.29\,hours, at the scan speed of 20 arcsec/s, and we used the blue band
(70\,$\mu$m) for the simulation. The total field area is a little over 3 square degrees.
The redundancy is limited to the minimum recommended, since the observation is made
of two scans.

The noise series are space calibration measurements taken during a staring observation,
and
{\ifgras \bf \fi
shall
}
be used by the PACS simulator
{\ifgras \bf \fi
(Herv\'e Aussel, private communication).
}
They are affected by glitches and very frequent brightness discontinuities.
This staring observation was not long enough to cover the duration of the simulation.
Therefore, we used all the data from the blue array, in both the 70 and 100\,$\mu$m bands
successively, for a total duration of 5.91\,hours. There is a slight bias in the average
white noise levels of both datasets, due to the change of filter, but only at the $0.6 \sigma$
level (3.5\% of the absolute white noise values), so we do not correct for it.

\begin{figure*}[!ht]
\centering
%
%
\hspace*{-0.5cm} \includegraphics[width=6cm, angle=90]{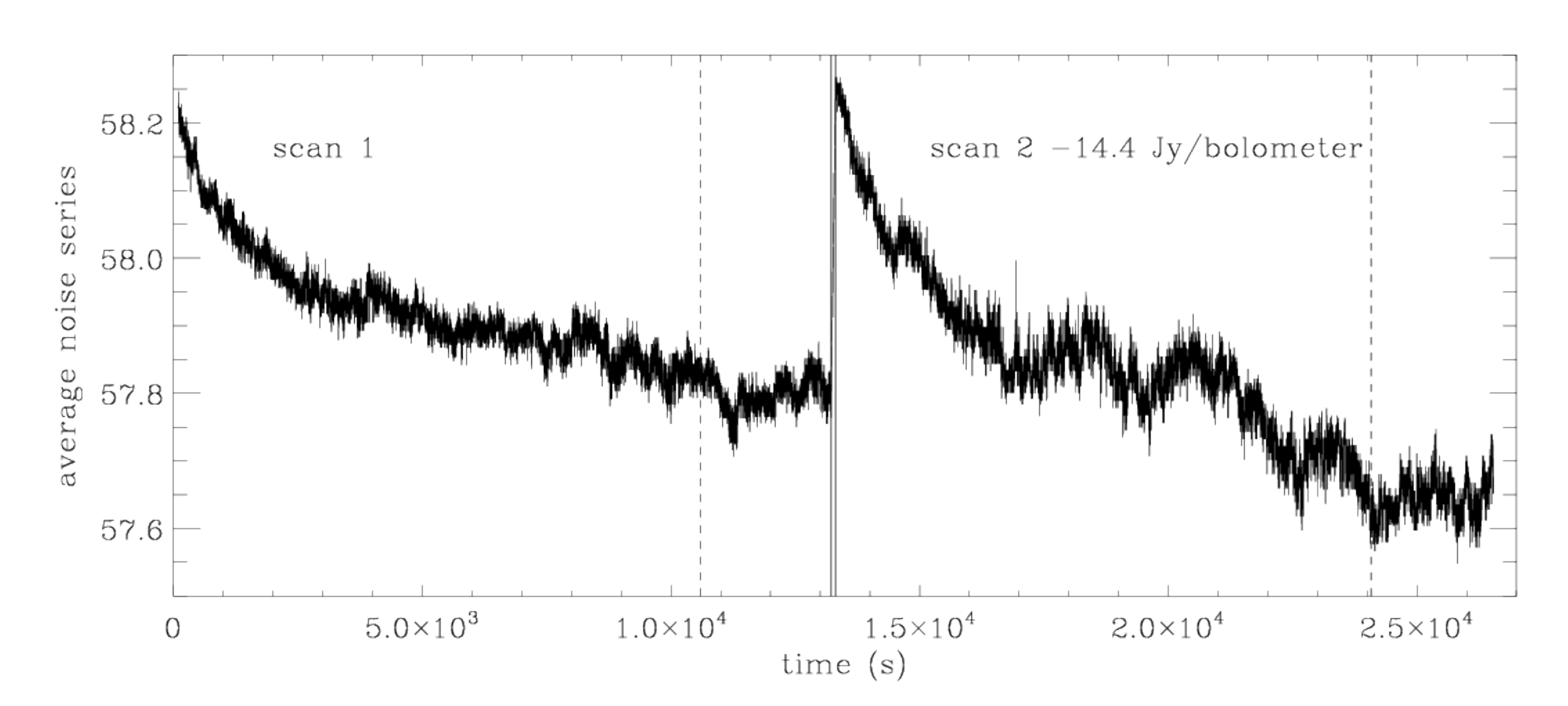}
\caption{Average over all valid bolometers of the noise series introduced in the
simulation (in Jy per bolometer).
The second scan is plotted with an offset for clarity. The dashed lines
indicate, for each scan, the time after which the series had to be extrapolated
(see the text).
}
\label{fig:avernoise}
\end{figure*}

To circumvent the fact that the duration of the noise observation still falls short,
we extrapolated the noise series in the following way: for each bolometer in turn,
we randomly select chunks of 100\,s in the observed series (excluding the start
affected by strong transients from the calibration sources systematically observed
before each science observation), and concatenate them to the signal after matching
their initial brightness with the final brightness of the preceding chunk. The original
time indices of the chunks are different for all bolometers, so that the average noise
cannot repeat itself. In addition to the five unstable rows that
are automatically excluded for Galactic fields
(Sect.\,\ref{preproc} and Fig.\,\ref{fig:pacs_jump}b),
we detected a total of 132 brightness
discontinuities affecting whole array rows, and 80 jumps affecting single bolometers,
but they were not corrected
(since this was impossible when using the {\it /galactic} option at the time the simulation
was made and analyzed).
Figure\,\ref{fig:avernoise} shows the average series of the noise, to demonstrate
that the simulation contains both a mild thermal drift and uncorrelated noise.

\begin{figure*}[!ht]
\centering
\includegraphics[width=14cm]{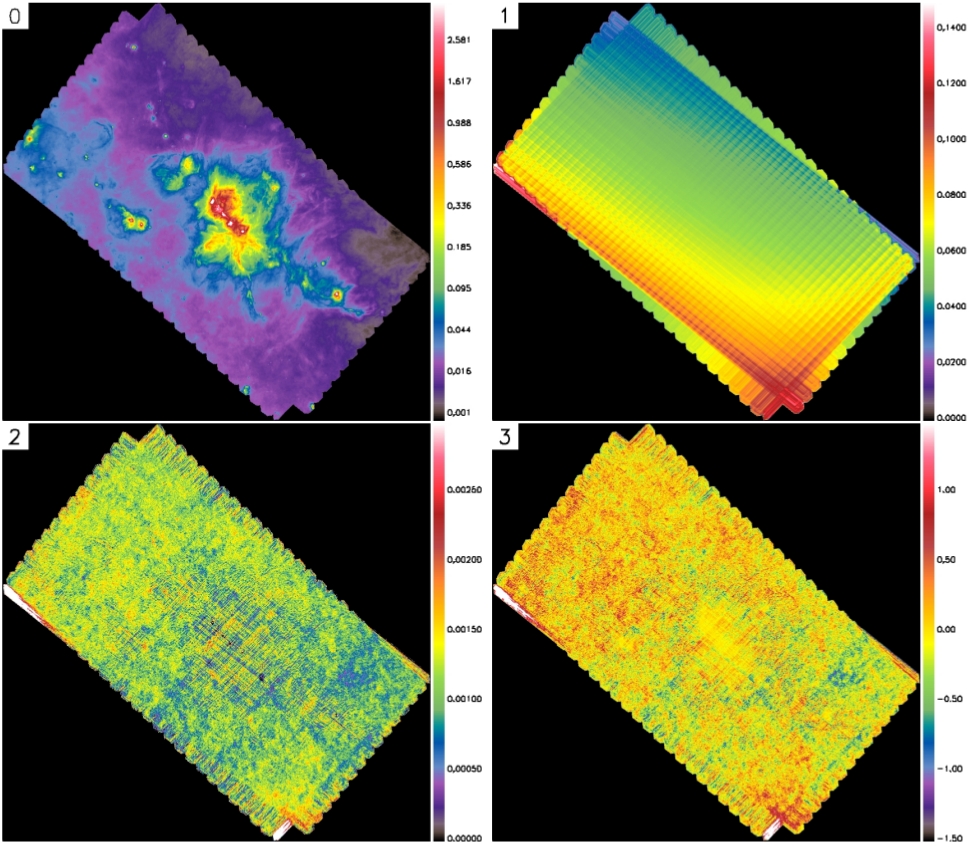}
\caption{
Simulation of the star formation region NGC\,6334.
0) Ideal map (with an applied brightness cut of 4\,Jy/pixel; the maximum
brightness is 33.6\,Jy/pixel).
1) Map of the noise added to the input sky, with the flux
calibration offsets removed. The displayed brightness range is 0 to 0.15 Jy/pixel.
2) Difference of the processed map and the ideal map (range 0 to 3 mJy/pixel).
3) Same difference map, with the median offset removed, and normalized by
the error map produced by {\it Scanamorphos} (range $-1.5 \sigma$ to $+1.5 \sigma$).
}
\label{fig:diffmaps}
\end{figure*}

We simulated time series from the map of NGC\,6334, added the noise series to them,
and then applied to the result the same digitization as that affecting the staring observation
(i.e. the quantized values of the result have the same minimum and the same digitization step).
In these data, the digitization noise represents 50\% of the white noise on average.
This simulation was processed with the {\it /galactic} option and the default choices
(deglitching, subtraction of the average drift, inclusion of turnaround data,
projection on pixels of the default size). We checked that disabling the
short-timescale
average drift
subtraction produces virtually identical results. We also projected on the same astrometric
grid two other maps: that of the noise from the staring observation, after subtracting
the flux calibration offsets to obtain the same median brightness for all bolometers;
and that of the idealized input signal from the map of NGC\,6334, after eliminating
the five unstable rows that are always excluded for Galactic fields, and without
any quantization nor weighting of the signal. We will refer to these three maps as
the processed map, the noise map and the ideal map, respectively.

\begin{figure*}[!ht]
%
%
\includegraphics[width=16.5cm]{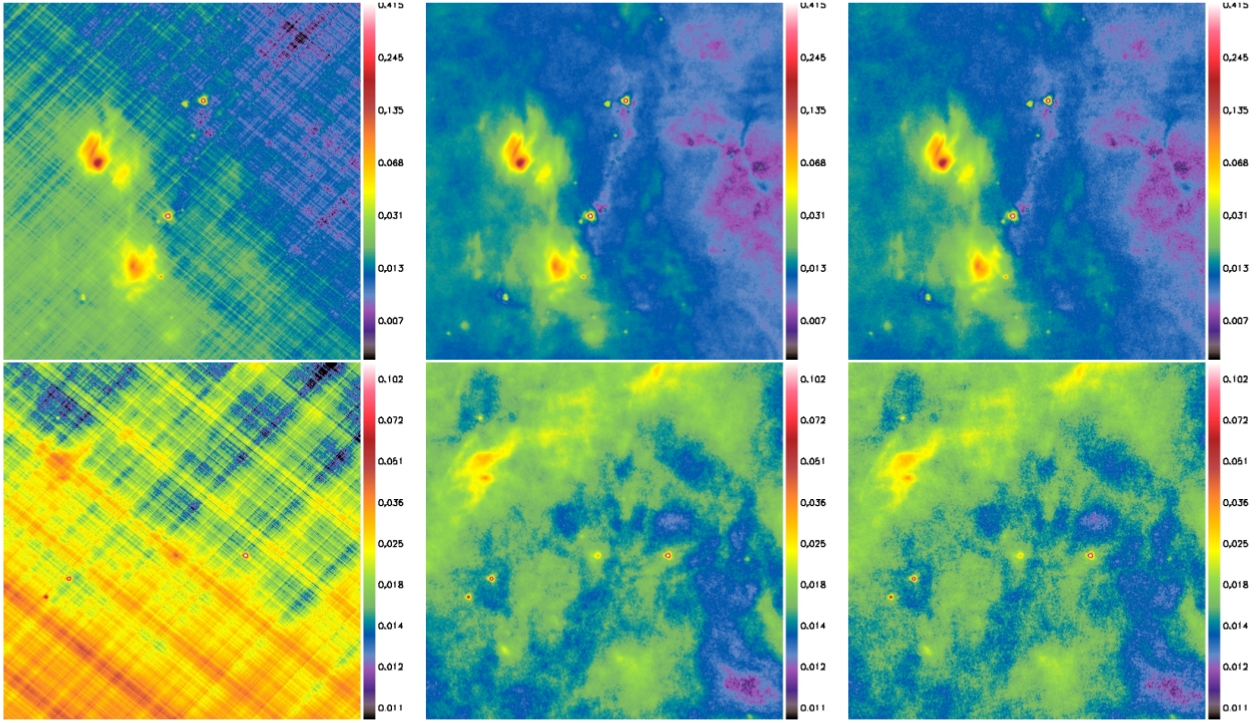}
\caption{Two small portions of the total field of view of the simulation
of NGC\,6334, $14^{\prime}$ on a side, are shown (in Jy/pixel).
From left to right: (ideal + noise) map, displayed with a brightness offset;
ideal map; processed map, displayed with the general offset of 1.1\,mJy/pixel. The flux
calibration offsets have been removed from the noise map, as above.
}
\label{fig:simu_zoom}
\end{figure*}

\begin{figure*}[!ht]
%
%
\vspace*{-0.3cm}
\hspace*{-0.5cm} \includegraphics[width=6.5cm, angle=90]{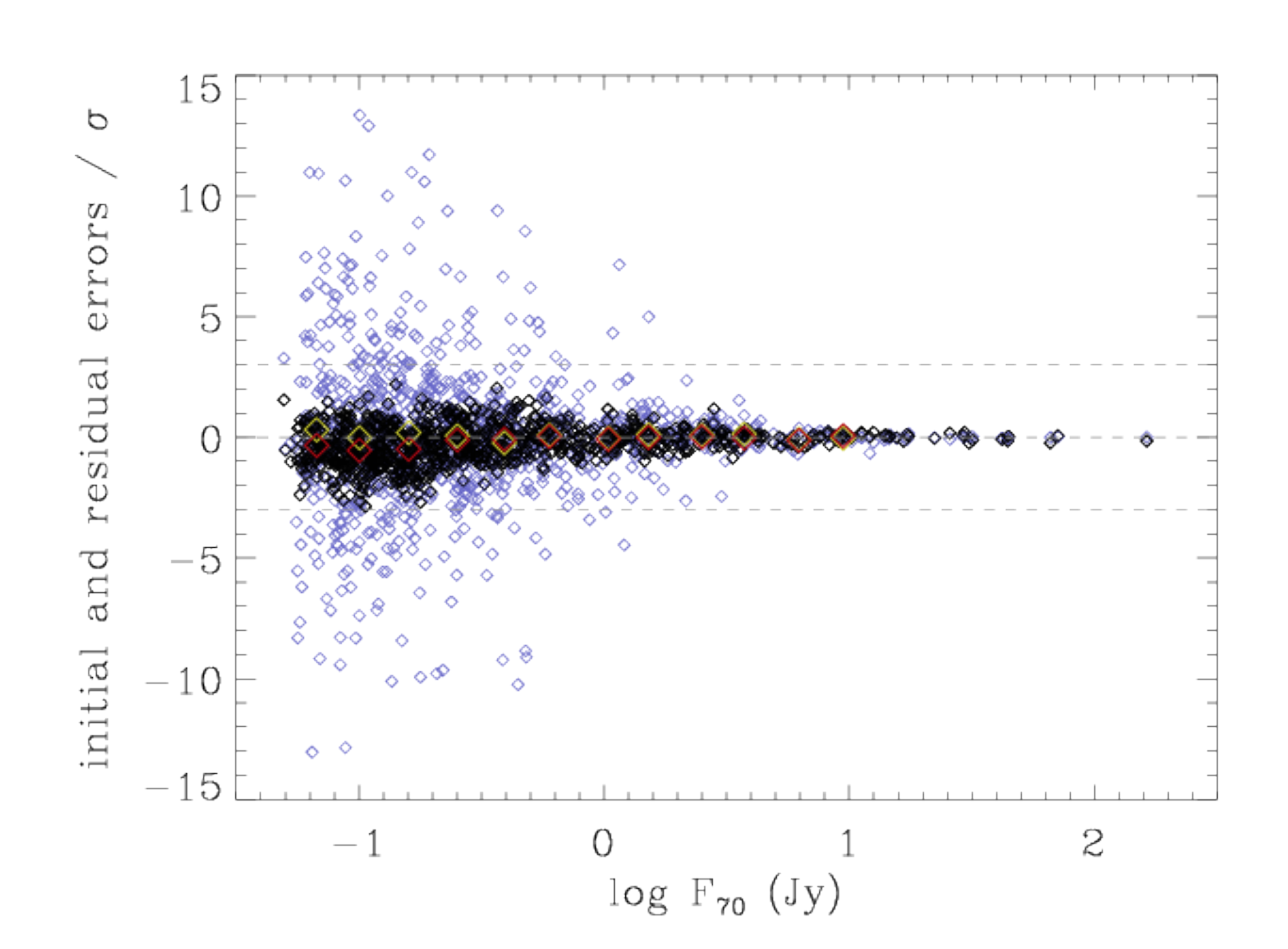}
\hspace*{-0.5cm} \includegraphics[width=6.5cm, angle=90]{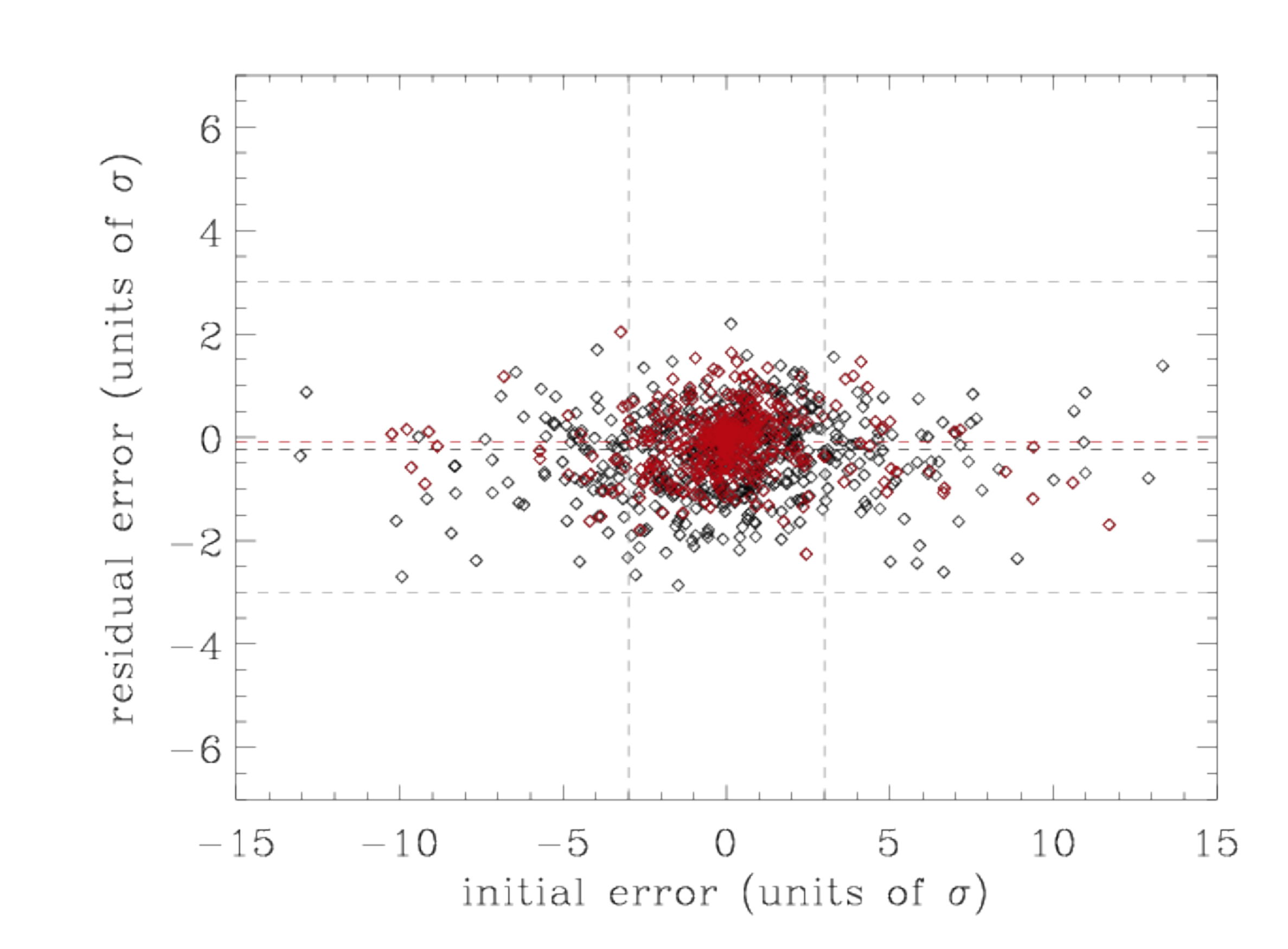}
\caption{{\bf Left:} The deviation from the flux measured in the ideal map, in units
of the photometric uncertainty $\sigma$, is plotted as a function of this flux. The deviation
introduced by the noise map (the initial error) is plotted in blue, and the deviation
in the processed map (the residual error) in black. The yellow diamonds are averages
computed in bins of 0.2\,dex for the initial error, and the red diamonds represent
the same for the residual error. To guide the eye, horizontal lines at $\pm 3 \sigma$
and zero are overplotted. {\bf Right:} The residual error is plotted as a fonction
of the initial error, both in units of $\sigma$ (notice the different scale for both axes).
Sources above $5 \sigma$ are in black, and sources above $15 \sigma$ in red.
The grey lines show the $\pm 3 \sigma$ limits, and the horizontal lines in the middle
show the average residual error.
}
\label{fig:simu_results}
\end{figure*}

\begin{figure*}[!ht]
\centering
%
%
\includegraphics[width=6.5cm, angle=90]{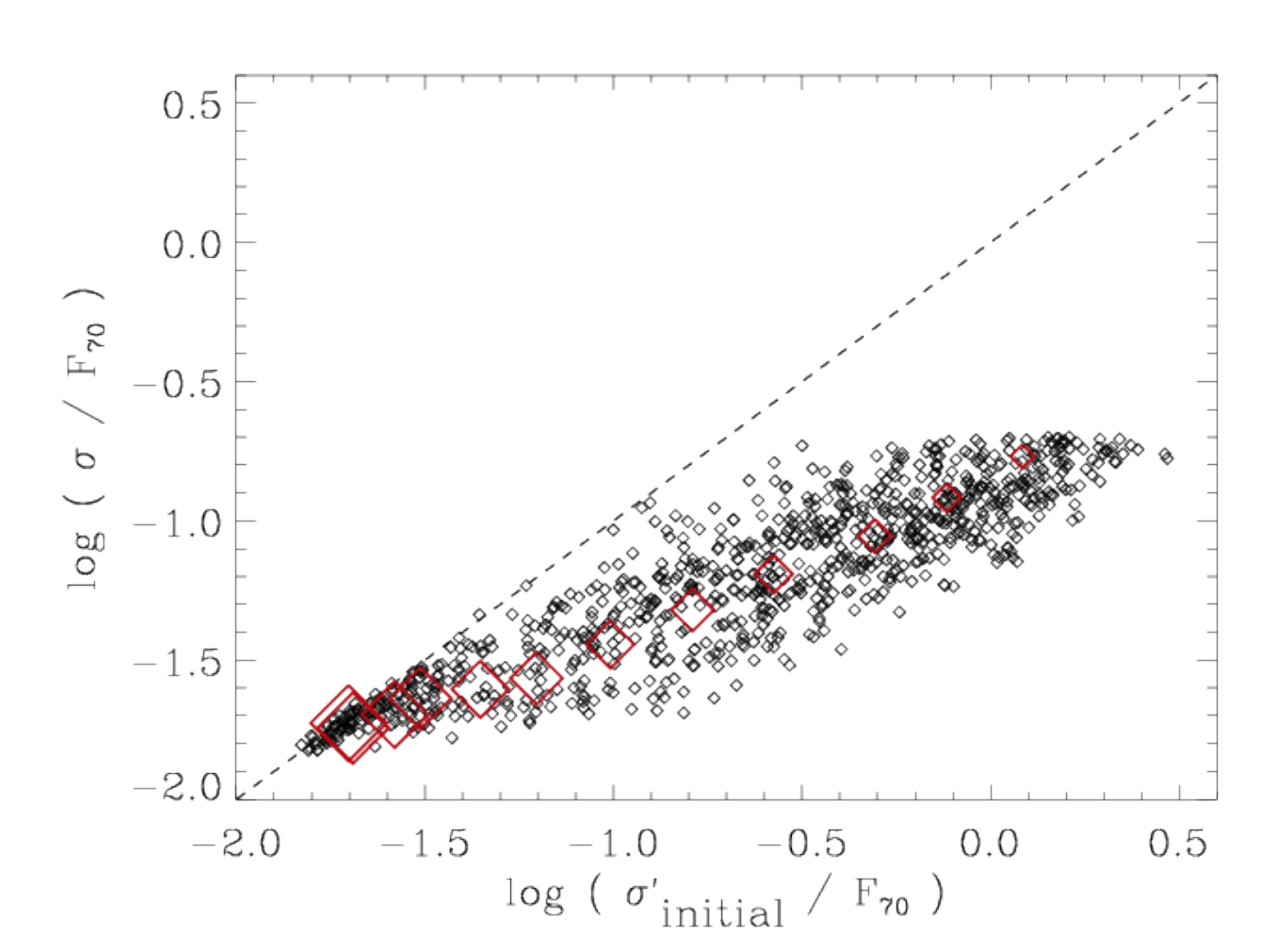}
\caption{The photometric uncertainty $\sigma$ (measured in the processed map)
is plotted as a function of the uncertainty $\sigma\prime_{\rm initial}$ measured
in the noisy map (sum of the ideal and noise maps). Both are normalized by the
true flux $F_{70}$.
The shift from the dashed line indicates by how much the noise has
been reduced in the map. The red diamonds indicate the average values in the
same flux bins as in Figure\,\ref{fig:simu_results} (left), and their size
increases proportionally with $\log F_{70}$. For high fluxes, the photometric
uncertainty is completely dominated by the binning noise that is present in the
ideal map; it results from projecting in finite pixels a signal that has a very
steep spatial gradient, i.e. a bright compact source.
}
\label{fig:simu_noise}
\end{figure*}

We analyze the results in several ways. First, we display the difference between
the processed map and the ideal map, to assess the effect of the noise and processing
on the very extended emission (Fig.\,\ref{fig:diffmaps}).
There is a slight offset between the processed and ideal maps (of 1.1\,mJy/pixel),
but offsets are of no relevance since Herschel instruments are not absolute photometers.
The zero-level of the sky emission can only be calibrated with other instruments.
Thus, we consider that the subtraction of this global offset is not a necessary or even
meaningful task for Herschel mapping tools. What is important to notice is that there
is no significant brightness gradient in the difference map, hence that the
{\ifgras \bf \fi
average
}
drift
was satisfactorily removed. The imprint of the sources is also not discernible, which
implies that the extended emission was preserved throughout the field. Low-level structure
due to imperfect drift subtraction is present, but remains within the photometric errors.
Figure\,\ref{fig:simu_zoom} shows details of the three maps in two small portions of
the field, chosen among the regions of relatively diffuse emission.

Then, we perform simple aperture photometry on all the compact sources (point sources
or slightly extended sources such as background galaxies) that we can find in the
ideal map wherever the coverage is nominal, and apply the same apertures to
the three maps. Although many sources sit on a very structured local background,
we make no attempt to correct for contamination from nearby sources or underlying
extended emission, since our aim is not to measure the absolute fluxes, but to test
the flux conservation. We use an aperture of 10 pixels in diameter, i.e. $14\arcsec$
or 2.5 FWHM, and an annulus for the local background of inner diameter 10 pixels and
outer diameter 24 pixels, i.e. $33.6\arcsec$ or 6 FWHM. We measure photometric uncertainties,
noted $\sigma$, by adding in quadrature the error on the mean brightness in each pixel
(from the error map produced by {\it Scanamorphos}). They are measured in the processed
map to obtain realistic values, and not in the ideal map, since by construction the latter
does not contain any white noise. If the drifts had been perfectly removed, $\sigma$
would be dominated by the white noise, but since residual errors are correlated
because of the small pixel size, these uncertainties are underestimated. We obtain
879 compact sources detected in the ideal map at the $5 \sigma$ level or more.

With these measurements, we test two properties at small spatial scales: whether
flux is conserved, and to which extent the noise is reduced. Figure\,\ref{fig:simu_results}
shows that the photometric errors initially introduced by the drifts, up to $20 \sigma$,
are greatly reduced by the processing, the residual errors being all within $3 \sigma$.
There is a slight residual average bias for faint sources (below 170\,mJy), but well
within the photometric error bars, of the order of $-0.5 \sigma$. This bias disappears
when the detection significance level increases, and would probably also disappear for
faint sources by increasing the redundancy in the observations, i.e. increasing
the number of scans or observing in nominal mode. Figure\,\ref{fig:simu_noise} shows
in addition by which amount the photometric uncertainty computed from the error map
is reduced, as a function of the initial uncertainty and of the source flux. We
conclude that {\it Scanamorphos} is both flux-conserving and efficient at reducing
the noise, including on small spatial scales.

\subsection{Nearby galaxy}

We now simulate a field with a much smaller dynamic range, typical of extragalactic
observations. To generate the input sky, we use an ancillary Spitzer observation
at 8\,$\mu$m of the face-on galaxy NGC\,3184, included in the SINGS and KINGFISH samples
\citep{Kennicutt03, Kennicutt11}. Since this map (showing mostly the emission from polycyclic
aromatic hydrocarbons in this band) contains both compact and very diffuse emission,
it provides a realistic test case for the far-infrared emission of galaxies.

We first subtract a tilted plane to rectify the background, then convolve the map to
the resolution of PACS at 160\,$\mu$m, using the kernel and tool provided by
\citet{Aniano11}, and rescale the peak brightness (of the nuclear regions) to that
of NGC\,3184 at 160\,$\mu$m.

The noise series are extracted, like those of the previous simulation, from a staring
observation (courtesy Marc Sauvage and PACS Instrument Control Center). They contain glitches
but only a very shallow thermal drift. Since their quantization step (Sect.\,\ref{highf})
is twice that of the extragalactic observations, we have to correct for this before adding
the noise series to the simulated sky series. Otherwise, the sensitivity would be artificially
decreased with respect to real observations. We thus average the noise series from two
successive observations
to divide the quantization step by two, and add back to the result $\sqrt{0.5}$
times the initial white noise. The (sky + noise) series were then digitized to match the
quantization step of the noise, as in Section\,\ref{simu_n6334}. The resultant digitization
noise to white noise ratio is of the order of 30\%, the same as in the real 160\,$\mu$m
observation of NGC\,3184.

This simulation was processed with the {\it /jumps\_pacs} option and the default choices.
In addition to the maps described for the previous simulation, we also produce an
(ideal + white noise) map, where only white noise has been added to the simulated sky
series, and the result digitized, to illustrate the erosion of the sensitivity that
results from this irreducible noise component. This map is not used for the analysis:
the brightnesses of the simulation are compared with those of the ideal map (without
any noise nor quantization). The results from the normal processing are also compared
with those obtained when the drifts are subtracted only on long timescales (i.e. the
processing is stopped after the baseline subtraction). The corresponding map is called
the noisy map or the deglitched noisy map, depending on whether the first iteration of
the deglitching after baseline subtraction was performed or not.

\begin{figure*}[!ht]
%
%
\hspace*{-0.8cm}
\includegraphics[width=18cm]{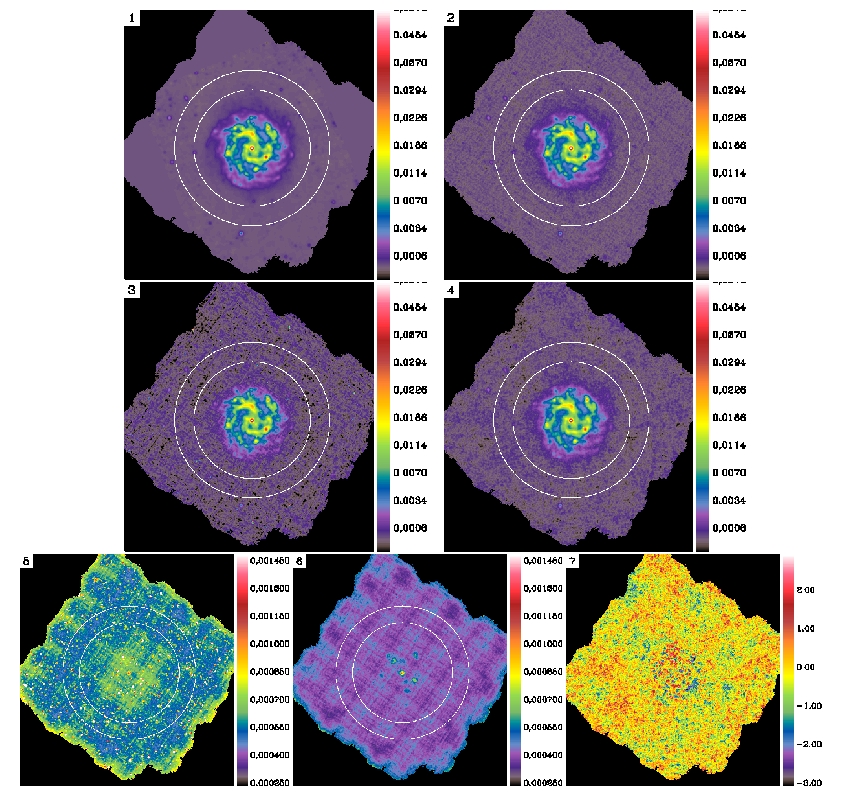}
\caption{
Maps of the galaxy simulation (in Jy/pixel).
1) Ideal map.
2) (Ideal + white noise) map (see text). 3) Noisy map, with only long-timescale
drifts (baselines) subtracted. 4) Processed map.
5) Error map associated with the noisy map. 6) Error map associated
with the processed map, displayed with the same brightness scale.
7) Difference between the processed and ideal maps, normalized by
the error map produced by {\it Scanamorphos} (range $-3 \sigma$ to $+3 \sigma$).
The white circles show the galaxy and background apertures.
}
\label{fig:maps_simu_gal}
\end{figure*}

\begin{figure*}[!ht]
%
%
\hspace*{-1.5cm} \includegraphics[width=4.5cm, angle=90]{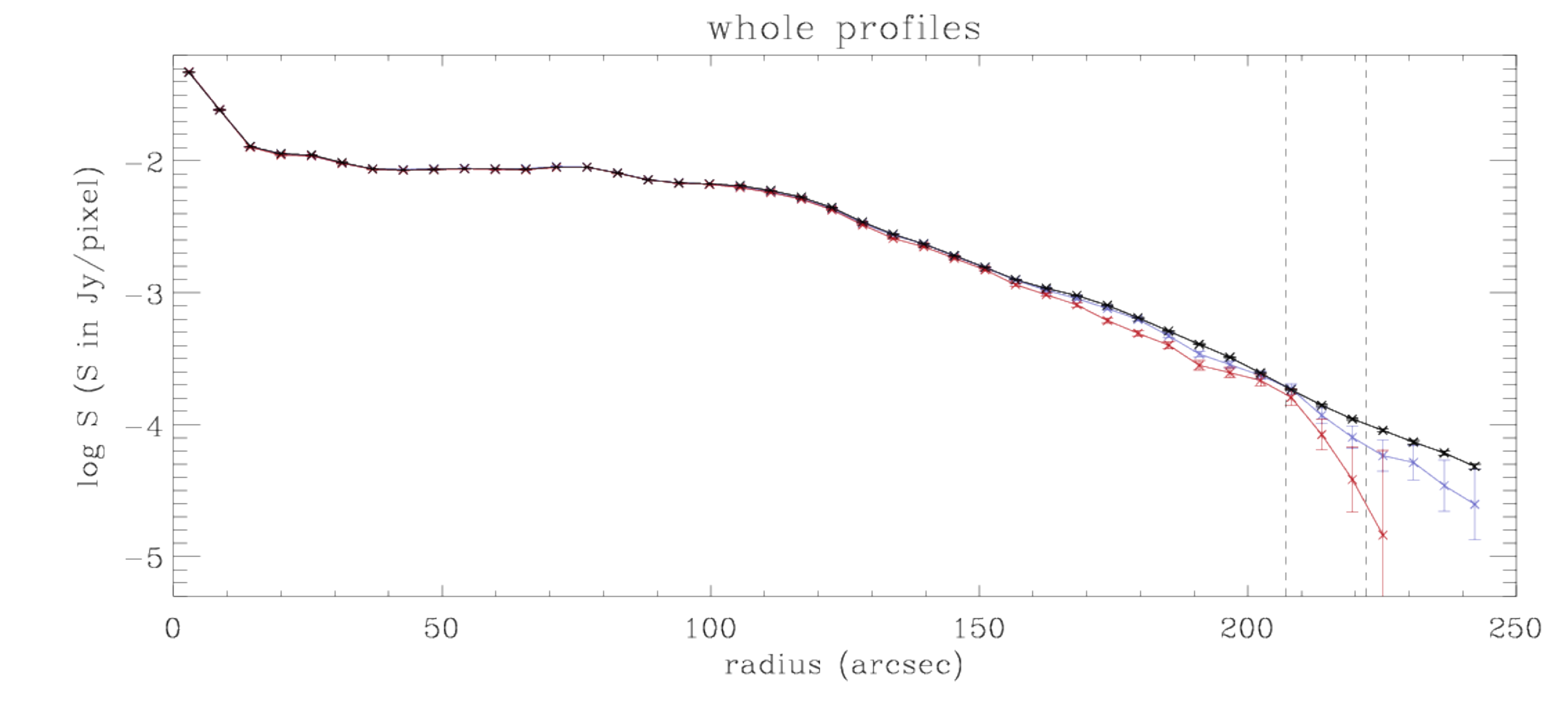}
\hspace*{-0.7cm} \includegraphics[width=4.5cm, angle=90]{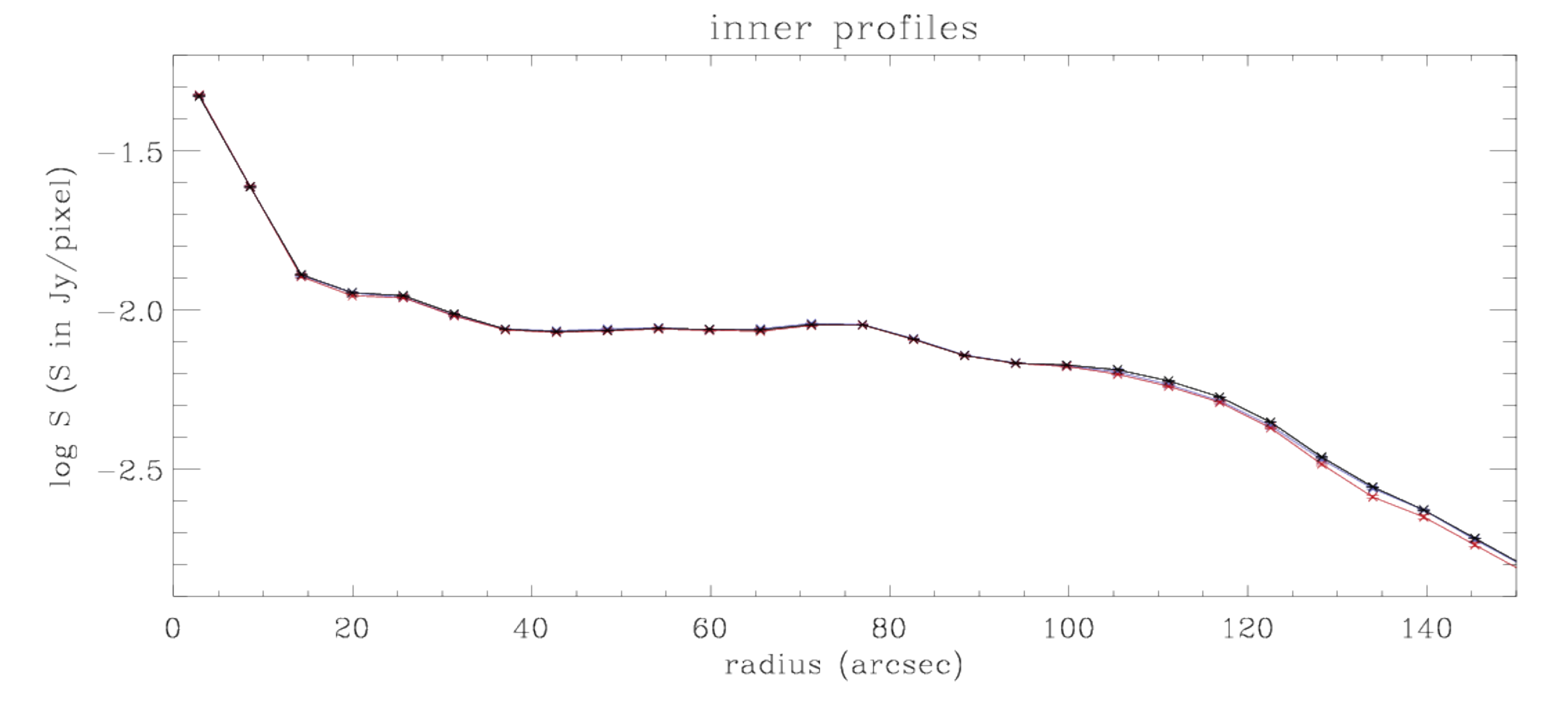}
\caption{Surface brightness profiles obtained from the ideal map (in black),
the deglitched noisy map (in red) and the processed map (in blue). Each annulus
has a width of 2 pixels, i.e. ${\rm FWHM} / 2$\,. The vertical dashed lines
in the first panel indicate the $R_{25}$ minor and major axis radii in the B band.
}
\label{fig:simu_profiles}
\end{figure*}

Figure\,\ref{fig:maps_simu_gal} shows the various signal and error maps, and
Figure\,\ref{fig:simu_profiles} the azimuthally-averaged surface brightness profiles.
The photometric uncertainties are reduced by the subtraction of the short-timescale
drifts, and the errors (the difference between the processed map and the ideal map)
are everywhere below the photometric uncertainties, as estimated from the error map
produced by {\it Scanamorphos} (Fig.\,\ref{fig:maps_simu_gal}). Some residual artefacts
are visible at the location of the galaxy, but their amplitude remains mostly below
the $2\sigma$ significance level. As demonstrated by Figure\,\ref{fig:simu_profiles},
the processing brings significant improvements in the recovery of diffuse emission,
at radii beyond about 150$^{\prime\prime}$.

\begin{figure*}[!ht]
\centering
%
%
\includegraphics[width=6.5cm, angle=90]{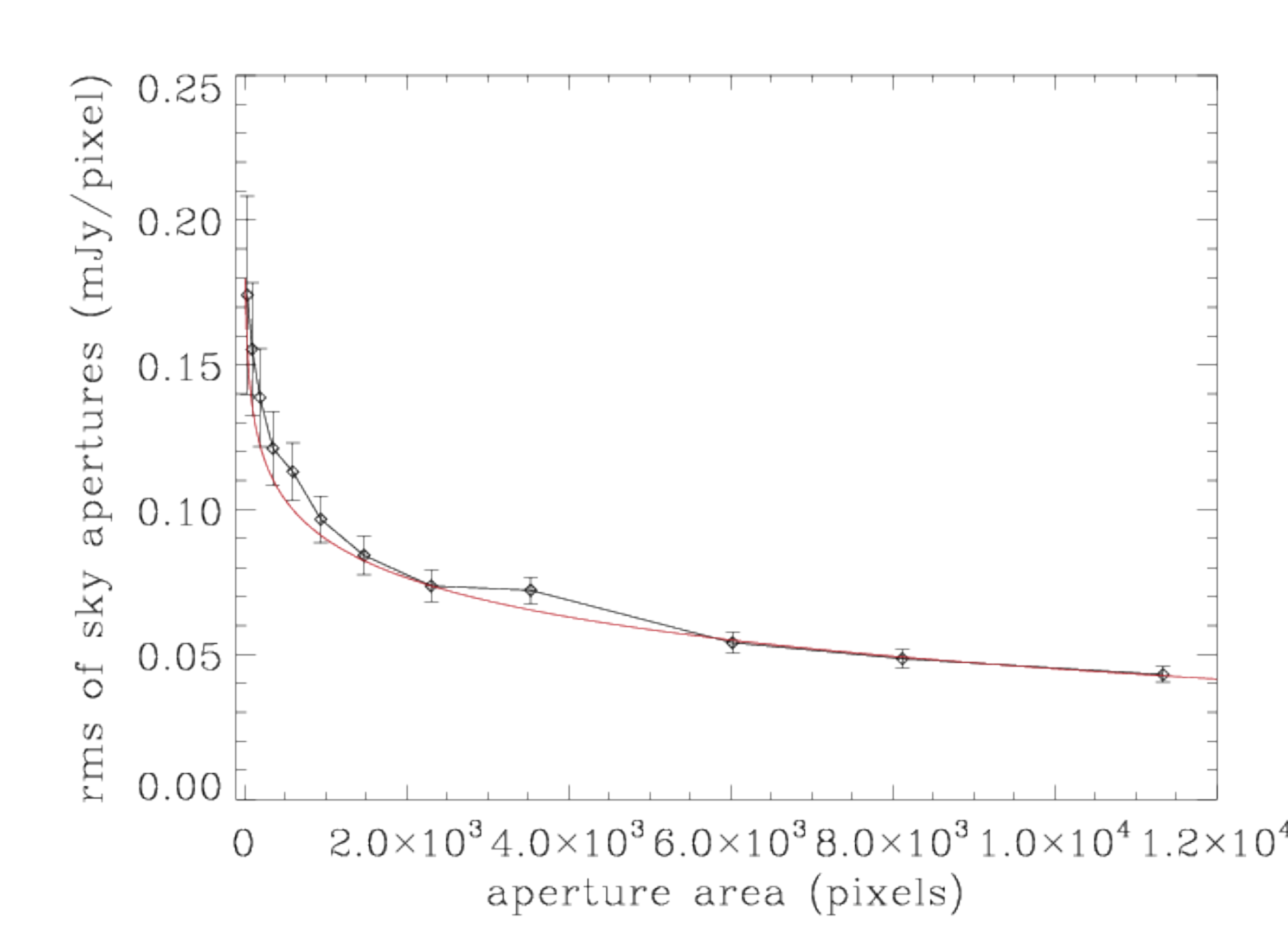}
\caption{Standard deviation of the average sky brightness in independent
apertures of equal area, as a function of aperture area. The vertical bars
represent the typical pixel-to-pixel standard deviations within individual
apertures. The function shown in red was used to extrapolate the results
to larger apertures, since the map is too small to define several background
apertures of the same size as the galaxy.
The dashed line represents the same quantity for a simulated pure Gaussian noise,
normalized to the smallest aperture area, showing that the measurements are
dominated by correlated errors.
}
\label{fig:simu_correrrors}
\end{figure*}

We also measure the global flux of the galaxy within a radius of 90 pixels,
and the background within an annulus of inner and outer radii of 90 and 120 pixels,
i.e. $256.5^{\prime\prime}$ and $342^{\prime\prime}$ (as indicated in Fig.\,\ref{fig:maps_simu_gal}).
To estimate a background uncertainty that takes into account the correlation of errors
in adjacent pixels, and in larger areas due to long-timescale drift residuals, we
use measurements of the average sky brightness in several sets of independent apertures.
For each set, an annulus around the galaxy is divided into equal-area angular sectors,
defining our sky apertures. From one set to the next, the widths of the annulus and
angular sectors are progressively increased. Fig.\,\ref{fig:simu_correrrors} shows
the standard deviation of the independent sky measurements as a function of aperture area.
The extrapolation of these results to the area of integration of the galaxy flux
gives us an estimate of the component of the background uncertainty that has to be summed
linearly. We define the total uncertainty as
$\Sigma_{\rm tot} = \sqrt{N^2~ \sigma_{\rm b~corr}^2 + N~ \sigma_{\rm b~rand}^2 + \Sigma_{\rm phot}^2}$
where N is the number of pixels of the galaxy aperture, $\sigma_{\rm b~corr}$ is the
correlated uncertainty
on the background just discussed, $\sigma_{\rm b~rand}$ is the random uncertainty
on the background (obtained from the pixel-to-pixel standard deviation of the sky
brightness within the background annulus), and $\Sigma_{\rm phot}$ is the photometric
uncertainty from the quadratically-summed error map. The error budget is completely
dominated by the $N~ \sigma_{\rm b~corr}$ term for this dataset (by 99\%). With
1-$\Sigma_{\rm tot}$ error bars, we obtain a total galaxy flux of ($58.81 \pm 0.05$)\,Jy
in the ideal map, ($56.31 \pm 0.70$)\,Jy in the deglitched noisy map, and
($58.06 \pm 0.69$)\,Jy in the processed map. The uncertainty increases from
0.7\,Jy to 1.1\,Jy if $\sigma_{\rm b~corr}$ is not extrapolated but assumed constant
for aperture areas greater than our last measurements in Fig.\,\ref{fig:simu_correrrors}.
In this simulation, the global flux is thus recovered within 1.2 to $0.8~ \Sigma_{\rm tot}$
(1.4\% of the true flux) in the processed map, while it is 3.7 to $2.4~ \Sigma_{\rm tot}$
below the true flux (4.4\% of the true flux) in the noisy map.

\section{Examples and visualization step by step}
\label{tests}

We apply below the algorithm to diverse datasets (with different types of structures
and brightnesses, in parallel-mode or nominal observations, with various scan speeds),
taken during the Science Demonstration Phase. We provide visualizations in terms of time
series and maps at each important step of the processing. We illustrate different details
on each dataset. All the maps displayed below are in units of Jy per beam for SPIRE
and Jy per array pixel for PACS, except the final maps for PACS whose brightness unit
is Jy per map pixel.
Only the final maps are displayed in the standard astronomical orientation (north to
the top and east to the left). Each dataset is identified by its observation day (OD)
and observation block numbers (obsid).

Table\,\ref{tab:runs} summarizes the field sizes, processing options, drifts amplitudes
and indicative processing times
(excluding the time needed for the initial data ingestion and reformatting)
for all observations presented here as examples, and a few others in addition.
The code has been reasonably optimized for speed, but gains shall still be possible.
The processing was made on a Linux machine with a CPU frequency of 2.93 GHz and
a total memory of 48 Gb; but it is possible to process most datasets, not extremely
deep, with much less memory.
Multithreading, used by default by IDL on multiprocessors, had to be deactivated
on this particular machine to improve performance (processing times decreased by 20 to 35\%
on the few tested datasets).
In this configuration, the ratio of the processing time to the observation duration is
of the order of 7\% at 250\,$\mu$m, 5\% at 350\,$\mu$m, and 2\% at 500\,$\mu$m for relatively
short observations ($< 10$\,hours) made with the SPIRE arrays; 25\%, 15\%, and 7\% at
250, 350, and 500\,$\mu$m, respectively, for the very deep GOODS-N observations;
for the PACS arrays, that are made of a much larger number of bolometers, this ratio
ranges from 50 to 75\% at 70\,$\mu$m and 15 to 20\% at 160\,$\mu$m, in the cases
presented in Table\,\ref{tab:runs}.
For large data volumes, the longest tasks are the baseline subtraction and average
drift subtraction. The individual drifts subtraction is comparatively very fast.
Note that we have adopted the shortest possible timescales for the drifts, and that
the processing times may be decreased by
enlarging the timescale for the average drift.

\begin{deluxetable}{l|l|l|r|r|r|r|l|l|l|l|l|l}
\tablecaption{Processing parameters for a few test observations
\label{tab:runs}}
\rotate
\tabletypesize{\scriptsize}
\tablewidth{0pt}
\startdata
\hline\hline
field  & size  & options  & N$_{\rm s}$ & t$_{\rm obs}$ & t$_{\rm proc}$ & T$_c$ & \multicolumn{2}{c}{N$_{\rm iter}$} & \multicolumn{2}{|c}{amp$_{\rm D}$ / HF} & \multicolumn{2}{|c}{amp$_{\rm D}$ (Jy/bol.)} \\
~      & ~     & ~        & ~           & (s)           & (s)            & (s)   & aver. & indiv.                     & aver. & indiv.                          & aver. & indiv.                               \\
\hline\hline
\multicolumn{13}{l}{~} \\
\multicolumn{13}{l}{SPIRE} \\
\hline
NGC\,6822 250\,$\mu$m    & $20^{\prime} \times 20^{\prime}$   & none       & 6  & 2453  & 182   & 0.59 & 1 & 2 & 141.  & 2.2 & 4.221 & 0.066 \\
NGC\,6822 350\,$\mu$m    & ~                                  & ~          & ~  & ~     & 112   & 0.81 & 1 & 2 & 149.  & 2.0 & 4.819 & 0.065 \\
NGC\,6822 500\,$\mu$m    & ~                                  & ~          & ~  & ~     & 54    & 1.18 & 1 & 2 & 169.  & 2.4 & 6.905 & 0.096 \\
\hline
NGC\,6946 250\,$\mu$m    & $30^{\prime} \times 30^{\prime}$   & none       & 4  & 2430  & 178   & 0.43 & 1 & 2 & 52.   & 1.6 & 1.591 & 0.048 \\
NGC\,6946 350\,$\mu$m    & ~                                  & ~          & ~  & ~     & 126   & 0.43 & 1 & 2 & 57.   & 1.5 & 1.860 & 0.050 \\
NGC\,6946 500\,$\mu$m    & ~                                  & ~          & ~  & ~     & 59    & 0.59 & 1 & 2 & 67.   & 1.6 & 2.762 & 0.065 \\
\hline
Rosette 250\,$\mu$m      & $88^{\prime} \times 69^{\prime}$   & /parallel, & 2  & 19317 & 1245  & 0.9  & 1 & 2 & 242.  & 11. & 6.109 & 0.285 \\
Rosette 350\,$\mu$m      & ~                                  & /galactic  & ~  & ~     & 790   & 1.3  & 1 & 2 & 255.  & 11. & 6.890 & 0.294 \\
Rosette 500\,$\mu$m      & ~                                  & ~          & ~  & ~     & 384   & 1.8  & 1 & 2 & 280.  & 17. & 9.646 & 0.598 \\
\hline
Atlas 250\,$\mu$m        & $245^{\prime} \times 224^{\prime}$ & /parallel  & 2  & 55209 & 10056 & 0.6  & 2 & 2 & 1332. & 58. & 33.11 & 1.442 \\
Atlas 350\,$\mu$m        & ~                                  & ~          & ~  & ~     & 8110  & 0.6  & 2 & 2 & 1457. & 55. & 38.70 & 1.456 \\
Atlas 500\,$\mu$m        & ~                                  & ~          & ~  & ~     & 6791  & 0.6  & 2 & 2 & 1704. & 73. & 57.73 & 2.460 \\
\hline
GOODS-N 250\,$\mu$m      & $33^{\prime} \times 33^{\prime}$   & none       & 60 & 47696 & 12135 & 0.59 & 2 & 2 & 743.  & 2.0 & 22.50 & 0.061 \\
GOODS-N 350\,$\mu$m      & ~                                  & ~          & ~  & ~     & 7200  & 0.81 & 2 & 2 & 812.  & 1.9 & 25.70 & 0.060 \\
GOODS-N 500\,$\mu$m      & ~                                  & ~          & ~  & ~     & 3354  & 1.18 & 2 & 2 & 976.  & 2.1 & 42.06 & 0.090 \\
\hline
\multicolumn{13}{l}{~} \\
\multicolumn{13}{l}{PACS} \\
\hline
NGC\,4559 70\,$\mu$m     & $15^{\prime} \times 15^{\prime}$   & /jumps\_pacs & 12 & 6838  & 4878  & 0.7  & 1 & 3+3 & 131. & 9.2 & 1.514 & 0.107 \\
NGC\,4559 100\,$\mu$m    & ~                                  & ~            & 12 & 6838  & 4934  & 0.7  & 1 & 3+3 & 90.  & 8.4 & 1.027 & 0.095 \\
NGC\,4559 160\,$\mu$m    & ~                                  & ~            & 24 & 13676 & 2650  & 0.9  & 1 & 3+3 & 95.  & 11. & 2.034 & 0.229 \\
\hline
Centaurus\,A 70\,$\mu$m  & $36^{\prime} \times 36^{\prime}$   & none         & 8  & 16633 & 12635 & 0.7  & 1 & 3+3 & 28.  & 9.0 & 0.314 & 0.100 \\
Centaurus\,A 160\,$\mu$m & ~                                  & ~            & 8  & 16638 & 3094  & 0.9  & 1 & 3+4 & 8.4  & 11. & 0.175 & 0.230 \\
\hline
M\,81 70\,$\mu$m         & $40^{\prime} \times 40^{\prime}$   & /jumps\_pacs & 8  & 22190 & 16199 & 0.7  & 1 & 3+3 & 24.  & 8.6 & 0.315 & 0.112 \\
M\,81 160\,$\mu$m        & ~                                  & ~            & 8  & 22190 & 3886  & 0.9  & 1 & 3+4 & 13.  & 11. & 0.387 & 0.320 \\
\hline
Rosette 70\,$\mu$m       & $91^{\prime} \times 72^{\prime}$   & /parallel,   & 2  & 19887 & 9814  & 1.2  & 1 & 3+4 & 102. & 38. & 1.165 & 0.436 \\
~                        & ~                                  & /galactic    & ~  & ~     & ~     & ~    & ~ & ~   & ~    & ~   & ~     & ~     \\
\hline
\enddata
\tablecomments{
The table entries are: name and wavelength; area with nominal coverage; options
on the command line; number of scans $N_{\rm s}$; observation duration; indicative processing time;
minimum timescale T$_c$ of the drifts; number of iterations
for average and individual drifts, respectively
(for PACS, the first 3 iterations for individual drifts are made with a time resolution
of 27, 9 and 3 times T$_c$, and the following of T$_c$);
amplitude of the total drifts (including
baselines) in units of high-frequency noise, for average and individual drifts, respectively;
amplitude of the total drifts in Jy per bolometer, also for average and individual drifts,
respectively.
}
\end{deluxetable}

\subsection{Application to SPIRE data}
\label{ex_spire}

\subsubsection{Rosette nebula (OD 159, obsid 1342186121 and 1342186122)}
\label{spire_rosette}

\begin{figure*}[!ht]
\centering
\vspace*{-3.2cm}
\includegraphics[width=15cm]{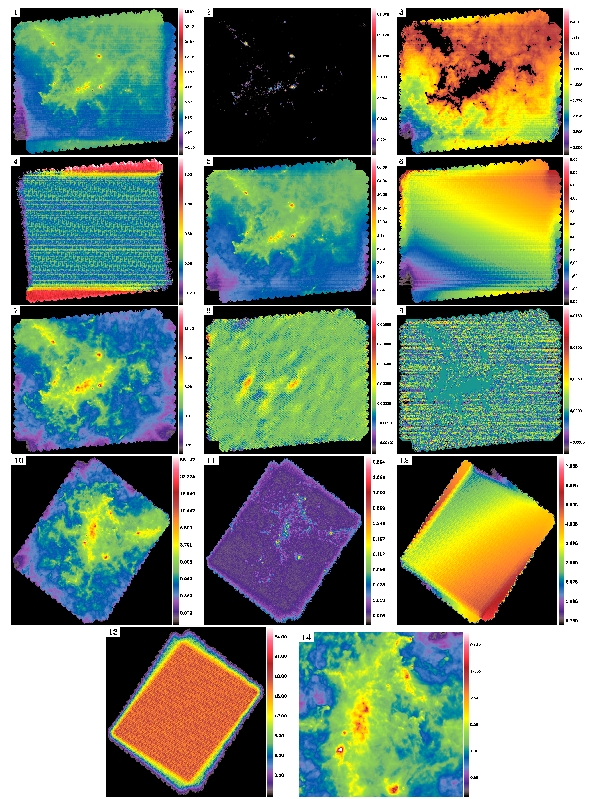}
\vspace*{-0.3cm}
\caption{Successive steps of the processing of the Rosette 250\,$\mu$m observations illustrated
by maps, showing the following: 1) raw level-1 data; 2) signal of compact sources
interpolated for the computation of high-frequency noise; 3) mask applied for the
computation of baselines; 4) simple baselines obtained by
zero-order fits (since the {\it /galactic} option is used);
5) data after
subtraction of the simple baselines; 6) residual baselines derived from the redundancy;
7) data after subtraction of the residual baselines; 8) average drift subtracted at the
first iteration; 9) individual drifts subtracted at the first iteration for these;
10) final map; 11) error map; 12) total drifts (dominated by baselines); 13) weight map;
14) enlargement of a central portion of the final map
(about $50^{\prime}$ on a side). Notice the very different brightness scales for the
signal and for the drifts.
Maps 1 to 9 are projected on a coarse spatial grid (with a pixel size equal to half the
stability length), and maps 10 to 14 are projected on a fine grid (with pixels of FWHM/4
on a side). The final map shows a wealth of filamentary structures at various spatial scales,
and compact sources, with no sign of residual striping.
}
\label{fig:rosette_maps}
\end{figure*}

The Rosette nebula was observed during the Science Demonstration Phase
\citep[][ and references therein]{Motte10} with two scans in parallel mode,
at low scan speed ($20\arcsec / {\rm s}$). Illustrations for the 250\,$\mu$m band are
provided in Figure\,\ref{fig:rosette_maps}.

On this dataset, we show the subtraction of baselines, i.e. long-timescale
drifts, when the {\it /galactic} option is used. The amplitude of the short-timescale
drifts is so small, compared with the signal brightness, that their removal is not
directly discernible in the maps.
Notice how simple baselines are only a first step in the subtraction of large-scale
drifts, and are insufficient to eliminate the striping caused by these drifts.
The module computing residual baselines from the redundancy is needed to completely
subtract large-scale drifts.

\begin{figure*}[!ht]
\centering
\includegraphics[width=10.3cm]{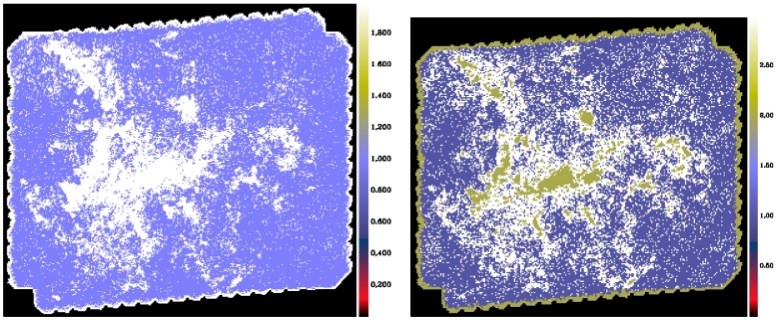}
\caption{Drift mode grids for the Rosette observation at 250\,$\mu$m (left)
and 350\,$\mu$m (right).
Whenever possible, the drift computation modules switch to a finer spatial grid
when they encounter coarse pixels within which the dispersion of the signal
is higher than a predefined threshold. At 250\,$\mu$m, the number of samples
per stability length is too small to be able to switch to a finer grid, whereas
it is sufficient at 350\,$\mu$m (for which the fine grid is coded by the value
3, in white).
}
\label{fig:rosette_driftmode}
\end{figure*}

Figure\,\ref{fig:rosette_driftmode} illustrates, on the same dataset, how the drift
subtraction modules automatically detect non-uniformities within the chosen stability
length, and reject the corresponding samples to avoid a contamination of the drifts
by sources and glitches (see Sect.\,\ref{averdrift}). The left image shows at
250\,$\mu$m, for each pixel of size equal to the stability length of the drifts,
whether the drift computation was enabled (in blue) or not (in white). The right
image shows the same thing at 350\,$\mu$m, where the stability length becomes large
enough to switch to a finer grid (with pixels half as wide) if necessary: it shows
whether the drift computation was enabled on the original grid (in blue), enabled
only on the finer grid (in white), or impossible (in yellow). Such maps (displayed
in visualization mode) delineate very well the location of compact sources and steep
brightness gradients.

\subsubsection{Atlas field (OD 192, obsid 1342187170 and 1342187171)}
\label{spire_atlas}

This low-brightness field, containing both very diffuse Galactic emission and distant
galaxies, was observed with two scans in parallel mode, at high scan speed ($60\arcsec / {\rm s}$).
The survey from which this observation was extracted is described by \cite{Eales10}.

On this dataset, we particularly illustrate the subtraction of the average drift,
compared with the temperature drift correction implemented in the pipeline,
and the masking of residual glitches amplified by Fourier filtering employed in some
pipeline modules. Each effect is illustrated in the array where it is the most
prominent, PMW (350\,$\mu$m) for the
{\ifgras \bf \fi
average
}
drift and PSW (250\,$\mu$m) for glitches.
The stability length was increased to 2 and 1.5 times the FWHM for PSW and PMW,
respectively.

\begin{figure*}[!ht]
\centering
\includegraphics[width=12cm]{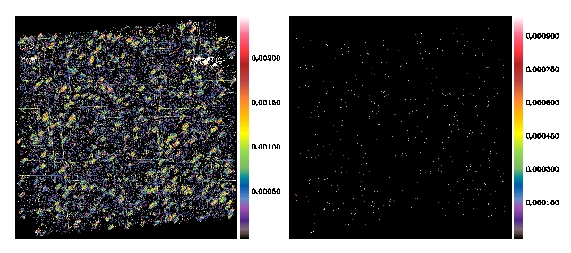}
\caption{Maps of glitches (in absolute value)
detected and masked within the high-frequency
noise computation module (left) and subsequently within the average drift
subtraction module (right). The dataset is that of the Atlas field, in the
PSW array. The first map shows examples of amplified glitches leaving long
trails, and glitches occuring simultaneously in many bolometers of the array.
The second deglitching (second map) detects only residual glitches that
do not exhibit these effects.
}
\label{fig:atlas_glitches}
\end{figure*}

\begin{figure*}[!ht]
%
%
\hspace*{-1.3cm} \includegraphics[width=4.2cm, angle=90]{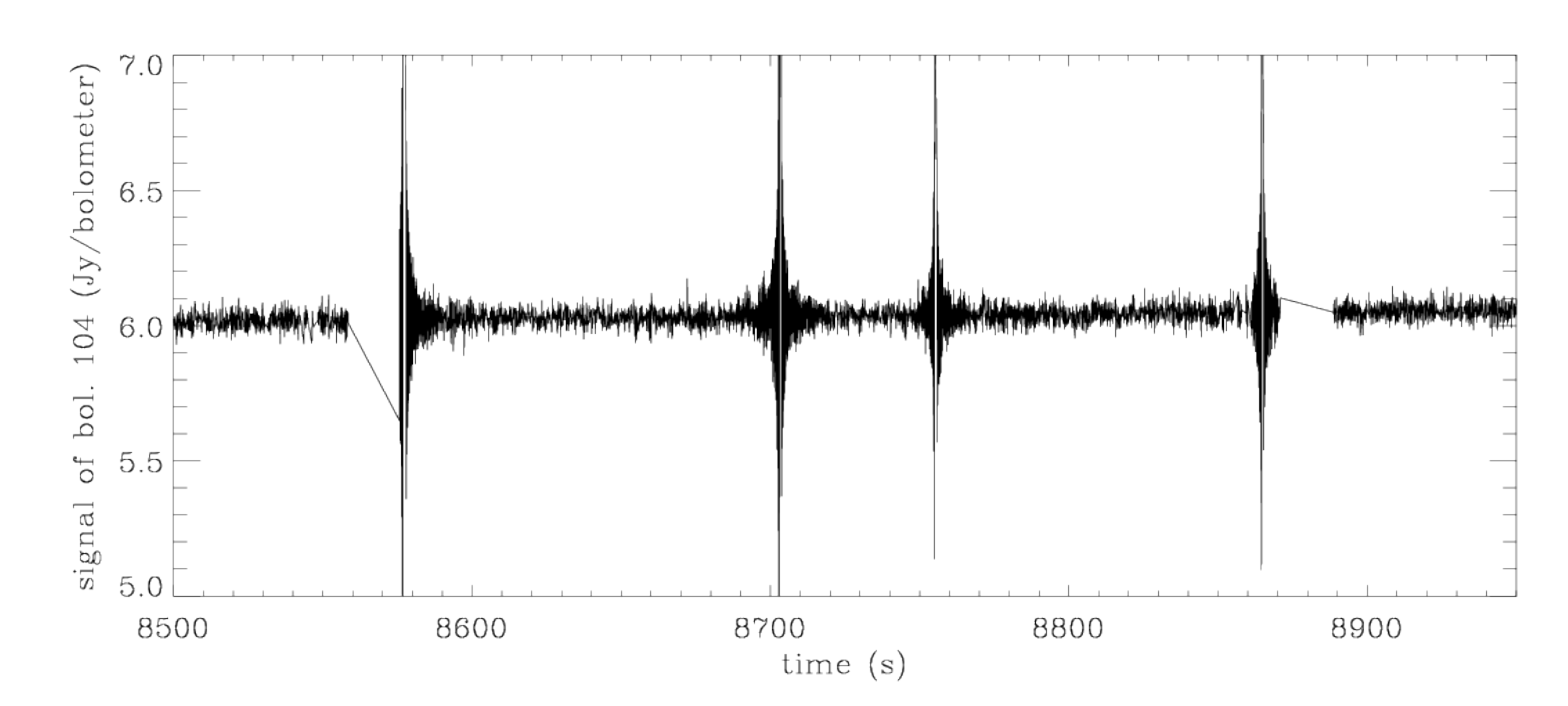}
\hspace*{-0.2cm} \includegraphics[width=4.2cm, angle=90]{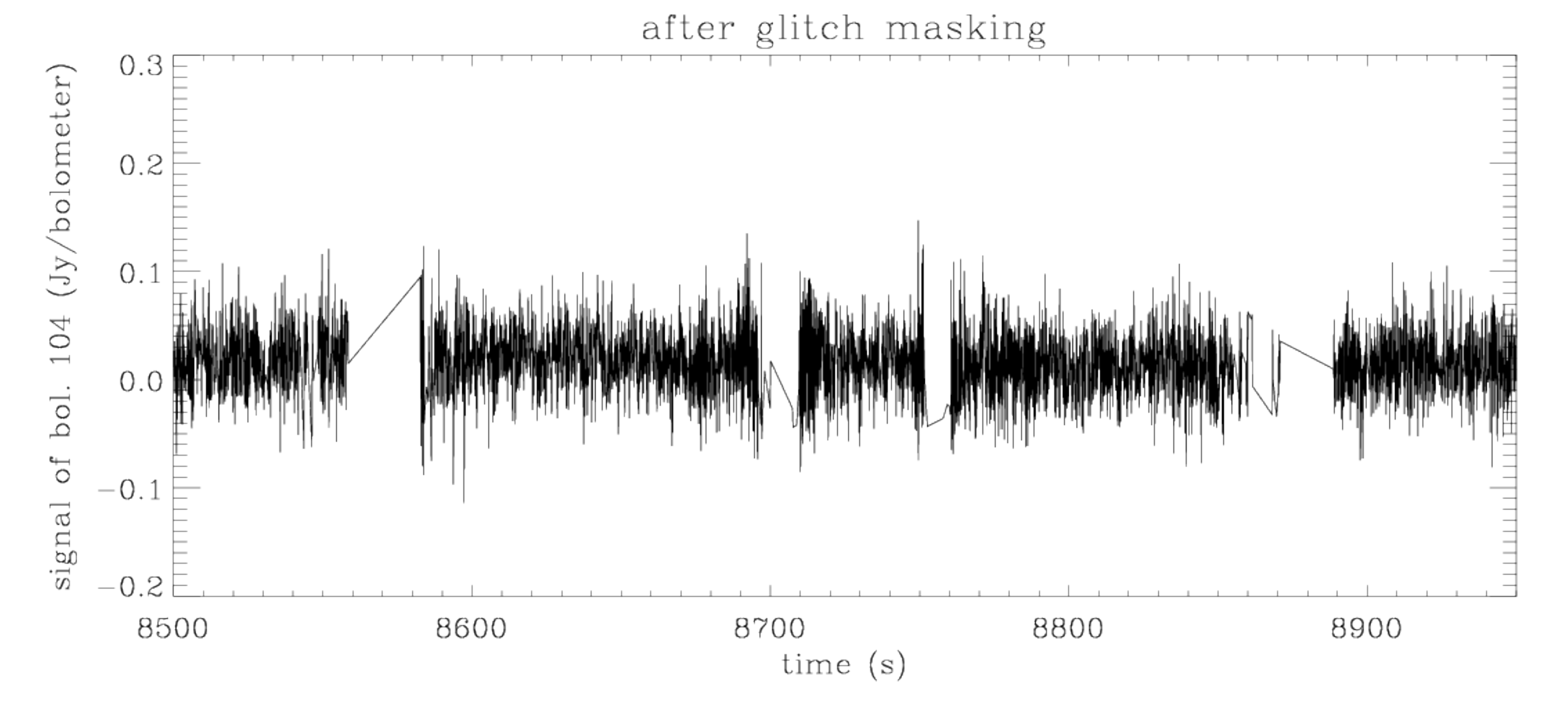} \\
\caption{Initial signal of a PSW bolometer, containing glitches amplified by some pipeline
modules (left), and signal of the same bolometer after running step 5 of the {\it Scanamorphos}
algorithm, as outlined in Sect.\,\ref{algooverview} (right). Notice that the ordinate range is not
the same in both plots.
}
\label{fig:ampl_glitch}
\end{figure*}

Two features are remarkable in the map of detected glitches (Fig.\,\ref{fig:atlas_glitches}).
It contains long trails corresponding to glitches not detected within the pipeline
and then amplified by the electrical filter response and the bolometer time response
corrections; and the imprint of the whole array is visible in many places,
corresponding to glitches that simultaneously affect the whole array.
This type of glitches is frequently seen in SPIRE \citep{Griffin10, Swinyard10}.
Figure\,\ref{fig:ampl_glitch} shows a striking example of the glitch amplification effect,
on a portion of the signal of a PSW bolometer, before and after the glitch masking module
of {\it Scanamorphos} has been run (Sect.\,\ref{glitches}).

\begin{figure*}[!ht]
%
%
\hspace*{-1.3cm} \includegraphics[width=4.2cm, angle=90]{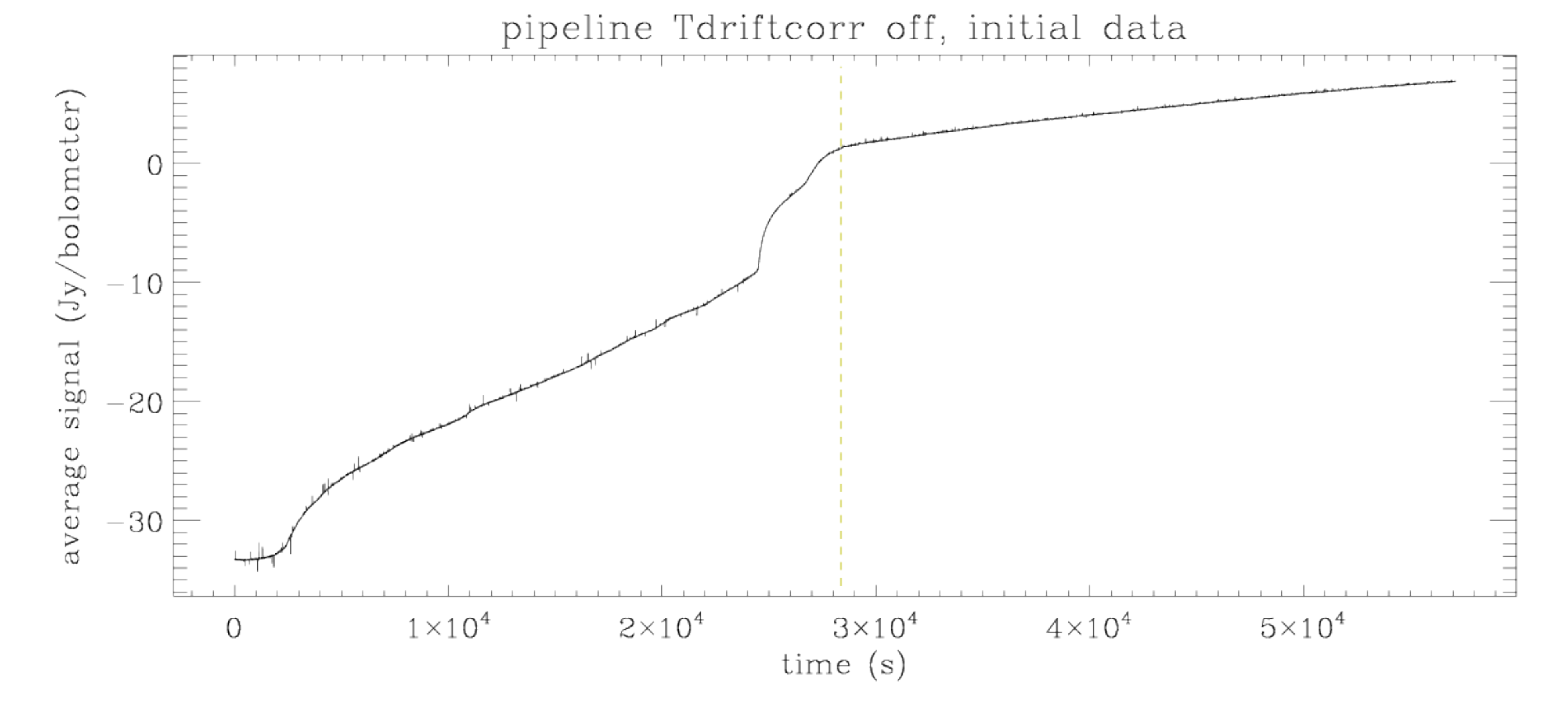}
\hspace*{-0.2cm} \includegraphics[width=4.2cm, angle=90]{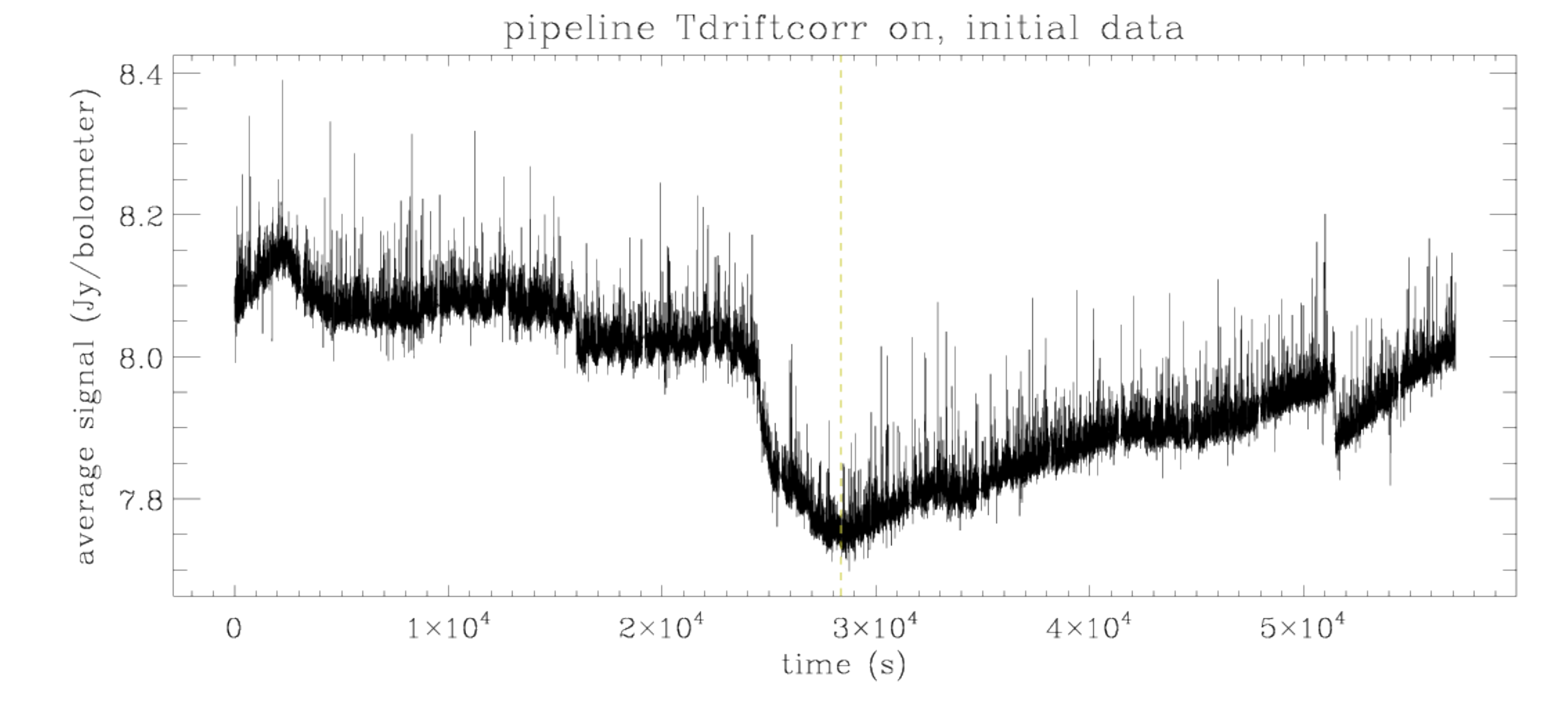} \\
\caption{Average signal of the Atlas 350\,$\mu$m observation, before baseline
subtraction, when the pipeline processing did not include the temperature
drift correction (left), and when it did include the temperature drift
correction (right). In this particular (but by no means isolated) case,
the SPIRE thermistors fail to provide an acceptable model for the thermal
drift. The vertical yellow line shows the limit between the two scans.
}
\label{fig:tdrift_atlas}
\end{figure*}

\begin{figure*}[!ht]
\centering
\vspace*{-3cm}
%
%
\includegraphics[width=17cm]{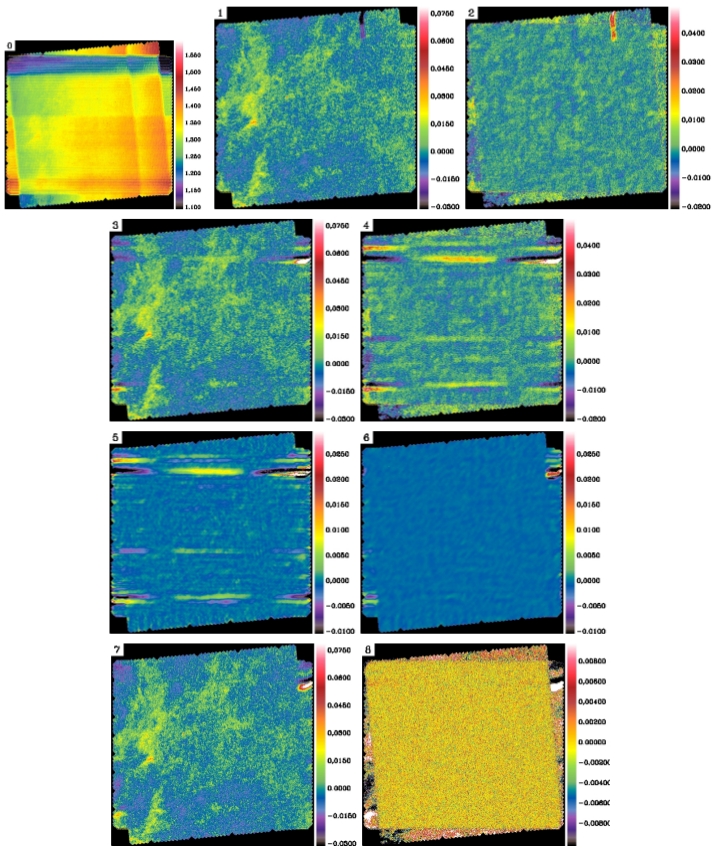}
\vspace*{-1.0cm}
\caption{Maps drawn from the Atlas 350\,$\mu$m observations to illustrate the
impact of the thermal drift and its subtraction by {\it Scanamorphos}:
0) data with the pipeline temperature drift correction applied, after simple
flux calibration offsets have been removed (the map should be similar to
the final map shown in 7 if the pipeline temperature drift correction were
of good quality);
1) data with the pipeline temperature drift correction applied, after full
baseline subtraction as implemented in {\it Scanamorphos}; 2) same, but mapping
the difference between the two scans;
3), 4) same as maps 1 and 2, but for data processed without the pipeline
temperature drift correction.
The results shown in the following maps were obtained from data processed
without the pipeline temperature drift correction. 5), 6) maps of the
average drift subtracted during the first two iterations; 7) data after
subtraction of the drifts; 8) same, but mapping the difference between the two scans.
Maps 0 and 2 immediately show the impact of the thermal drift left by
the pipeline. Map 8 shows that the empirical drift subtraction carried out by
{\it Scanamorphos} worked perfectly, except on the edges with non-nominal coverage
(only one scan). Notice the different brightness ranges of maps 2-4 and 8.
}
\label{fig:atlas_maps}
\end{figure*}

Figures\,\ref{fig:tdrift_atlas} and \ref{fig:atlas_maps}
illustrate the subtraction of the
{\ifgras \bf \fi
average
}
drift, which is particularly challenging
for this field, because
{\ifgras \bf \fi
its time derivative
}
varies abruptly
during the first scan, and does not conform to the thermal noise model implemented
in the pipeline; in addition, the redundancy is minimal, with only two scans acquired
at high scan speed and with the lowest sampling rate. The subtraction of baselines
removes the linear component of the drift, but leaves strong residuals, that are
essentially confined to the first scan. The comparison between the two scans mapped
separately emphasizes these residuals. Figure\,\ref{fig:atlas_maps} shows that after
the drift subtraction, the two scans agree almost perfectly in the region where they
overlap (i.e. with no sign of residual
{\ifgras \bf \fi
average
}
noise, only
differences caused by the white
noise and pointing errors). Note that the only residuals from the temperature drift 
are found on the edges of the map that are covered by only one scan, i.e. outside
the field of view with nominal coverage. For optimum results, when the thermal drift
varies so abruptly, it is also possible to perform the average drift subtraction within
{\it Scanamorphos} after the thermal drift has been attenuated by the relevant pipeline module.

\begin{figure*}[!ht]
%
%
\hspace*{3.3cm} \includegraphics[width=4.2cm, angle=90]{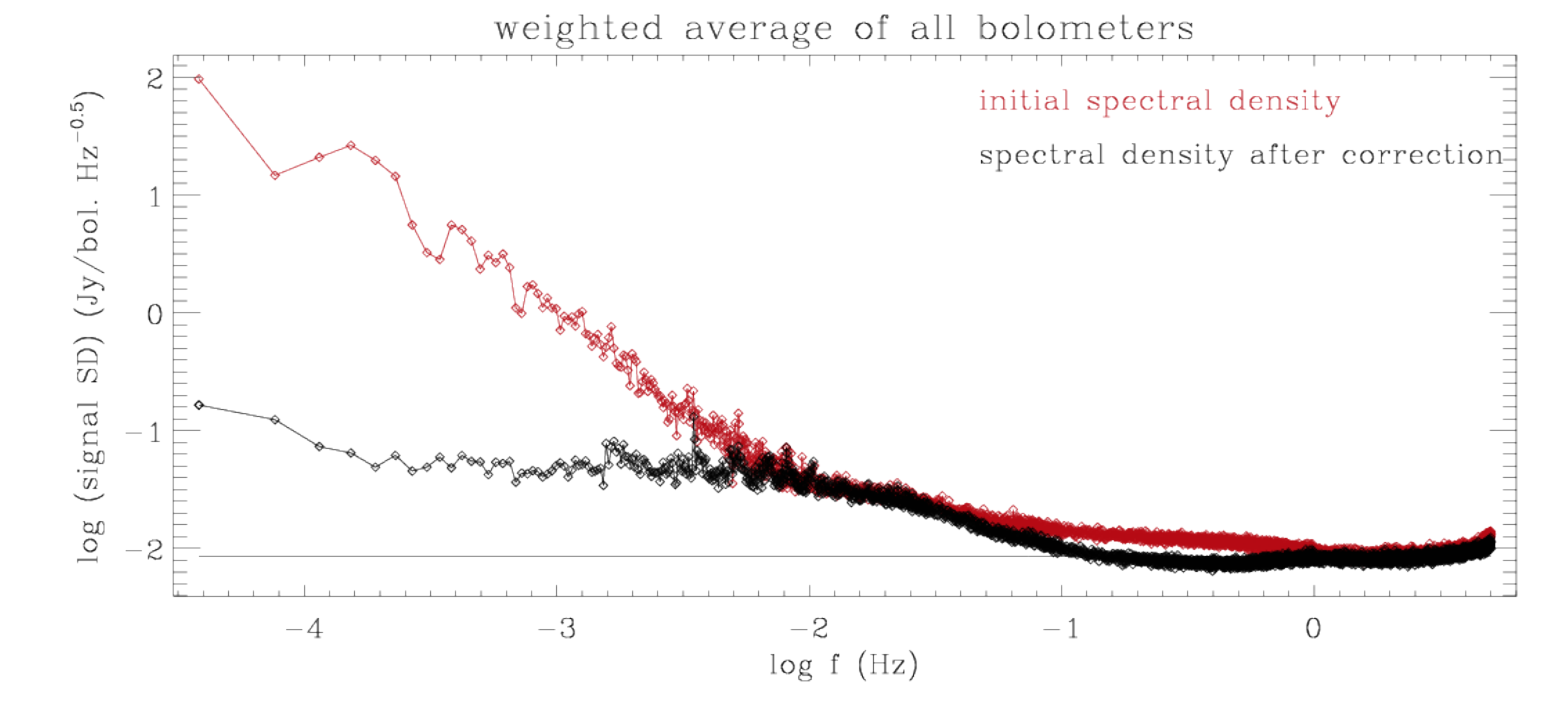} \\
\hspace*{-1.3cm} \includegraphics[width=4.2cm, angle=90]{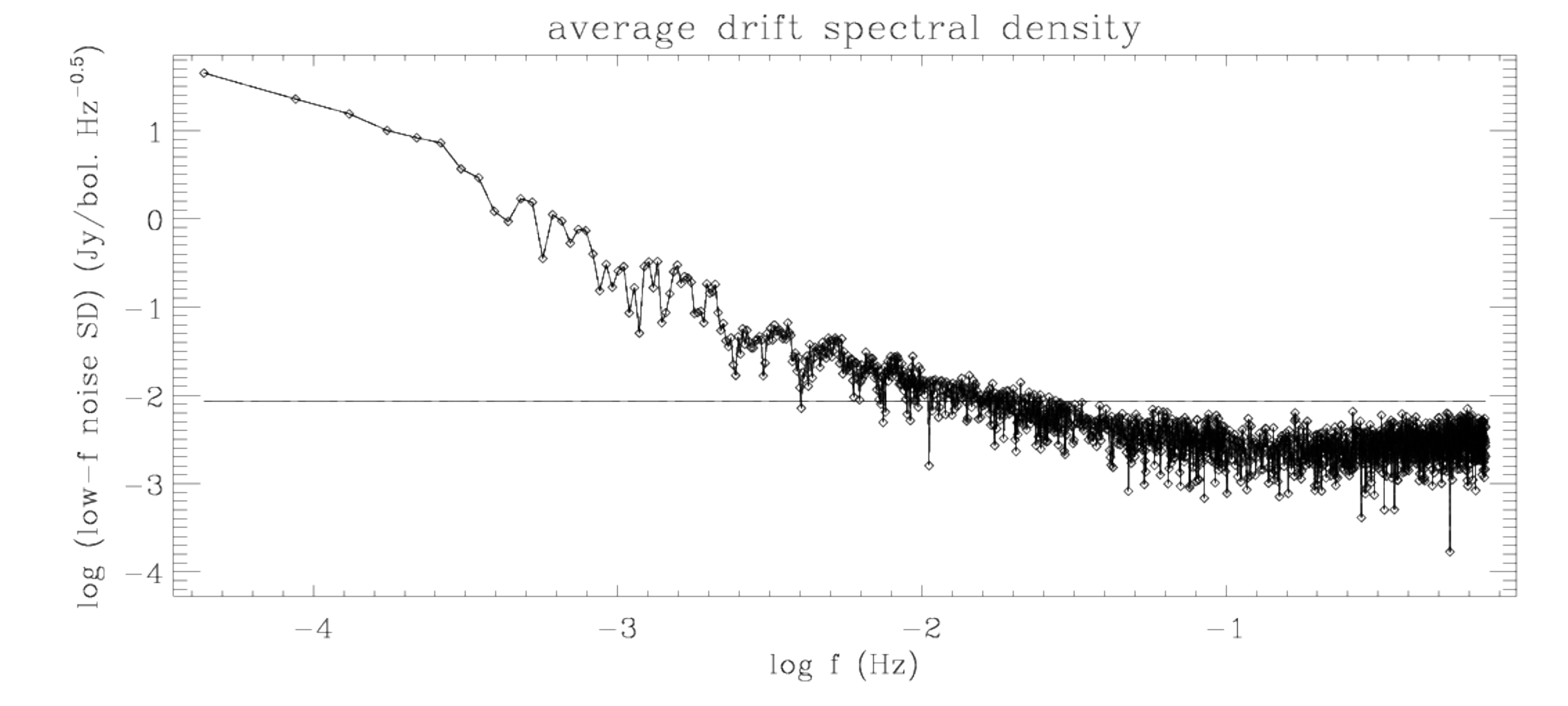}
\hspace*{-0.5cm} \includegraphics[width=4.2cm, angle=90]{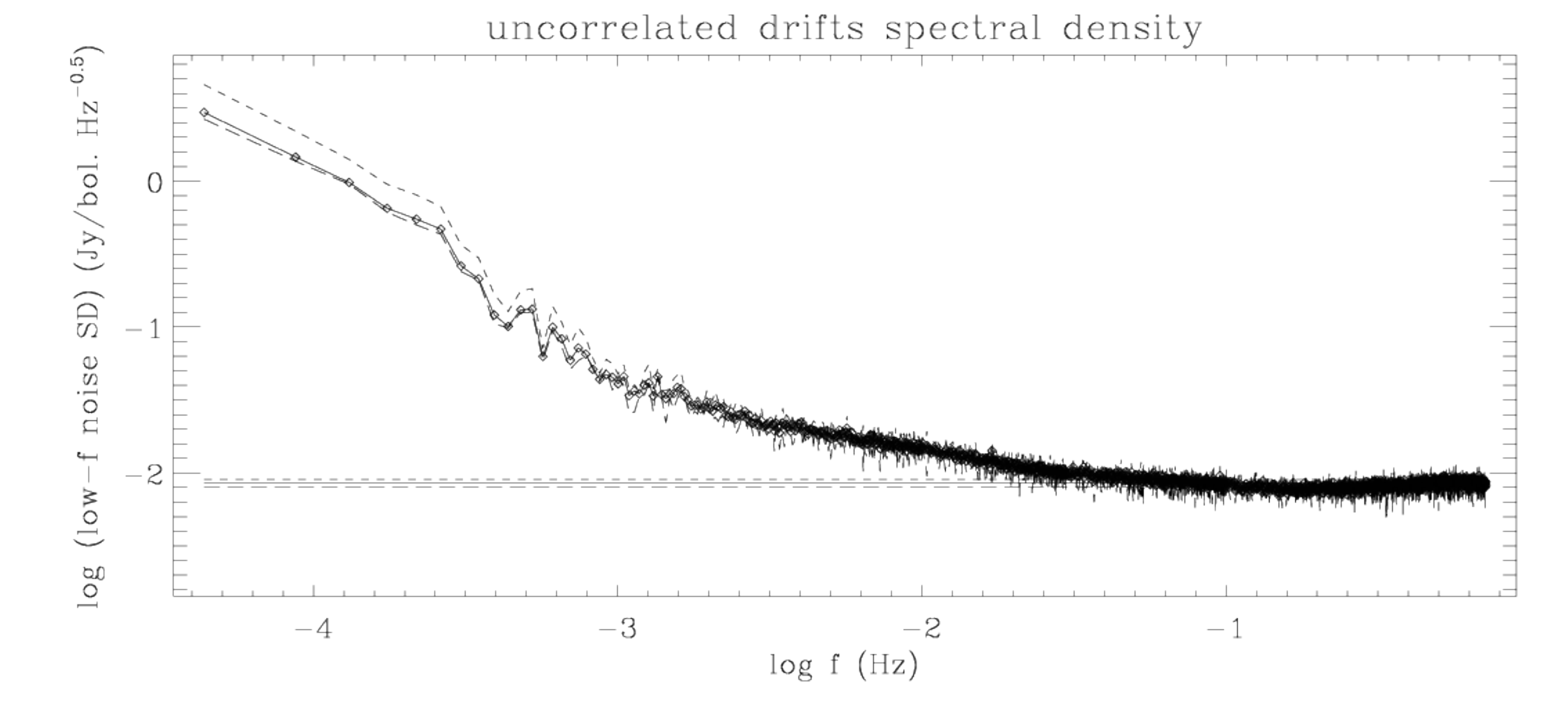}
\caption{Spectral densities computed from the Atlas 350\,$\mu$m observation.
{\bf Top:} Spectral density of the signal, computed on the length of a scan
(weighted average of the spectral densities of all valid bolometers), after the
processing (in black), compared to the initial spectral density (in red).
{\bf Bottom:} Spectral densities of the subtracted average drift (left),
and of the individual drifts (right), computed on the length of a scan
and on the coarse time grid (hence the different frequency scale). They are
simple averages of the spectral densities of all valid bolometers. 
In each panel, the horizontal lines represent the spectral density of the
high-frequency noise. The dashed lines are 25\% and 75\% quartiles.
Before computing the Fourier transform, the longest baselines were removed
from the time series, and the rectified time series were apodized with a $\sin^2$
function at both ends, on each scan separately.
}
\label{fig:atlas_spd}
\end{figure*}

Spectral densities of the initial signal, of the signal corrected for drifts, and of
the drifts themselves are shown in Figure\,\ref{fig:atlas_spd}. The spectral index of
the drift spectral density is close to $-1$\,.

\subsubsection{NGC\,6822 (OD 145, obsid 1342185533)}
\label{spire_n6822}

This dwarf galaxy, containing both compact H{\small II} complexes and diffuse emission,
is located behind relatively bright cirrus filaments. It was observed with 6 scans at
the nominal scan speed of 30 arcsec/s \citep{Galametz10}.

\begin{figure*}[!ht]
\vspace*{-3cm}
%
%
\hspace*{-0.5cm} \includegraphics[width=18cm]{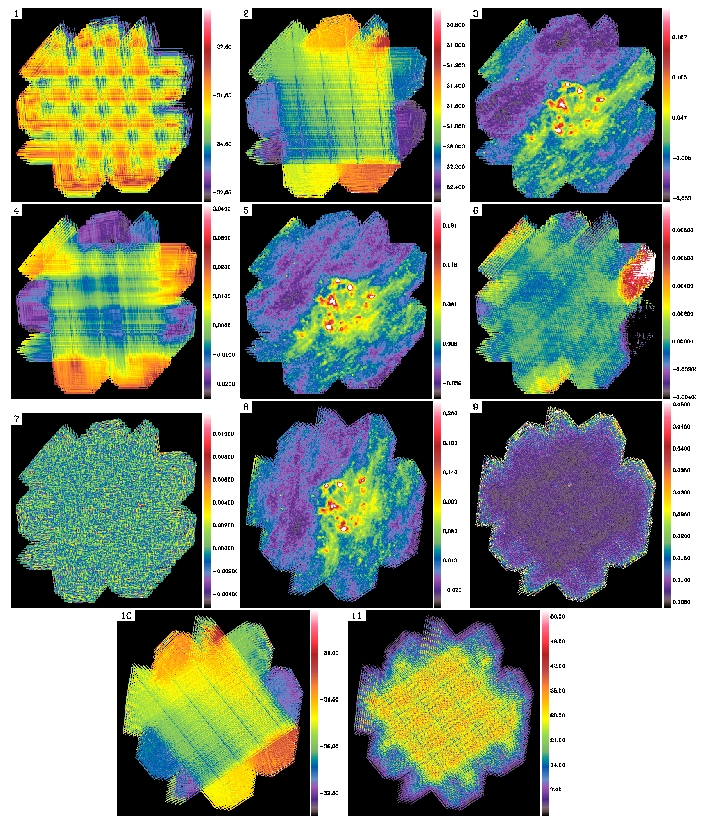}
\vspace*{-1.0cm}
\caption{Successive steps of the processing of the NGC\,6822 250\,$\mu$m observations illustrated
by maps, showing the following: 1) raw level-1 data; 2) simple baselines obtained by linear fits;
3) data after subtraction of the simple baselines; 4) residual baselines derived from the redundancy;
5) data after subtraction of the residual baselines; 6) average drift subtracted at the
first iteration (and glitches masked); 7) individual drifts subtracted at the second iteration;
8) final map; 9) error map; 10) total drifts (dominated by baselines); 11) weight map.
Maps 1, 2, 4, 6 and 7 are projected on a coarse spatial grid, and the other maps are projected
on a fine grid (with pixels of FWHM/4 on a side). Note that the very faint filamentary structures
superposed on the westernmost part of NGC\,6822 (to the right)
are not caused by residual drifts: they are not
aligned with any of the two scanning directions, but with the general direction of the cirrus filaments.
}
\label{fig:n6822_maps}
\end{figure*}

\begin{figure*}[!ht]
%
%
\vspace*{-1cm}
\includegraphics[width=16cm]{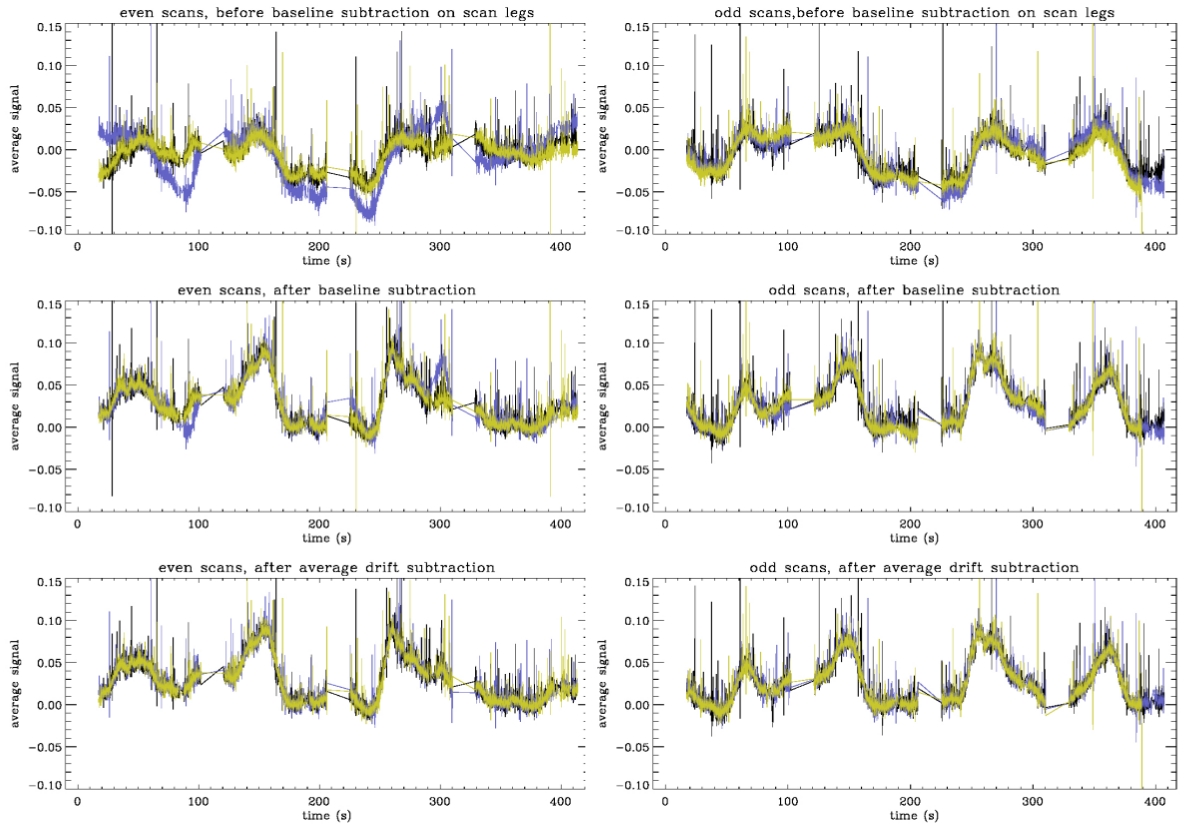}
\caption{Another way of looking at the average drift subtraction, in the time domain (from the NGC\,6822
250\,$\mu$m observation). Average signal of even-number scans (left; in the same scanning
direction) and odd-number scans (right; in the orthogonal direction),
in units of Jy/beam,
before the baseline
subtraction performed on scan legs (top), after full baseline subtraction (middle), and after
subtraction of the average drift (bottom). Different scans coded by different colors are
superposed after applying a time shift, such that a given time corresponds approximately
to the same spatial coordinates. After the baseline subtraction, there are still significant
differences between successive scans (mostly in the even-scan direction, at times close to
90\,s and 300\,s) that vanish after applying the average drift subtraction.
}
\label{fig:n6822_averdrift}
\end{figure*}

This example shows how baselines and drifts (both the average and individual components)
can be robustly derived even in the presence of complex structures on the largest
scales of the map (such as cirrus emission here), when the {\it /galactic}
option is {\it not} used. Figure\,\ref{fig:n6822_maps} also shows that the raw data and total
drifts are completely dominated by the linear baselines. Figure\,\ref{fig:n6822_averdrift}
demonstrates in addition that it is possible to separate the residual non-linear average
drift from the complex sky signal, by the method implemented in {\it Scanamorphos}.

\subsubsection{NGC\,6946 (OD 116, obsid 1342183366)}
\label{spire_n6946}

This large spiral galaxy was mapped with 4 scans at high scan speed (60 arcsec/s)
during the Performance Verification phase, and is part of the KINGFISH program
\citep{Kennicutt11}.

\begin{figure*}[!ht]
%
%
\hspace*{0.8cm} \includegraphics[width=15.cm]{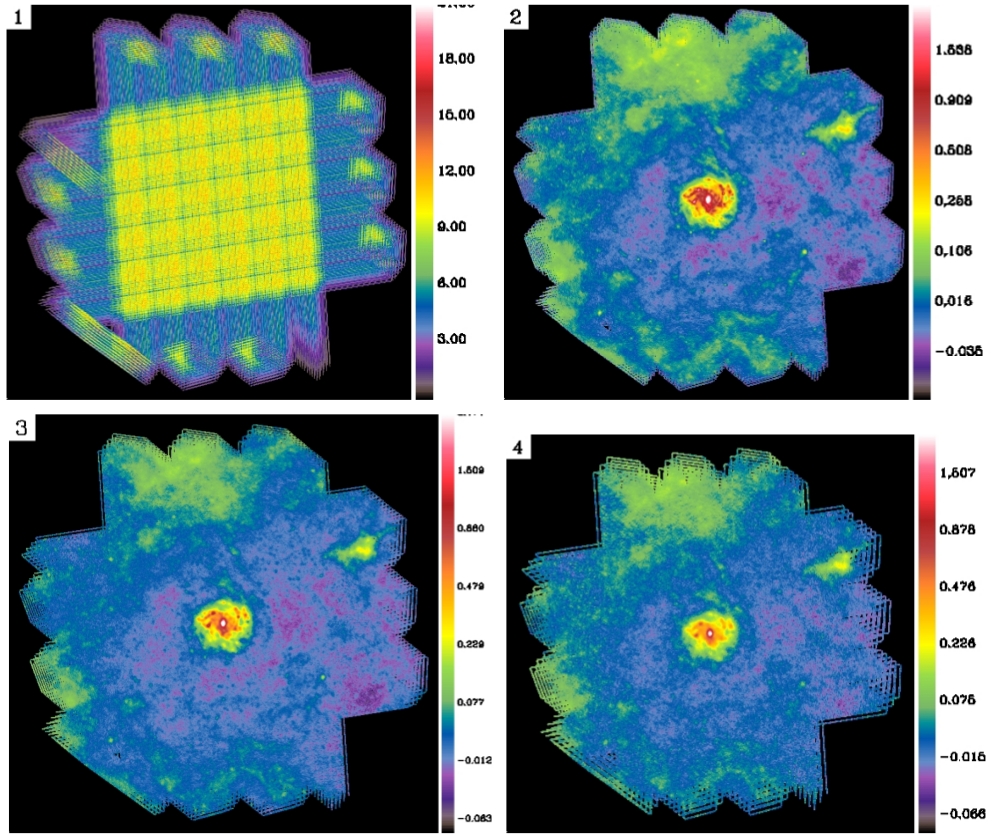}
\caption{Final maps obtained after processing the NGC\,6946 observations:
1) weight map at 250\,$\mu$m, showing the large turnaround regions with sparser
coverage, due to the high scan speed; 2, 3, and 4) final maps at 250\,$\mu$m,
350\,$\mu$m, and 500\,$\mu$m, respectively.
}
\label{fig:n6946_maps}
\end{figure*}

The stability length was automatically increased to 1.5 times the FWHM for PSW.
Since the average drift is very close to being purely linear for this particular
observation, making it easy to subtract with simple baselines, we do not show
the intermediate results, but only the final maps (Fig.\,\ref{fig:n6946_maps}).
Even though wide parts of the map have non-nominal coverage, due
to the high scan speed and thus the longer deceleration distance between
successive scan legs, it is possible to accurately recover the sky structure in
these edges, as shown by the maps, as long as they are covered by two scans
(in the same scanning direction).

\subsection{Application to PACS data}
\label{ex_pacs}

\subsubsection{NGC\,4559 (OD 188, obsid 1342187067, 1342187068, 1342187069 and 1342187070)}
\label{pacs_n4559}

This spiral galaxy was observed as part of the KINGFISH program \citep{Kennicutt11}
at nominal scan speed ($20\arcsec / {\rm s}$).
The observations were split into four distinct obsid. The blue (70\,$\mu$m) and red (160\,$\mu$m)
filters were used simultaneously during the first and third obsid, and the green (100\,$\mu$m)
and red filters during the second and fourth obsid. The blue and green maps are each made of
12 scans, and the red map of 24 scans.

One word of caution: since observations in the blue and green filters use the same PACS
array, commonly referred to as the blue array, some confusion might arise between the two.
When formatting the HIPE data for input to {\it Scanamorphos}, one should always check that the selected
filter is correct. The module allows this by printing the name of the filter and issuing
an error message if necessary.

The scans almost always
need reordering before input to {\it Scanamorphos}, in such a way that
the scanning directions are alternating from one scan to the next. The reason is that
this is the default ordering for SPIRE, and that the baseline subtraction module
always assumes this order. This is explained in detail in the examples directory of the
{\it Scanamorphos} distribution.
The {\it /vis\_traject} option allows one to visualize the scan pattern and to check that
scanning directions are alternating correctly. In addition, all the scans of a given
obsid must have been used before turning to the next obsid made with the same scanning
direction.

\begin{figure*}[!ht]
%
%
\hspace*{-0.8cm} \includegraphics[width=18cm]{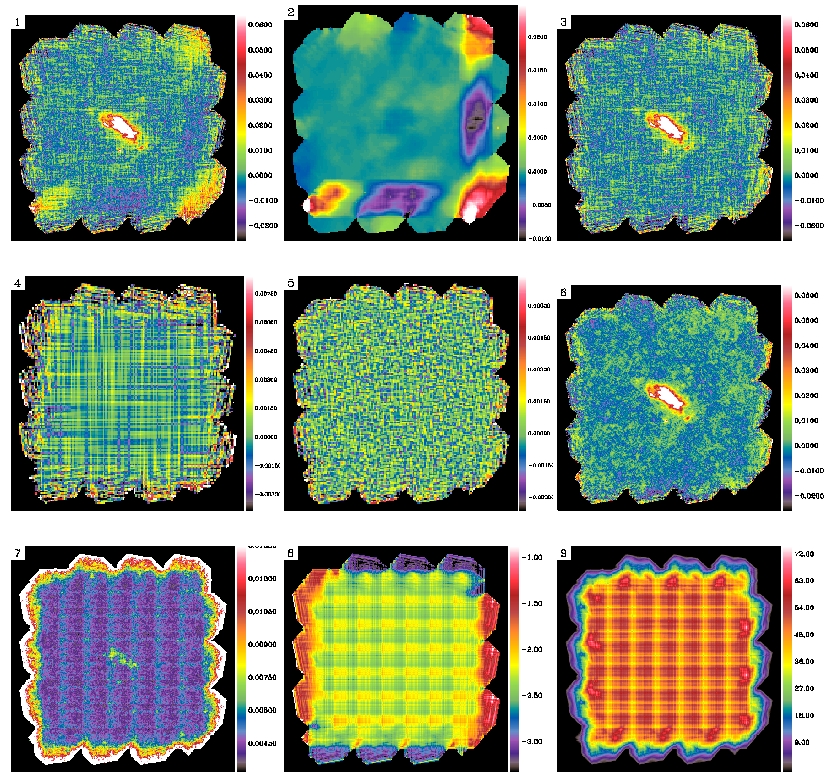}
\caption{Successive steps of the processing of the NGC\,4559 160\,$\mu$m observations
illustrated by maps.
To accentuate the effects of the thermal drift in these illustrations, we have retained
only the two obsids, and the first two scans in each, that immediately follow calibration
blocks \citep{Poglitsch10},
thus containing the strongest transients, on a total of four obsids with six scans each
(but the numbers given in Table\,\ref{tab:runs} apply to the full observations).
The maps show the following: 1) data after subtraction of the full baselines;
2) average drift subtracted at the first iteration (notice the imprint of the transient
following the observation of calibration sources
at the start of each obsid, visible in the bottom and right
edges of the map); 3) data after subtraction of the average drift; 4) individual drifts
subtracted on a timescale of $27 \times T_c$\,; 5) individual drifts subtracted on the
minimum timescale $T_c$\,;
6) final map; 7) error map; 8) total drifts (dominated by baselines);
9) weight map. All maps are projected on a fine spatial grid (with pixels of FWHM/4
on a side), except maps 2, 4 and 5 that are projected on the coarser grid used for the processing.
}
\label{fig:n4559_maps}
\end{figure*}

The stability length was increased to 2.5, 2, and 1.5 times the FWHM at 70\,$\mu$m,
100\,$\mu$m, and 160\,$\mu$m, respectively. Intermediate results of the drifts subtraction
for the 160\,$\mu$m array can be seen in Figure\,\ref{fig:n4559_maps}. For these illustrations,
the intermediate maps were reprojected on the final pixel grid (instead of the mapping
grid used for the processing) in order to make the effect of the low-frequency noise
easier to see in the maps. The maps of the drifts were not reprojected, and are thus
shown on a coarser grid (with pixels of $3/4$ FWHM on a side).

\subsubsection{Rosette (OD 159, obsid 1342186121 and 1342186122)}
\label{pacs_rosette}

These observations are the PACS component of the parallel-mode observations
already described in Section\,\ref{spire_rosette}. We processed only the 70\,$\mu$m data.
In this example, the high-frequency noise is completely dominated by the
quantization noise (Sect.\,\ref{highf}). The stability length was increased
to 4.5 times the FWHM. Notwithstanding the low redundancy, we used a final pixel size
of FWHM/4.

\begin{figure*}[!ht]
\centering
%
%
\includegraphics[width=15cm]{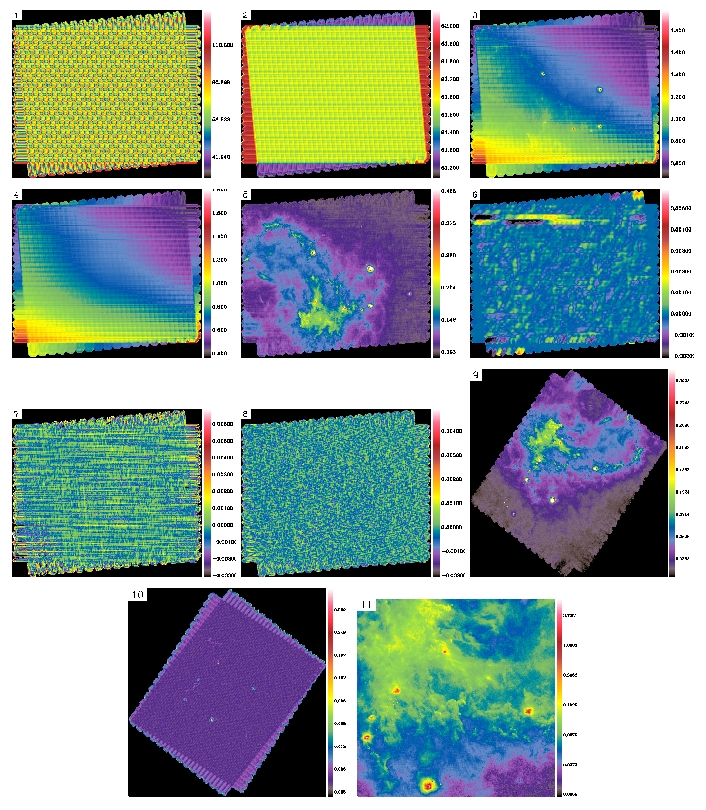}
\caption{Successive steps of the processing of the Rosette 70\,$\mu$m observations illustrated
by maps, showing the following: 1) raw level-1 data; 2) simple
zero-order baselines (flux calibration offsets);
3) data after subtraction of the zero-order baselines; 4) residual linear
baselines derived from the redundancy;
5) data after subtraction of the residual baselines; 6) average drift subtracted at the
first iteration (and glitches masked); 7) individual drifts subtracted
on a timescale of $27 \times T_c$\,; 8) individual drifts subtracted on a timescale
of $3 \times T_c$\,;
9) final map; 10) error map; 11) enlargement of a central portion of the
final map (about $51^{\prime}$ on a side).
Maps 1 to 8 are projected on a coarse spatial grid (with a pixel size equal to half the
stability length), and maps 9 to 11 are projected on a fine grid (with pixels of FWHM/4
on a side). All the very compact sources are real.
Notice that Scanamorphos was able to pick up the non-linear thermal drift
near the end of the first scan, even though it has a low amplitude with respect to
the sky signal (map 6).
}
\label{fig:rosette_pacs_maps}
\end{figure*}

Despite the unfavorable observing parameters (high quantization noise and low
sampling rate of 5\,Hz), Figure\,\ref{fig:rosette_pacs_maps} shows that the different
processing steps perform as expected, and that the final map quality is very good,
compared with maps produced within the pipeline (not shown here).
{\it Scanamorphos} preserves both diffuse
emission and compact sources, whereas the use of polynomial baselines or Fourier
filtering in the pipeline strongly distorts extended emission.

\subsubsection{Centaurus\,A (OD 233, obsid 1342188855 and 1342188856)}
\label{pacs_cena}

Centaurus\,A was observed in nominal mode as part of the Very Nearby Galaxies Survey
(P.I. C. Wilson) and the spatial variations of its dust to gas mass ratio were
analyzed by \citet{Parkin12}.
Taking advantage of the fact that, for this galaxy, maps built from various methods
were already available (courtesy of Marc Sauvage), we present a brief comparison of
{\it Scanamorphos} with the two softwares fully implemented in HIPE for PACS:
{\it PhotProject}, subtracting low-frequency noise by means of a high-pass filter,
and MADmap, subtracting the uncorrelated noise with a
matrix-inversion
algorithm.

\begin{figure*}[!ht]
%
%
\hspace*{-0.8cm} \includegraphics[width=18cm]{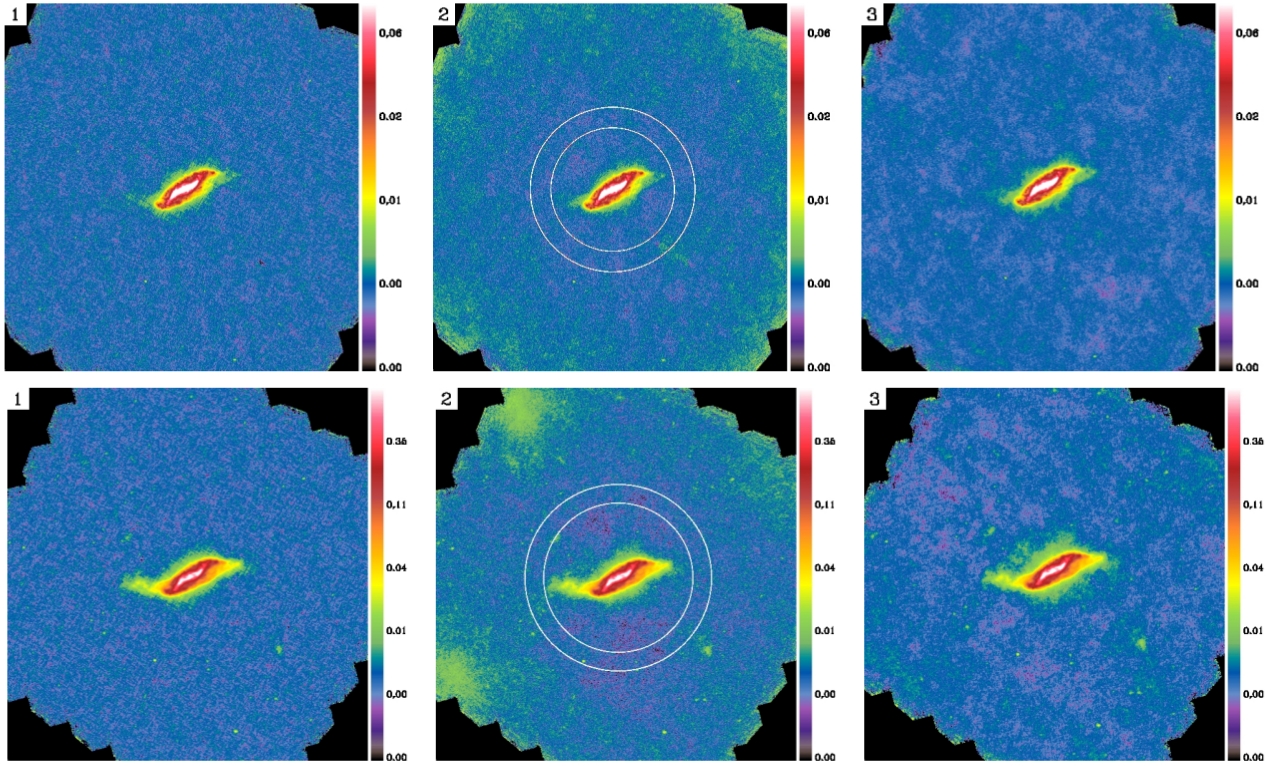}
\caption{Maps produced by: 1) {\it PhotProject} at 70\,$\mu$m; 2) MADmap at 70\,$\mu$m;
3) {\it Scanamorphos} at 70\,$\mu$m; 4) {\it PhotProject} at 160\,$\mu$m; 5) MADmap
at 160\,$\mu$m; 6) {\it Scanamorphos} at 160\,$\mu$m, in logarithmic scale.
The pixel size is $2^{\prime\prime}$ at 70\,$\mu$m and $4^{\prime\prime}$ at 160\,$\mu$m.
A brightness cut was applied to emphasize fainter structures (the results of the
three softwares at higher surface brightnesses being similar). The concentric circles
superimposed on the MADmap image indicate the region used to compute the background level
in all maps.
}
\label{fig:cena_pacs_maps}
\end{figure*}

\begin{figure*}[!ht]
%
%
\hspace*{-1.3cm} \includegraphics[width=4.2cm, angle=90]{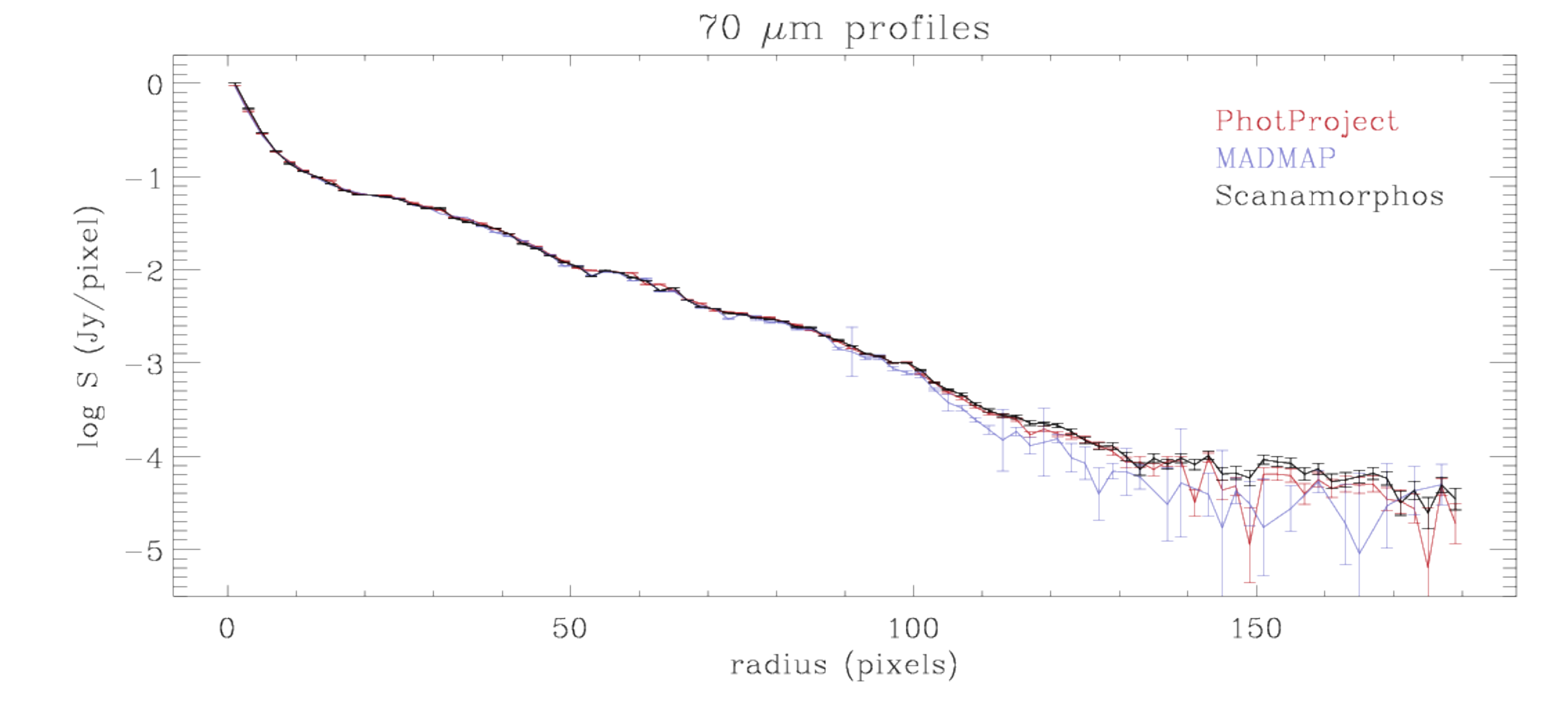}
\hspace*{-0.3cm} \includegraphics[width=4.2cm, angle=90]{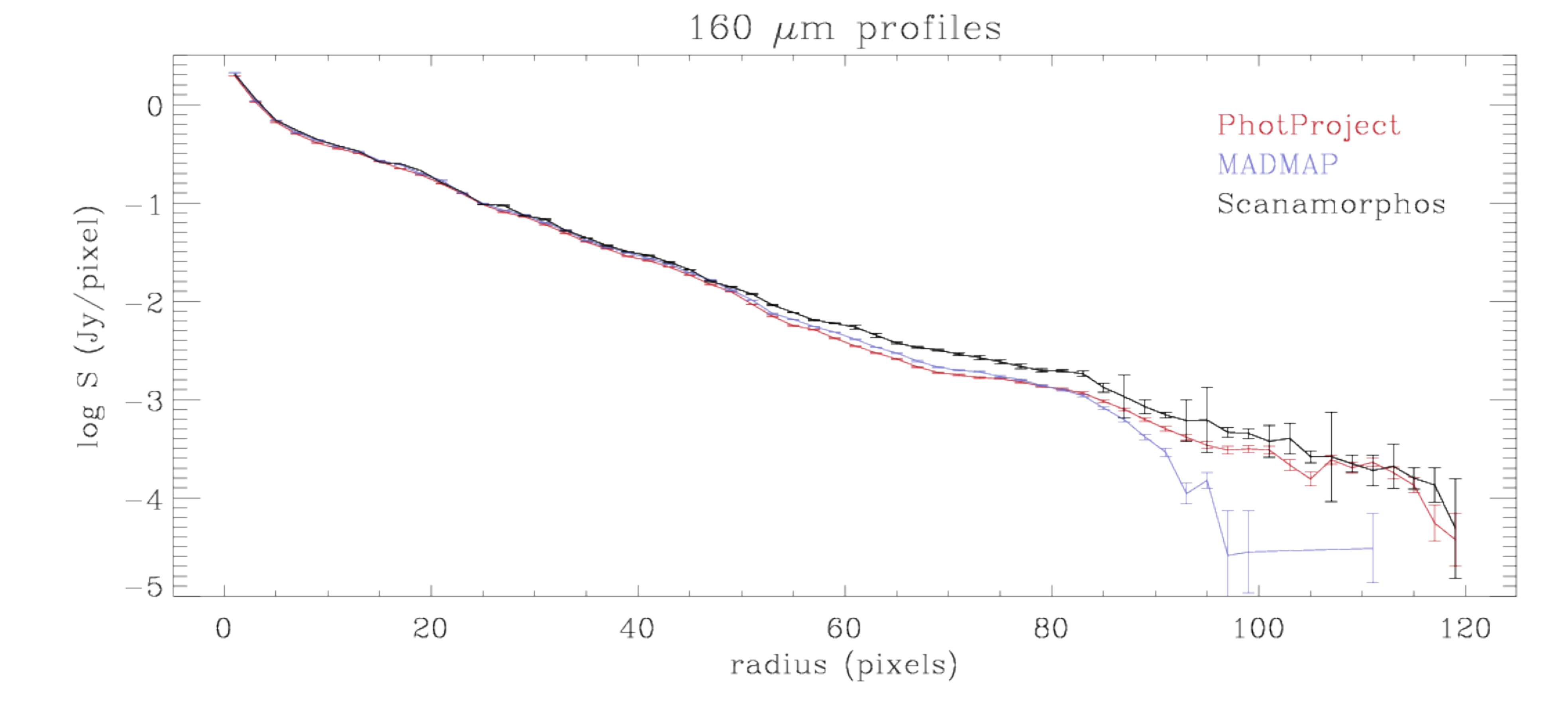}
\caption{Surface brightness profiles of the {\it Scanamorphos} map (in black), the
{\it PhotProject} map (in red) and the MADmap map (in blue). The radii of 180
pixels at 70\,$\mu$m and 120 pixels at 160\,$\mu$m correspond to the inner radius
of the annulus shown in Fig.\,\ref{fig:cena_pacs_maps}.
}
\label{fig:cena_pacs_profiles}
\end{figure*}

Figure\,\ref{fig:cena_pacs_maps} shows the final maps obtained for the two observed
PACS bands, 70 and 160\,$\mu$m, and Figure\,\ref{fig:cena_pacs_profiles} shows
the azimuthally-averaged surface brightness profiles. The respective advantages and
disadvantages of the three methods are readily perceptible. We summarize them as follows: \\
1) {\it PhotProject} has no difficulty preserving compact sources, but filters out
a non-negligible fraction of the extended emission. Even though a protective mask
is applied to sources in the field before running the high-pass filter, this is
insufficient to fully preserve the diffuse emission, because the brightness threshold
of the mask cannot be set arbitrarily low. One can also notice residual short-timescale
noise in the form of stripes, due to the fact that the filtering can be applied only
to relatively large scales. \\
2) MADmap also preserves compact sources (except very bright ones, for which
it is known to introduce artefacts in the shape of a cross-hair). Since it does
not handle thermal drifts, however, it has difficulty restoring extended emission,
and the sky structure on large spatial scales is not correct. On the other hand,
the background noise on small scales is reduced to a very low level. \\
3) {\it Scanamorphos} preserves both compact and extended emission, and subtracts
the low-frequency noise down to lower levels than the white noise on all timescales
above $T_c$. The sky structure is less smooth than with the other two softwares,
but this residual noise pattern remains well below the photometric errors, as shown
in Section\,\ref{simul}. Figure\,\ref{fig:cena_pacs_profiles} shows that while there
is essentially no difference between the three softwares at high surface brightnesses,
{\it Scanamorphos} recovers more of the diffuse emission, especially at 160\,$\mu$m.

\section{Final remarks}
\label{conclu}

The foremost task of the algorithm described above is to subtract the low-frequency noise
in scan observations made with bolometer arrays, or any photometers affected by the
same type of noise. Its core principle is to use the redundancy in the observations,
stemming from both the scanning strategy and the array structure, to derive the drifts
from the data themselves, without using any noise model. It copes with both the thermal
and non-thermal components of the noise
{\ifgras \bf \fi
(or more exactly, from the point of view of the algorithm, the average drift and the
individual drifts; see Sect.\,\ref{algooverview}),
}
that are subtracted one after the other.

For each observation, the map size determines a boundary between what we have called
long timescales (longer than the scan leg crossing time $T_{\rm leg}$) and short timescales
(shorter than $T_{\rm leg}$). There is no fundamental difference between the drifts
in these two frequency windows, but they call for distinct methods, because the
long-timescale component is not adequately sampled by the data: subtracting it
on the same principles as the short-timescale component would require a larger map.

The level of redundancy that is necessary to achieve our goals can be estimated in view
of the coverage maps of Herschel observations. For those discussed in Section\,\ref{tests},
the redundancy varies between 75 and 530 samples per square FWHM per scan pair (depending
on the array structure and map parameters, and on the scan speed and sampling rate).
The coverage can be rather inhomogeneous. Using pixels of size FWHM/4, a convenient fiducial
value to remember is 10 samples per pixel per scan pair. The choice of the unit ``scan pair''
highlights the necessity to combine at least two scans separated by a large position angle
difference in order to separate the drifts from the signal.

The simulations and examples of application presented here, in the context of Herschel
observations, demonstrate that {\it Scanamorphos} is efficient at reducing the noise
while preserving the flux and morphology of both compact sources and extended emission.
It is robust with respect to small pointing and calibration errors, and we did not find
any dataset from which it would not be able to produce a scientifically correct map.
But since the redundancy is not infinite, and in some cases quite limited, correlated noise
residuals can create artificial fluctuations of the sky emission (that remain well below
the $3 \sigma$ significance level). {\it Scanamorphos} is easily adaptable to progressively
include corrections for new instrumental effects, as they become better understood, and has
in particular been used to determine relative gain corrections for the SPIRE arrays
(Appendix\,\ref{spire_gains}).

Finally, we wish to remark on the impact of scanning strategies and on some adaptations
that may be required for other instrument and telescope combinations:
1) The method to subtract the long-timescale drifts depends on the scan pattern. The
algorithm in its current form is adapted when each scan can be decomposed into distinct
scan legs (spanning the whole extent of the map in one direction), but would need to be
modified for other types of scan patterns, for example rasters of spirals.
2) The thermal noise could be more complex for other instruments. For example, it could
be significantly different from one subarray to the next, or among the parts of the array
that are connected to different readout circuits. This is already partly accounted for
in the subtraction of the long-timescale drifts, since this correction is made individually
for each bolometer. The average drift subtraction on short timescales could also easily
accommodate this level of complexity by splitting the signal into as many components as
necessary, and doing the correction separately for each.
It has in fact already been attempted, but was not found to bring any improvement for Herschel.
3) Our algorithm to subtract the thermal noise is in principle applicable to ground-based
observations, and would be more effective than the customarily used filtering techniques.
But to this effect the detectors need to be sampled at very high rates, at least a few times
the knee frequency of the atmospheric noise. In other words, the unit length on which the
correction can be achieved should be similar to or smaller than the atmospheric stability
length.

\section{Distribution}
\label{distrib}

The {\it Scanamorphos} package can be retrieved from this URL:
\url{http://www2.iap.fr/users/roussel/herschel}\,. It is possible to run the code
with a minimal knowledge of IDL. Instructions for the installation and usage are
included in the distribution, and example HIPE scripts and IDL commands to format
the level-1 data and run {\it Scanamorphos} on the observations presented in
Section\,\ref{tests} can be found in a subdirectory. Some advice and basic documentation,
as well as an archive of all successive versions, can be found at the same URL.

The software can be run in visualization mode or in non-interactive mode.
To avoid typing input parameters for each run, and process sequentially a large
number of observations, a batch mode is also enabled. It is recommended to use
this functionality only after gaining familiarity with the software.

{\it Scanamorphos} will be maintained for the duration of the Herschel mission
and the archive phase.

\clearpage

\appendix

\section{Gain corrections for SPIRE}
\label{spire_gains}

From our tests made on various datasets with extended sources, and following up
on an original idea by Darren Dowell (private communication), we have found necessary
to apply some corrections to the gains of SPIRE bolometers. While the flux
calibration of each array taken as a whole is exact, individual bolometers
do not all have the same response (with a gain of one).
This is because the flux calibration is built on point sources, and thus cannot account
for the fact that the coupling of extended sources to each beam (with beam areas that are
not uniform over the array) is different from that of point sources.
The photometric corrections for extended sources that are given in the SPIRE Observers'
manual (in the ``Flux density calibration'' chapter) assume a unique beam, and neglect
variations among bolometers. Corrections for non-uniform beams were first
derived from a particular Galactic field
(the Rosette nebula; see Sect.\,\ref{spire_rosette}). The large brightness dynamic range
over all spatial scales in this field enables the calibration of relative gains
without having to worry about low-frequency drifts on short timescales, whose
amplitude is completely negligible with respect to the effect of differential gains.
We repeated the computation of gains, as explained below, using
17 different observations of bright complex fields, observed at low or nominal scan speed,
and found consistent results in all cases.
We could also check that there is an excellent correlation between the gains
that we derived empirically and the measured beam areas of the individual bolometers,
obtained from special calibration observations. The uncertainties associated with
the latter observations are however much larger, which is why we adopt the empirical
gains. These corrections have been successfully tested by the SPIRE ICC and
have been included in branch 8 of HIPE. They should be applied in only one of the two
softwares, HIPE or {\it Scanamorphos}, not in both.

\begin{figure*}[!ht]
%
%
\hspace*{-1.3cm} \includegraphics[width=4.2cm, angle=90]{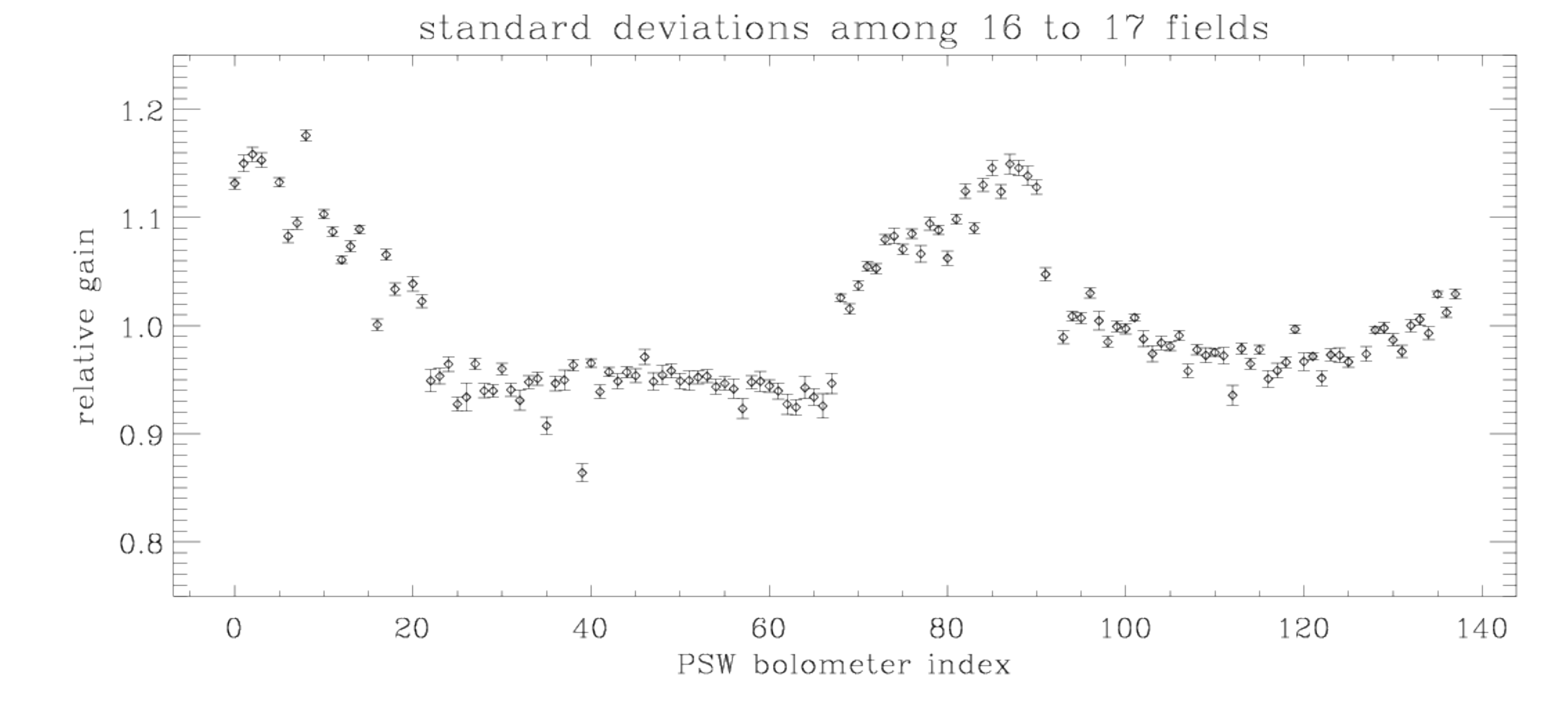}
\hspace*{-0.5cm} \includegraphics[width=4.2cm, angle=90]{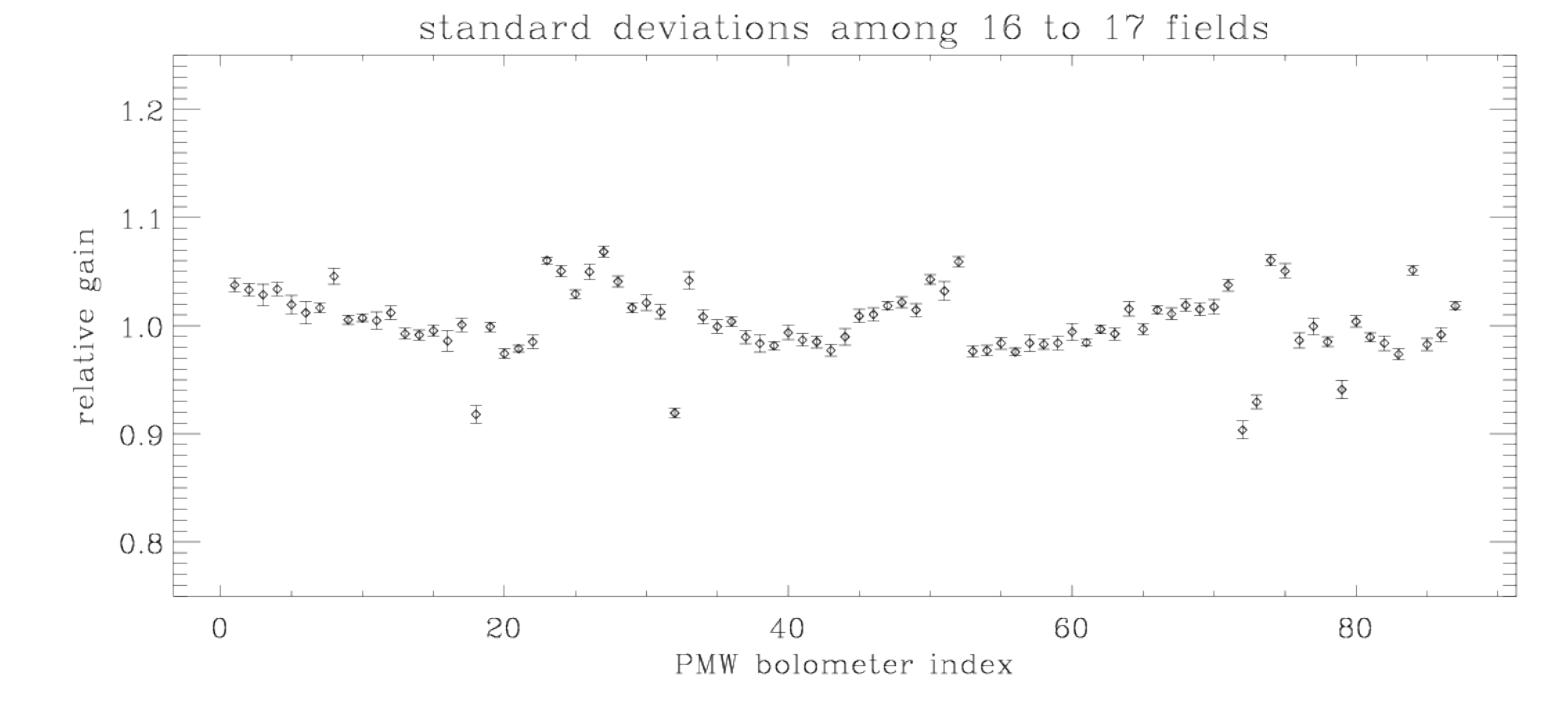} \\
\hspace*{3.2cm} \includegraphics[width=4.2cm, angle=90]{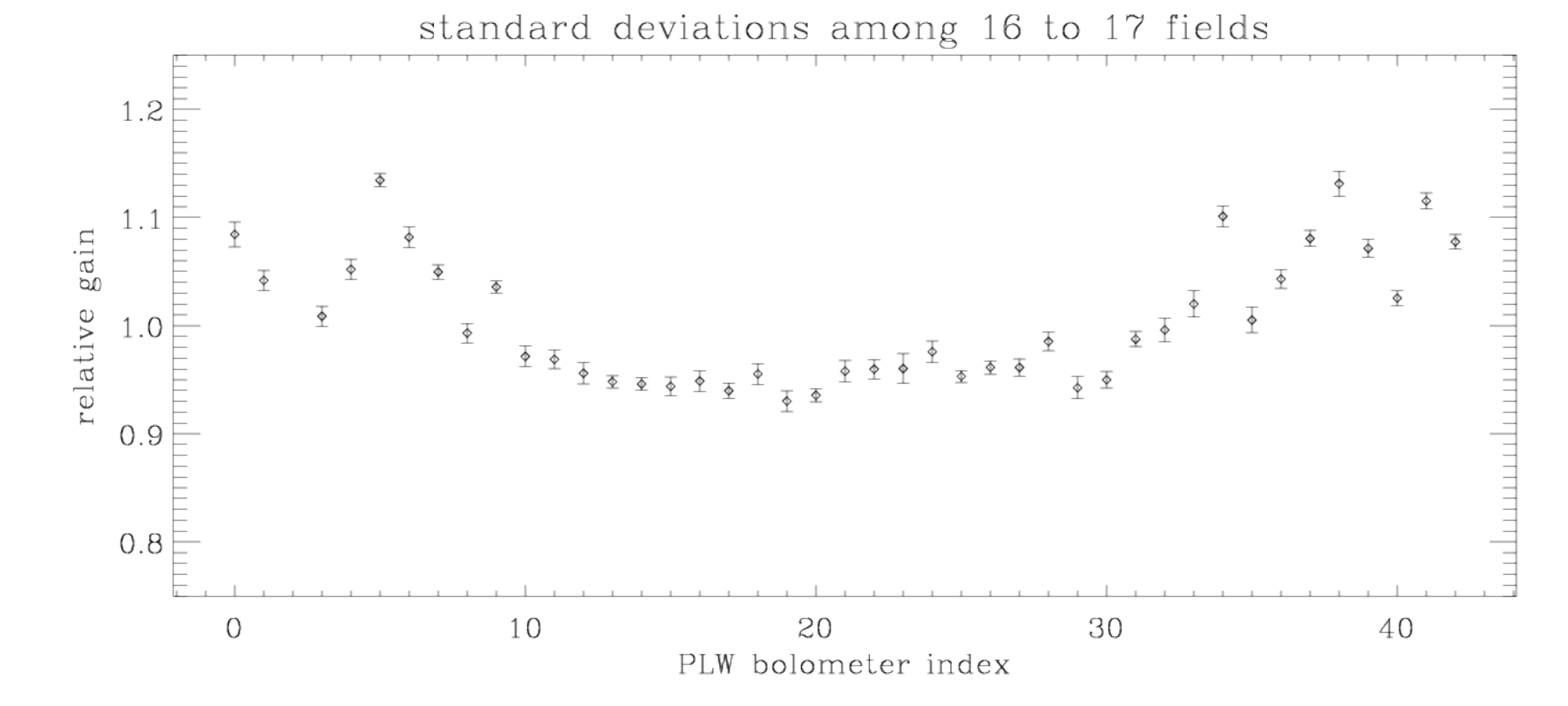}
\caption{Relative gains of valid SPIRE bolometers of the 250, 350, and 500\,$\mu$m arrays,
respectively, as a function of bolometer index, for the flux calibration set embedded
in HIPE 5 and subsequent branches. The gains were derived from the observation of the
Rosette nebula (Sect.\,\ref{spire_rosette}) and 16 other bright complex fields. The error
bars show the standard deviations among the fields.
}
\label{fig:relgains}
\end{figure*}

\begin{figure*}[!ht]
%
%
\hspace*{0.5cm} \includegraphics[width=15.cm, angle=0]{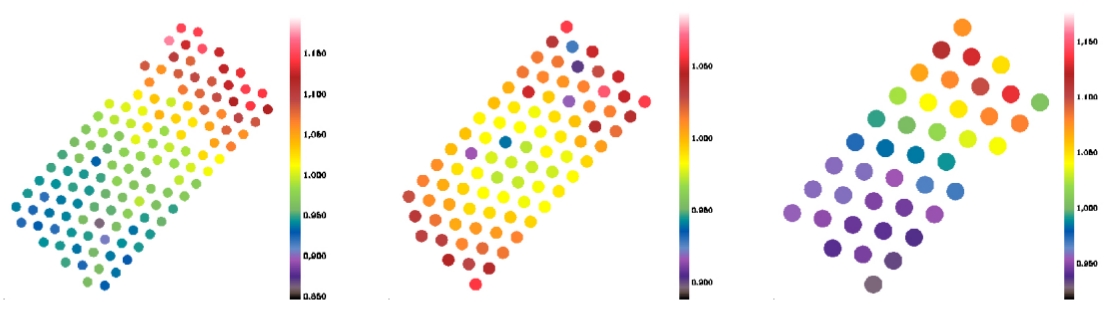}
\caption{Maps of the relative gains of valid SPIRE bolometers of the 250, 350, and 500\,$\mu$m
arrays, respectively, for the flux calibration set embedded in HIPE 5 and subsequent branches.
The arrays are oriented in the same way as when the scan direction is horizontal.
}
\label{fig:relgains_map}
\end{figure*}

The method to calibrate relative gains from science observations is the following.
Baselines are first subtracted from the data, as explained in Section \ref{baselines},
and the resulting data are projected onto a map (with a pixel size of FWHM/4).
Then, for each bolometer, we fit
its brightness series as a linear function of the series simulated from the map,
within a predefined brightness range. If the bolometer were perfectly calibrated,
the fitted slope should be equal to one. The series of the bolometer is divided
by this slope. The subtraction of baselines and division by slopes are then
iterated a few times.
The slopes are determined for each scan independently, and then averaged over
all scans before being applied.
At the end of the procedure, a relative gain is computed for each bolometer
as the product of the slopes obtained in all iterations.
The corrections are stored in files
in the calibration directory and are applied by default to SPIRE data.
Figure\,\ref{fig:relgains} shows the gain corrections as a function of bolometer
index and Figure\,\ref{fig:relgains_map} as a function of bolometer position within
the array. These gains are valid for branch 5 of the pipeline,
and for all subsequent branches as long as the flux calibration changes only by
a global factor. They are also available for branch 4, and will continue
to be updated for each new flux calibration set if necessary (if the relative
calibration is affected).

Global fluxes of extended sources are not changed by the application of the gains,
since the map used as a reference for all bolometers is that initially built with the
default flux calibration. On small scales, however, the effect may be non-negligible.
For a given source, its exact amplitude is bound to depend on the position of the source
relative to the array while it is being scanned, because the gains are a non-random
function of bolometer position (Fig.\,\ref{fig:relgains_map}). Although fluxes
are not or only slightly changed, photometric errors have been found to be significantly
decreased by the application of the gains.

\clearpage

\section{Options}
\label{options}

All the inputs and options are summarized in Table\,\ref{tab:options}.
They are described below. A full user guide, explaining how to set up the input data and
describing the structure of the output, can be found in the online distribution.

\subsection{Main options}

\noindent
- {\it /jumps\_pacs}: If PACS data are affected by brightness discontinuities
(see Sect.\,\ref{jumps}),
this option can be used to detect the discontinuities and mask the affected samples.
When {\it Scanamorphos} is run in iteractive mode, the user is required to visually check,
and confirm or infirm each detection. The very unstable rows or bolometers are automatically
detected and pre-filtered by the same module. \\
- {\it /nothermal}: If the thermal drift subtraction performed within HIPE is believed
to be correct, the corresponding step in {\it Scanamorphos} can be bypassed. \\
- {\it /noglitch}: Likewise, if the deglitching performed within HIPE is trusted, it is
possible to not perform it in {\it Scanamorphos}. \\
- {\it /nogains} (for SPIRE data only): If the HIPE version used locally does not match the
flux calibration version for which the relative gains were derived, or if the gain correction
was already performed within HIPE, the gain correction must not be performed in {\it Scanamorphos}. \\
- {\it /parallel}: For observations acquired in parallel mode, it is necessary to select the
relevant option, because the smaller sampling rate implies that the frequency intervals used
to compute the high-frequency noise amplitudes have to be adapted. \\
- {\it /galactic}: For observations of bright Galactic fields, it is highly recommended
to use the relevant option, so that no sky structure can be removed by the baselines, as explained
in Section\,\ref{baselines}. It should {\it not} be used for observations of diffuse
Galactic structures, and in general for fields where the brightness gradients induced
by the low-frequency noise are completely dominant over genuine brightness gradients of the sky.
In case of doubt, it is advised to produce maps both with and without the {\it /galactic}
option, and to compare them to decide which is the best strategy. \\
- {\it /minimap}: For fields observed in the mini-scan or small-map mode, that produces very
small maps by definition, the relevant option should be selected in order to deactivate
the destriping and average drift subtraction modules, since there are not enough resolution
elements with nominal coverage across the map to make these corrections necessary and accurate.
These corrections may also need to be deactivated (by selecting the {\it /minimap} option)
when there are only two or three scan legs per scan. As above, in case of doubt, it is advised
to produce maps both with and without the {\it /minimap} option and to compare them. \\
- {\it /nocross}: For observations in which the nominal coverage consists of scans taken in
only one direction, some adaptations are necessary, and in particular no destriping can be
performed. Note that such observations are not recommended. This option is provided to
qualitatively assess observations that are not completed yet, but is not guaranteed
to produce science-grade maps.

In visualization mode (with the {\it /visu} or {\it /debug} option), it is possible
to inspect the average drift series and choose to apply the correction or not.
When the nonlinear component of the thermal drift is negligible, and consequently the average
drift amplitude very small (with respect to the high-frequency noise), the correction may
introduce noise. In case of doubt, it is best to do the processing once with the average
drift subtraction, and a second time without this correction, to compare the final maps
and to select the best one.

\subsection{Non-default astrometry}

By default, the astrometry and spatial grid are determined from the input scans.
The user has however the latitude to change the spatial grid or request that only
a portion of the entire field be processed and mapped: \\
1) by supplying a reference fits header (in the form of an IDL string array).
In that case, the astrometry is taken entirely from the header. With this option,
only one array should be processed at a time. Note that it is usable in batch mode,
and then requires the header to be in the form of an IDL save file, containing
a variable named exactly {\it hdr\_ref}\,. \\
2) by supplying a three-element array containing the central coordinates and minimum
radius of a subfield to be excised from the data (all in degrees):
$cutout=[ra_{center}, dec_{center}, radius]$

It is also possible to specify astrometric offsets for each scan (angular distances
in arcsec) with the {\it offset\_ra\_as} and {\it offset\_dec\_as} parameters.

\begin{deluxetable}{l|l}
\tablecaption{Summary of inputs and options
\label{tab:options}}
\tabletypesize{\scriptsize}
\tablewidth{0pt}
\startdata
\hline\hline
\multicolumn{2}{c}{auxiliary files} \\
\hline
scanlist\_spire & ascii file containing the directory and file names of the input scans \\
scanlist\_pacs  & idem \\
\hline\hline
\multicolumn{2}{c}{command line keywords and parameters} \\
\hline
spire        & select to process SPIRE observations \\
pacs         & select to process PACS observations \\
nobs         & for PACS: number of distinct obsids \\
visu         & select to visualize intermediate results and stop after each major step \\
vis\_traject & select to check that scan directions are alternating correctly \\
debug        & detailed visualizations, for experts only \\
version      & select to print the software and SPIRE relative gains versions \\
jumps\_pacs  & select to detect and mask brightness discontinuities in PACS data\tablenotemark{a} \\
nothermal    & select to bypass the
short-timescale
average drift correction \\
noglitch     & select to bypass glitch masking and asteroid detection \\
nogains      & select if the gain version does not match the local flux calibration version \\
~            & or if the relative gain correction has already been applied in HIPE (SPIRE) \\
parallel     & select in case of parallel-mode observations \\
galactic     & select in case of very bright Galactic field or similar dataset \\
minimap      & select in case of mini-scan or small-map mode \\
flat         & select to force the sky background to be flat \\
nocross      & select for observations consisting of scans in a single direction or non-overlapping \\
hdr\_ref     & reference header to enforce the same astrometry \\
~            & (IDL string array or IDL save file containing a variable named ``hdr\_ref'') \\
cutout       & central RA and DEC coordinates and radius of the subfield to be processed and mapped \\
nblocks      & number of spatial blocks the field of view will be sliced into\tablenotemark{b} \\
block\_start & index of the first spatial block to be processed, when nblocks $> 1$ \\
one\_plane\_fits & select to save each plane in a separate fits file \\
offset\_ra\_as   & right ascension offsets (angular distances) to be applied to the map coordinates (one for each scan) \\
offset\_dec\_as  & declination offsets (angular distances) to be applied to the map coordinates (one for each scan) \\
\hline\hline
\multicolumn{2}{c}{inputs at the prompt} \\
\hline
output directory      & directory where final maps and intermediate variables will be stored \\
output FITS file root & full file name = file root + instrument + wavelength + block index + ``.fits'' \\
detector arrays       & choose only one for PACS and between one and three for SPIRE \\
map orientation       & binary choice: orientation of the first scan (default), or standard astronomical orientation \\
pixel size            & in arcsec, equal to FWHM/4 by default \\
turnaround data       & option to reject turnaround data from final map \\
nobs                  & for PACS: if not supplied at the command line \\
\hline
\enddata
\tablenotetext{a}{
In interactive mode (with {\it /visu}),
the user is allowed
to reject any false detections.}
\tablenotetext{b}{The option to slice the field of view should be used only in very specific cases:
see Section\,\ref{slicing}.}
\end{deluxetable}

\acknowledgements
Marc Sauvage has developed some code to format the PACS data produced by HIPE
for input to {\it Scanamorphos}, and in addition has reduced many PACS datasets up to
level 1 and shared useful insights. Pierre Chanial has provided an IDL utility
to read HIPE FITS files, that is used for both SPIRE and PACS. Their generous help
and computer support from Gilles Missonnier are gratefully acknowledged.
I also thank Herv\'e Aussel and Marc Sauvage for providing the noise measurements
used in the PACS simulations. Excellent IDL programming advice was found at:
{\it http://www.idlcoyote.com/documents/tips.php}\,.
This work indirectly benefited from regular interaction within the Scan Map
Pipeline Validation Group of the SPIRE Instrument Control Center (ICC).
The members of both the PACS and SPIRE ICC are here collectively thanked
for their dedication to the instruments.
Finally, I thank the colleagues who allowed me to conduct tests
on various observing configurations, using data from their programs, who
reduced PACS data up to level 1 before ingestion by {\it Scanamorphos},
and/or who reported bugs and suggested useful improvements in early versions
of the code: Nicolas Billot, Loren Anderson, Ekkehard Wieprecht and Martin
Hennemann, as well as many others who sent useful reports at later stages. \\
SPIRE has been developed by a consortium of institutes led by Cardiff University (UK)
and including Univ. Lethbridge (Canada); NAOC (China); CEA, LAM (France);
IFSI, Univ. Padua (Italy); IAC (Spain); Stockholm Observatory (Sweden);
Imperial College London, RAL, UCL-MSSL, UKATC, Univ. Sussex (UK);
and Caltech, JPL, NHSC, Univ. Colorado (USA).
This development has been supported by national funding agencies: CSA (Canada);
NAOC (China); CEA, CNES, CNRS (France); ASI (Italy); MCINN (Spain);
SNSB (Sweden); STFC, UKSA (UK); and NASA (USA). \\
PACS has been developed by a consortium of institutes led by MPE (Germany)
and including UVIE (Austria); KUL, CSL, IMEC (Belgium); CEA, OAMP (France);
MPIA (Germany); IFSI, OAP/AOT, OAA/CAISMI, LENS, SISSA (Italy); IAC (Spain).
This development has been supported by the funding agencies BMVIT (Austria),
ESA-PRODEX (Belgium), CEA/CNES (France), DLR (Germany), ASI (Italy), and CICT/MCT (Spain). \\
This research has made use of HIPE, a joint development by the
Herschel Science Ground Segment Consortium, consisting of ESA, the NASA
Herschel Science Center, and the HIFI, PACS and SPIRE consortia, and of
the NASA/IPAC Extragalactic Database (NED) which is operated by the Jet
Propulsion Laboratory, Caltech, under contract with NASA.

\end{document}